\documentclass[a4paper,11pt,x11names]{article}
\pdfoutput=1
\usepackage{jheppub}
\usepackage{graphicx}
\usepackage{amsmath}
\usepackage{amsfonts}
\usepackage{mathrsfs}
\usepackage{amssymb}
\usepackage{amstext}
\usepackage{placeins}
\usepackage{slashed}
\usepackage{braket}
\usepackage{color}
\usepackage[colorlinks=true,linkcolor=blue]{hyperref}
\usepackage{verbatimbox}
\usepackage[x11names]{xcolor}
\usepackage{bm}
\usepackage{empheq}
\usepackage[super]{nth}
\usepackage{caption}
\usepackage{subcaption}
\usepackage{tabularx}
\usepackage{wrapfig}
\usepackage{float}

\usepackage{relsize}
\DeclareFontFamily{OT1}{pzc}{}
\DeclareFontShape{OT1}{pzc}{m}{it}{<-> s * [1.10] pzcmi7t}{}
\DeclareMathAlphabet{\mathpzc}{OT1}{pzc}{m}{it}

\newif\ifcomments
\commentsfalse
\newcommand{\sus}{\ensuremath{\chi_\text{top}}}

\newcommand{\ab}{\ensuremath{\text{d}}}

\newcommand{\tr}{\ensuremath{\text{tr}}}
\newcommand{\Tr}{\ensuremath{\text{Tr}^{\,}}}

\newcommand{\e}[1]{\ensuremath{\hat{e}_{#1}}}

\DeclareMathOperator{\Lag}{\mathcal{L}}

\DeclareMathOperator{\is}{\, = \,}
\newcommand{\dash}{$\,$-$\,$}

\newcommand\ncoverline[1]{\mkern1mu\overline{\mkern-1mu#1\mkern-1mu}\mkern1mu}

\newcommand{\istr}{\ensuremath{\, =_\tr \,}}
\newcommand{\isnotr}{\ensuremath{\,=_{\setminus\tr}\,}}

\newcommand{\Det}{\ensuremath{\det{}^{\!}{}'}}

\newcommand{\spt}{\ensuremath{\mathbb{R}^3\times S^1_{\text{rad}{}^{\,}={}^{\,1\!}/{}_{2\pi}}}}

\newcommand{\intspt}{\ensuremath{\int^1\!\text{d}^4 x\,}}

\newcommand{\fferm}{\ensuremath{\mathpzc{f}_{\,\text{ferm}}}}

\newcommand{\gammas}{\ensuremath{\gamma_{\text{s,}-}}}

\title{Finite -- $T$ topological Susceptibility with heavy Quarks}
\author{Bruno H\"ogl, Guy D. Moore}
\affiliation{Institut f\"ur Kernphysik, Technische Universit\"at Darmstadt\\
Schlossgartenstra{\ss}e 2, D-64289 Darmstadt, Germany}
\emailAdd{bruno.hoegl@web.de,guy.moore@physik.tu-darmstadt.de}

\abstract{
Axion cosmology needs the QCD topological susceptibility between 400 and $1100\,\text{MeV}$.
In this range the bottom quark is inconvenient to include in lattice simulations, but not heavy enough to ignore.
We estimate its effect on the susceptibility by computing the ratio of the 4\dash quark susceptibility and the $4+1$\dash quark susceptibility in the caloron gas approximation.
We do so by computing small\dash mass and large\dash mass expansions of the finite\dash mass and -temperature fluctuation determinant and connecting them with a Pad\'e approximant.
}


\date{\today}

\graphicspath{{images/}}

\begin{document}
\begin{myverbbox}{\link}git.rwth-aachen.de/qcd/ancillary_files_finite_t_top_suscep_heavy_quarks\end{myverbbox}
\maketitle

\section{Introduction and main Result}
\label{sec:intro}
An important feature of non\dash abelian gauge theories in general and the theory of the strong interactions, Quantum Chromodynamics (QCD), in particular is the presence of topologically non\dash trivial gauge field configurations.
Regarding QCD, this makes the true $\theta$\dash vacuum a superposition of infinitely many topologically\dash different gluon vacuum states which differ only by their topological properties. These otherwise distinct vacua are connected by the presence of topological configurations.
Semi\dash classically, these are the \textsl{instantons} at zero temperature and the \textsl{calorons} for finite $T>0$.
These local, topological gauge field configurations change the global, topological properties of gluonic vacua \cite{BPST_instanton, calorons_exist}.

The caloron's topological charge density
\begin{equation}
\label{eq:top_charge_density}
    q(x) \is \frac{1}{32\pi^2} \epsilon^{\mu\nu\alpha\beta} \tr \!\left( G^{\mu\nu\!}(x)^{\,} G^{\alpha\beta\!}(x)\right) \is \frac{1}{16\pi^2}\tr \!\left( G^{\mu\nu\!}(x)^{\,} \widetilde{G}^{\mu\nu}(x)\right)
\end{equation}
is given by the its (dual) field strength $G$ ($\widetilde{G}$). Integrating this density over Euclidean spacetime with periodic time boundaries of extent $\beta\is T^{-1}$ (a compact spacetime without boundary), as arises in the standard Euclidean thermal path integral \cite{Kapusta:2006pm,Bellac:2011kqa}, yields an integer called the caloron number or topological charge $\mathbb{Z}\in n \is \int_0^\beta\! \ab \tau\!\int_{\mathbb{R}^3 \!}\ab^3x\, q(\vec{x},\tau)$ \cite{qcd_at_finite_T}.

The physical $\theta$\dash vacuum and the presence of calorons require one to add the topological term
$\Lag_{\text{top}} \is -i\theta q(x)$, $\theta\in [-\pi,\pi]$
to the QCD Lagrangian.
Because $q\propto \vec{E}\cdot\vec{B}$, the pure phase $\Lag_{\text{top}}$ gives rise to a violation of the $\mathcal{CP}\,$- or $\mathcal{T}$\dash symmetry in the path integral.
From studying the neutron electric dipole moment, for example, $\mathcal{CP}$\dash violating strong interaction effects are known experimentally to be extremely small, setting an upper bound \mbox{$-1.52(71)\cdot 10^{-18}\,\theta\,e\cdot\text{m} \leq 1.8\cdot 10^{-28}\,e\cdot \text{m}$,} i.e., $|\theta|\leq 1.2\cdot 10^{-10}$ \cite{neutron_dipole_moment_theo, neutron_dipole_moment_exp}.
This especially tight bound on $\theta$ is known as the strong $\mathcal{CP}$ problem: since neither $\mathcal{CP}$ nor $\mathcal{T}$ are fundamental symmetries of nature, there is no fundamental reason for this fine\dash tuning of $\theta$.

A very promising solution of the strong $\mathcal{CP}$ problem is the extension of the standard model in terms of the axion \cite{axion_origin_1, axion_origin_2}.
For this, a high temperature $U(1)_\text{Peccei Quinn}$\dash symmetry \cite{Peccei_Quinn_1, Peccei_Quinn_2} is introduced, which is spontaneously broken at some very high energy scale $10^8\,\text{GeV}\lesssim f_a \lesssim 10^{17}\,\text{GeV}$ \cite{axion_dm_what_why, axion_precisely, axion_cosmo_f_range}. The axion $a$ is the associated Nambu\dash Goldstone boson. At low temperatures $T\lesssim \Lambda_\text{QCD}$, the axion obtains a very weak coupling $\propto f_a^{-1}$ (``invisible axion'') to gluons via topological terms and settles into its vacuum expectation value $\langle a\rangle \is -\theta f_a$, thus also picking up a very small mass $m_a\propto f_a^{-1}$.
Thereby, the axion modifies $\theta$ as $\theta\rightarrow \theta_\text{eff}\is\theta + \frac{\langle a\rangle}{f_a}\is 0$.
This means that $\mathcal{CP}$\dash conservation of QCD is dynamically ensured at the cost of introducing the axion into the theory.

The strong $f_a^{-1}$\dash suppression of all axion interactions makes it a promising dark matter candidate%
\footnote{As of yet, the axion is still a hypothetical particle lacking verification by observation, but many experiments focusing on axion dark matter are ongoing or planned \cite{axion_experiments}.}
\cite{axion_book, axion_landscape, axion_cosmo, diga_and_axion}.
Determining its mass and cosmological abundance requires understanding the temperature dependence of the topological susceptibility
\begin{equation}
\label{eq:top_suscep_def}
    \sus(T) \is \int\!\ab^4 x\,\langle q(x) q(0) \rangle_{T} \is -\frac{1}{\beta V}\left.\frac{\partial^2\ln\!\big(Z(\theta)\big)}{\partial\theta^2}\right|_{\theta^{\,} =^{\,} 0}\,,
\end{equation}
where $Z$ is the partition function and $\beta V$ is the volume of spacetime. Indeed, the axion mass depends on $\sus$ via
\begin{equation}
\label{eq:axion_mass}
    m_a(T) \is \frac{\sqrt{\sus(T)}}{f_a}\,.
\end{equation}

For $T\is 0$, chiral perturbation theory gives a reliable tool for computing $\sus$ and precise results are available:
$\sqrt[4]{\sus(0)}\is (75.5\pm 0.5)\,\text{MeV}$
\cite{axion_mass_1,axion_chiral_perturbation,axion_precisely,top_suscep_lattice}.
As a result, the axion mass is $m_a(0)\is (5.69 \pm 0.05)\,\mu\text{eV}\left(\frac{10^{12}\,\text{GeV}}{f_a}\right)$ \cite{axion_mass_1}.

Evaluating the topological susceptibility well above the QCD crossover temperature $T_{c} \simeq 155\,\text{MeV}$ \cite{Borsanyi:2013bia,HotQCD:2014kol} is more challenging, but axion cosmology requires precise results:
\cite{axion_mass_2} shows that the cosmological history of the axion depends critically on $\sus(T)$ in the temperature range $400\,\text{MeV}\lesssim T \lesssim 1.1\,\text{GeV}$.
In this temperature range, $\sus$ is small and dominated by isolated topological objects.
Semiclassically, these would be the \textsl{Harrington\dash Shepard} (HS) \textsl{\mbox{(anti$\,$-)} calorons} \cite{HS_caloron} with $n\is \pm 1$. Unfortunately, the HS caloron density cannot reliably be determined perturbatively, requiring a lattice investigation instead.
Both existing lattice investigations in this temperature range \cite{top_suscep_lattice} and any future investigations using topology reweighting techniques \cite{Jahn:2018dke,Jahn:2020oqf} will be performed in so\dash called $2+1+1$ simulations, meaning that the up, down, strange, and charm quarks are included, but the bottom quark is not.
Indeed, we are not aware of any lattice simulations at physical quark masses which include dynamical bottom quarks - and adding them to the existing simulation framework would require very significant additional work.
At the highest temperatures mentioned above, however, it is not clear that the bottom quark can be considered heavy compared to the thermal scale; it may influence the topological susceptibility.

Therefore, we address the question of how adding a dynamical $b$ quark alters the topological susceptibility of finite\dash temperature $2+1+1+1$ theory compared to the $2+1+1$ case accessible to lattice QCD.
The two theories should be compared keeping the infrared physics fixed, e.g., at the same value of the strong coupling in the $2+1+1$\dash flavor effective infrared theory.
We do this by computing the ratio of the topological susceptibility in the dilute caloron gas model with a $b$ quark at the physical mass, and the susceptibility in the same model but with the $b$ quark taken to be asymptotically heavy.
We match the coupling so that the two theories coincide in the IR.
The assumption is then that the effect of a heavy quark on a caloron suitably captures its effect on the somewhat messier topological objects which matter at physical coupling values.

\begin{figure}
    \centering
    \includegraphics[width=\textwidth]{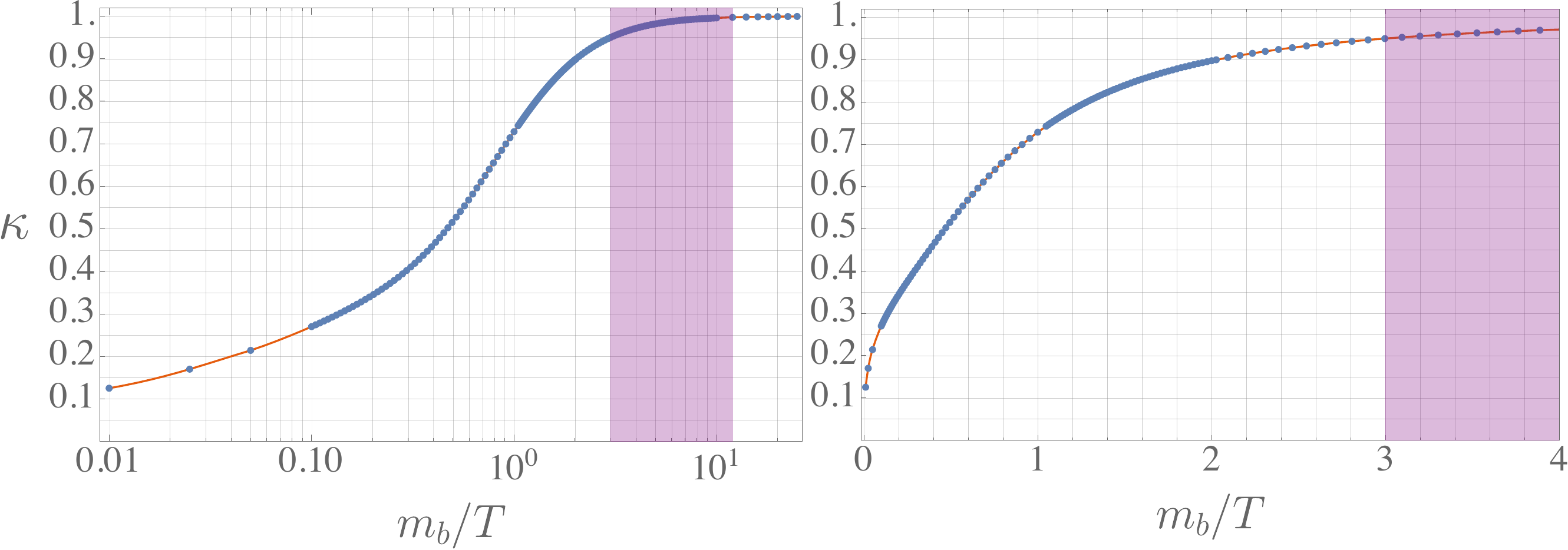}
    \caption{The ratio of topological susceptibilities $\kappa(m_b/T, 4, 1, 3)$ (\ref{eq:top_suscep_ratio}) comparing a theory with four light and a physical $b$ quark in $SU(3)$ gauge theory to lattice QCD, where the $b$ quark is asymptotically heavy. Due to the modification of the running coupling (\ref{eq:coupling_modified}), $\kappa$ depends only on the physical quark mass $m_b$. The interesting mass range $3\lesssim m_b/T \lesssim 12$, marked in purple, is chosen to be slightly wider than the physically relevant range, which is determined by the
    temperature range \mbox{$400\,\text{MeV}\lesssim T\lesssim 1.1\,\text{GeV}$}
    and the bottom mass $m_{b}\approx 4.2\,\text{GeV}$ \cite{particle_data}.
    (This is the $\overline{\mathrm{MS}}$ mass at the renormalization point $\overline{\mu} = m_b$, which is the quantity relevant in a perturbative calculation. 
    The pole mass is somewhat heavier.)}
    \label{fig:suscep_ratios}
\end{figure}

\hyperref[fig:suscep_ratios]{Figure \ref{fig:suscep_ratios}} shows our main result concerning this question.
We show the temperature\dash dependent ratio $\kappa\is \frac{\sus(m_b/T)}{\sus(m_{b\text{, asy}}/T)}$ of the topological susceptibilities for theories with physical and asymptotic $b$\dash masses, respectively.
We see that for the physically interesting temperature range $400\,\text{MeV}\lesssim T \lesssim 1.1\,\text{GeV}$ (i.e., the temperature most important for the cosmological history of axions, see \cite{axion_mass_2}) and thus the mass range $4\lesssim \frac{m_b}{T} \lesssim 10$, the difference between the physically heavy $b$ and its asymptotically heavy counterpart used in lattice QCD is $\,\lesssim 5\%$.
Only for high temperatures with $\frac{m_b}{T}\lesssim 2$ do we see an appreciable ($\gtrsim 10\%$) difference between the topological susceptibilities of finite\dash temperature $2+1+1+1$ theory (including a dynamical $b$ quark) and the $2+1+1$ case of lattice QCD.

The remaining article shows our derivation of this and other results. It is organized as follows: in the \hyperref[sec:prelim]{preliminaries} we give a short overview over calorons and how they enter into QCD, how the partition function and consequently the topological susceptibility are related to the caloron density, and how we incorporate heavy quarks at finite temperature.
In \hyperref[sec:strategy]{section \ref{sec:strategy}} we outline in more detail our strategy for dealing with different quark mass regimes and the interpolation between them.
In \hyperref[sec:smallmass]{section \ref{sec:smallmass}} we then calculate the caloron density for light, but non\dash vanishing quark masses and in \hyperref[sec:largemass]{section \ref{sec:largemass}} for heavy quarks.
In \hyperref[sec:Pade]{section \ref{sec:Pade}} we perform the interpolation between these mass regimes to obtain the caloron density for general quark masses and finally calculate the topological susceptibility ratio.
At \href{https://git.rwth-aachen.de/qcd/ancillary_files_finite_t_top_suscep_heavy_quarks}{\link} we\linebreak provide our data files and numerical calculations.\newline

\textbf{Conventions: }We employ a system of natural units with $c\is\hbar\is k_\text{B}\is 1$.
We work in Euclidean, finite\dash $T$ spacetime $\spt$ with coordinates $x^\mu \is (\vec{x},\tau )$, $\mu\is 1,...,4$ and $i\is 1,2,3$. 
In the remaining sections, we will rescale all length scales, masses, energies, etc.\ by appropriate factors of $\beta\is T^{-1}$ to be dimensionless, e.g., $m\;\widehat{=}\;\frac{m}{T}$ is the dimensionful mass rescaled by the temperature.
By $[q]$ we denote the mass\dash dimension of the corresponding dimensionful quantities $q$.
Furthermore, when there is a caloron present, we choose coordinates for $\spt$ so that the caloron is at the center, i.e., the caloron center is located at $(\vec{x},\tau)\is (\vec{0},0)$.
All properties tied to the appropriately dimensionless, Euclidean $T\is 0$\dash spacetime $\mathbb{R}^4$ are denoted as barred, e.g., $\ncoverline{x}^\mu\is (\vec{\ncoverline{x}},\ncoverline{t})$.
The $T>0$ and $T\is 0$ radii are defined as $r\is \sqrt{x^ix^i}$ and $\ncoverline{r}\is \sqrt{\ncoverline{x}^\mu\ncoverline{x}^\mu}$, respectively.
The unit vector in the $\mu$\dash direction is written as $\e{\mu }$.
We abbreviate the spacetime integral $\intspt \is \int_0^1\!\ab\tau\int_{\mathbb{R}^3\!}\ab x^3$ and the operator trace $\Tr (\cdot)\is\intspt\,\tr_\text{Dirac, color, etc.}(\cdot)$.
We consider $SU(N)$ gauge theory and $N_{\! f}$ quark flavors and specify $N$ and $N_{\! f}$ only when computing concrete results.
We employ the geometrical normalization of the gauge covariant derivative $D\is \partial -iA$ and occasionally use the short\dash hand notation: $D^\mu q^{\nu_{1}...\nu_{n}} \is q^{\nu_1...\nu_n;\mu}$.
The four Pauli matrices are denoted as $\sigma^\mu \is (\vec{\sigma},i)$ with roman indices $\sigma^a$ running from 1 to 3. We abbreviate $\vec{x}\cdot\vec{\sigma}\is r\, \e{r}(x)\cdot\vec{\sigma} \is r\,\sigma^r(\theta,\varphi)$.
We choose anti\dash Hermitian Euclidean $\gamma$\dash matrices ${\gamma^\mu}^\dagger \is -\gamma^\mu$ which satisfy the Clifford algebra $\lbrace\gamma^\mu,\gamma^\nu\rbrace\is -2\delta^{\mu\nu}$.
To avoid confusion with the gamma function, we denote the quantum effective action by $\widetilde{\Gamma}$.
$\gamma_\text{E}\is 0.57721...$ is the Euler\dash Mascheroni constant.
Lastly, we include 0 in the natural numbers $\mathbb{N}\is \lbrace 0,1,2,...\rbrace$ (mainly used for summation).

\section{Preliminaries}
\label{sec:prelim}

As stated in \hyperref[sec:intro]{section \ref{sec:intro}}, we are interested in obtaining the topological susceptibility (\ref{eq:top_suscep_def}) $\sus(m_f,T)\is -\frac{1}{V_4}\left.\frac{\partial^2\ln(Z(\theta))}{\partial\theta^2}\right|_{\theta{\,} =^{\,} 0}$, with $V_4\is \text{vol}\!\left(\spt\right)$ the volume of dimensionless spacetime (the spacetime $\spt$ is periodic in the temperature direction with fermions experiencing anti\dash periodic boundary conditions.), for $2+1+1+1$ theory at finite temperature including a dynamical $b$ quark.

As we discuss later, it suffices to consider only the aforementioned HS calorons \cite{HS_caloron}:
\begin{align}
    & A_\text{HS} \is -\ncoverline{\eta}^{\,a\mu\nu}\partial^{\nu\!}\ln\!\big(\phi(x)\big)\frac{\sigma^a}{2}\quad\text{with} \label{eq:HS_caloron}\\
    & \phi(x)\is 1 + \frac{\pi\varrho^2 \sinh(2\pi r)}{r(\cosh(2\pi r)-\cos(2\pi\tau)} \label{eq:phi_function}
\end{align}
given in singular gauge (i.e., all topological information is ``stored'' at the caloron center). Here $\ncoverline{\eta}$ is the anti\dash 't Hooft symbol \cite{tHooft, tHooft_erratum} and $\varrho$ gives the ($\beta$\dash rescaled) size of the caloron placed at $(\vec{0},0)$.
For the HS anti\dash caloron the anti\dash 't Hooft symbol $\ncoverline{\eta}$ in (\ref{eq:HS_caloron}) is replaced by the 't Hooft symbol $\eta$.

Calorons can be written as sums of infinitely many, uniformly spaced instantons placed at $\mathbb{R}^4\ni 0+j\e{4}$, $j\in\mathbb{Z}$. In particular, HS calorons can be written as a sum over infinitely many \textsl{Belavin\dash Polyakov\dash Schwartz\dash Tyupkin} (BPST) \textsl{instantons}.\footnote{BPST instantons \cite{BPST_instanton} are of the same form as (\ref{eq:HS_caloron}), but with $\phi_\text{BPST}(\ncoverline{x},\ncoverline{z})\is 1+\frac{\varrho^2}{(\ncoverline{x}-\ncoverline{z})^2}$, where $\ncoverline{z}$ is the instanton center. The HS caloron is then constructed by $\phi(x)\is \sum_{j^{\,}\in^{\,}\mathbb{Z}}\phi_\text{BPST}\!\left(\ncoverline{x},0+j\e{4}\right)$.}
Instantons approach pure gauge form at infinity $\lim_{{}^{\,}\ncoverline{r}^{\,}\rightarrow{\,}\infty}A_\text{inst}\is i\Omega\partial\Omega^{-1}\in\mathfrak{su}(N)$ with $\Omega\in SU(N)$, which serves to restrict the gauge configurations $\lim_{{}^{\,}\ncoverline{r}^{\,}\rightarrow{\,}\infty}\Omega\is 1$ and compactify $\mathbb{R}^4$ to an infinite 3\dash sphere $S^3$. According to a theorem by Bott \cite{Bott}, the ``topologically active'' part of such maps $\Omega:S^3\rightarrow SU(N)$ (or any simple Lie group) is only a subgroup $SU(2)\subset SU(N)$.
Therefore, HS calorons are fully determined by the $\mathfrak{su}(2)$\dash object (\ref{eq:HS_caloron}), embedded in an $\mathfrak{su}(N)$\dash matrix
\begin{equation}
\label{eq:caloron_embedding_SU(N)}
A_{\text{HS, }\mathfrak{su}(N)} \is \left(\begin{array}{c|c} A_\text{HS} & 0_{2\times(N-2)}\\ \hline 0_{(N-2)\times 2} & 0_{(N-2)\times (N-2)} \\ \end{array}\right),
\end{equation}
together with two residual symmetry groups: one group of global (``rigid'') $SU(2)$\dash transfor- mations acting on $A_\text{HS}$ that leave the caloron invariant and one group of $\frac{SU(N)}{SU(N-2)\times U(1)}$\dash transformations of $A_{\text{HS, }\mathfrak{su}(N)}$ that only change the embedding \cite{inst_zero_modes, lectures}.

In the limit of small distances $|(\vec{x},\tau)|\ll 1$ the HS caloron (\ref{eq:HS_caloron}) takes the form of a BPST instanton
\begin{equation}
\label{eq:HS_small_R}
A^\mu_\text{HS}\stackrel{|x|^{\,} \ll^{\,} 1}{\cong}\ncoverline{\eta}{}^{\,a\mu\nu}\frac{2\tilde{\varrho}^2}{x^2} \frac{x^\nu}{x^2+\tilde{\varrho}^2}\frac{\sigma^a}{2}\Big(1+\mathcal{O}(|x|^4)\Big)
\end{equation}
with modified, reduced size
\begin{equation}
\label{eq:caloron_mod_size}
\tilde{\varrho}\is\frac{\varrho}{\sqrt{1+\frac{\pi^2\varrho^2}{3}}}\,.
\end{equation}
This means that on length scales much smaller than the temperature/periodicity scale the caloron is identical to an instanton with modified size $\tilde{\rho}$ and the actual periodicity of $\spt$ is concealed in the far distance \cite{qcd_at_finite_T}.

(Anti$\,$-)Calorons are (anti$\,$-)self\dash dual ($\widetilde{G}^{\mu\nu}G^{\mu\nu}\is \pm G^{\mu\nu}$) solutions to the classical equations of motion $D^\mu G^{\mu\nu}\is 0$ and can be treated as a classical, stationary background for quantization, i.e, the gauge, quark, and Faddeev\dash Popov ghost fields of the QCD\dash system are of the form $A^\mu \is A^\mu_{\text{HS, }\mathfrak{su}(N)} + A^\mu_\text{quantum}$, $\psi\is\psi_\text{quantum}$, and $c\is c_\text{quantum}$, respectively.
One then also enforces the background gauge condition $D^\mu(A_{\text{HS, }\mathfrak{su}(N)})A^\mu_\text{qm} \is 0$ to eliminate residual gauge freedom in the fluctuations \cite{tHooft, tHooft_erratum}.

Consider the contribution to the partition function from configurations with topology~1, $Z_1(\theta)$, due to a single topological object.
Using Laplace's method for the integration with respect to the quantum fluctuations, the regularized, vacuum\dash normalized partition function $Z_1$ and effective action $\widetilde{\Gamma}$ up to first loop order in the quantum fluctuations $A_\text{qm}$, $\psi_\text{qm}$, $c_\text{qm}$ in a classical HS caloron background - i.e., up to $\mathcal{O}\big((\text{qm. fluct.})^2\big)$ -, including the topological term (\ref{eq:top_charge_density}), are \cite{tHooft, tHooft_erratum, callan_dashen_gross, qcd_at_finite_T}
\begin{align}
& Z_1(\theta)\is e^{i\theta}\,V_4 \, \mathpzc{D} \is \exp\!\big(-\widetilde{\Gamma}(\theta)\big)\,,\quad \mathpzc{D}\is \int_0^\infty \!\!\ab\varrho\,\mathpzc{d}(\varrho) \is \int_0^\infty \!\!\ab\varrho\,e^{-\gamma(\varrho)}\,, \label{eq:partitioN_func_cal_density}\\
& \begin{aligned}
\mathpzc{d}(m_f,\varrho,\lambda) \is & \frac{2^{2-2N}\,e^{-\frac{8\pi^2}{g^2}}}{\pi^2(N-1)!(N-2)!}\left(\frac{8\pi^2}{g^2}\right)^{2N}\varrho^{4N-5}\lambda^{4N}\sqrt{\frac{\Det\mathfrak{M}_A(\lambda)\,\Det\mathfrak{M}_{A,\,0}}{\Det\mathfrak{M}_A\,\Det\mathfrak{M}_{A,\,0}(\lambda)}}\,\times \\
& \times \frac{\det\mathfrak{M}_\text{gh}\,\det\mathfrak{M}_{\text{gh},\,0}(\lambda)}{\det\mathfrak{M}_\text{gh}(\lambda)\,\det\mathfrak{M}_{\text{gh},\,0}}\,\prod_{f}\frac{m_f}{\lambda}\left(\frac{\det(-D_-^2+m^2_f)\,\det(-\partial_-^2+\lambda^2)}{\det(-D_-^2+\lambda^2)\,\det(-\partial_-^2+m^2_f)}\right)^2,
\end{aligned} \label{eq:cal_density_general}
\end{align}
where $\mathpzc{D}$ and $\mathpzc{d}(\varrho)$ are the caloron density and caloron density/likelihood per size, respectively, and $\gamma(\varrho)$ is the negative logarithmic caloron density.\footnote{denoted by $\gamma$ since $\exp\!\big(-\widetilde{\Gamma}(\theta)\big)\is e^{i\theta}\,V_4\,\int_0^\infty\!\ab\varrho\,\exp(-\gamma)$}
The caloron density $\mathpzc{D}$ can be understood as the mean\dash squared topology per unit volume, and $\mathpzc{d}$ its integrand when $\mathpzc{D}$ is expressed as an integral over caloron size $\varrho$.
The gluon and ghost field differential operators are $\mathfrak{M}^{\,\mu\nu}_A\is \Big(-D^2_+\big(A^\mu_{\text{HS, }\mathfrak{su}(N)}\big)\delta^{\mu\nu}-2G^{\mu\nu}_{\mathfrak{su}(N)}\Big)_\text{adj}$ and $\mathfrak{M}_\text{gh}\is -D^2_{+\text{, adj}}\big(A^\mu_{\text{HS, }\mathfrak{su}(N)}\big)$, respectively, where ``adj'' stands for the adjoint representation.
The subscript ``$0$'' denotes the free, vacuum case without a caloron background and the subscripts ``$+/-$'' denote $\tau$\dash periodicity/anti\dash periodicity of the respective eigenfunctions.
The fermionic differential operator $i\slashed{D}\big(A^\mu_{\text{HS, }\mathfrak{su}(N)}\big)+m_f$ was translated to a Klein\dash Gordon operator $-D^2_-+m_f^2$ of an anti\dash periodic scalar in the determinant, which is unique to (anti\dash)self dual gauge fields \cite{propagators_pseudoparticle_fields, fermion_prop_with_mass}.
$\lambda$ is the renormalization energy scale.

Since the BPST and thus the HS solution to the self\dash duality condition as well as the general \textsl{Atiyah}\dash \textsl{Drinfeld}\dash \textsl{Hitchin}\dash \textsl{Manin} (ADHM) construction of such solutions \cite{ADHM} are all limited to 4\dash dimensional spacetime, dimensional regularization methods are not available in a theory with a classical caloron background, as there is no known expression for calorons in $4-2\varepsilon$\dash dimensional spacetime.
Instead, regularization is achieved by employing the Pauli\dash Villars method which introduces additional copies of all quantum fields with large mass $\lambda$ and minimal coupling only to the background. This means adding mass terms $\mathfrak{M}_{A,\,\text{gh}}(\lambda)\is \mathfrak{M}_{A,\,\text{gh}}+\lambda^2$.
An alternative interpretation is that we are comparing the topology density to the topology density at an extremely large quark mass $\lambda$.
Ultraviolet divergences cancel in the difference between these two theories.

In \cite{tHooft, tHooft_erratum} the determinant ratios in (\ref{eq:cal_density_general}) were obtained for the simplified case of zero temperature and vanishingly light fermions $0<m\varrho\ll 1$ and the connection to dimensional regularization was established. The instanton density for $N_{\! f}$ vanishingly light quarks reads
\begin{align}
& \mathpzc{d} \is \frac{2 e^{-\alpha(1)+4\alpha\left(\frac{1}{2}\right)+\ln 2-N\left(2\alpha\left(\frac{1}{2}\right)+2\ln 2\right)+2N_{\! f}\alpha\left(\frac{1}{2}\right)}}{\pi^2(N-1)!(N-2)!}\left(\frac{8\pi^2}{g^2}\right)^{\!2N}\!e^{-\frac{8\pi^2}{g^2(1/\varrho)}}\,\frac{\prod_{f}m_f\varrho}{\varrho^5} 
\label{eq:cal_density_T=m=0} \\
&\text{with the running coupling }\quad\frac{8\pi^2}{g^2\big(1/\varrho)} \is -\frac{1}{3}\ln(\lambda\varrho)(11N-2N_{\! f})\,,\label{eq:coupling}
\end{align}
$\alpha\!\left(\frac{1}{2}\right)\approx 0.145873$, and $\alpha(1) \approx 0.443307$. Here $m_f$, $\lambda$, and $\varrho$ can be understood as the non\dash rescaled, dimensionful quantities at $T\is 0$.
Usually, the factor $g^{-4N}$ in (\ref{eq:cal_density_T=m=0}) is ``manually'' replaced by the running coupling at scale $\varrho^{-1}$ (\ref{eq:coupling}), which is identified in the exponent during the calculation of (\ref{eq:cal_density_T=m=0}). However, as we argue below (\ref{eq:correction_factor_massless_quarks}) for $T\is 0$ and as we show in \hyperref[fig:density_ratios]{figure \ref{fig:density_ratios}} for $T>0$, this is a higher order effect and we thus choose to neglect this correction when calculating the topological susceptibility, since $\sus$ in its known form is itself based on calculations only up to $\mathcal{O}\big((\text{qm. fluct.})^2\big)$, i.e., 1\dash loop order (cf. (\ref{eq:cal_density_general})).

In order to extend (\ref{eq:cal_density_T=m=0}) to the physically interesting case of $T>0$ and $m_f\varrho\not\ll 1$ as well as $m_f\not\ll 1$, the determinant relations must be reevaluated for these new parameters. For that the determinant relations in (\ref{eq:cal_density_general}) are split into the parts with $T\is 0$ and $m_f\varrho \ll 1$ as well as a correction factor $\mathpzc{f}$:
\begin{equation}
\label{eq:caloron_density_with_correction_factor}
\mathpzc{d}(m_f,\varrho,\lambda)\is \left.\mathpzc{d}(m_f\rho\ll 1,\varrho,\lambda)\right|_{T^{\,}=^{\,} 0}\,\left.\mathpzc{f}(m_f,\varrho)\right|_{T^{\,}\geq^{\,} 0}\,.
\end{equation}
This was established in \cite{qcd_at_finite_T}, where is was also shown that (\ref{eq:cal_density_T=m=0}) still holds at $T>0$.

For finite temperatures this correction factor due to gluons and $N_{\!f_f}$ vanishingly light fermions was calculated in \cite{qcd_at_finite_T} and reads $\mathpzc{f}(m_f\ll 1 \land m_f\varrho\ll 1,\varrho) \is \mathpzc{f}(m_f\is 0,\varrho)$:
\begin{equation}
\label{eq:correction_factor_massless_quarks}
\left.\mathpzc{f}(0,\varrho)\right|_{T\,>\,0}\is \exp\!\left(-\frac{(\pi\varrho)^2}{3}(2N+N_{\! f})-A(\pi\varrho)\big(12+2(N-N_{\! f})\big)\right)
\end{equation}
with $A(x)\approx -\frac{1}{12}\ln\big(1+\frac{x^2}{3}\big)+a_1\big(1+\frac{a_2}{x^{3/2}}\big)^{-8}$, $a_1\approx 0.01289764$, and $a_2\approx 0.15858$. From (\ref{eq:correction_factor_massless_quarks}) one can deduce that large calorons $\varrho\gtrsim 0.8$ are exponentially unfavorable, while (\ref{eq:cal_density_T=m=0}) shows that small calorons $\varrho\lesssim 0.1$ are polynomially suppressed. The preferred caloron density size is $\varrho\approx 0.42$ in pure glue with $N\is 3$ and goes down to $\varrho\approx 0.34$ for $N_{\! f_l}\is 4$.
From this preferred caloron size we can deduce that the replacement $g^{-4N}\rightarrow g^{-4N\!}(1/\varrho)$ in (\ref{eq:cal_density_T=m=0}) adds only small corrections: the large Pauli\dash Villars mass $\lambda\gg 1$ (the regularization energy scale in $\ncoverline{\text{MS}}$) yields $\big(\ln(\lambda\varrho)\big)^{2N} \is (\ln\lambda)^{2N}\!\left(1+2N\epsilon+\mathcal{O}(\epsilon^2)\right)$, $\epsilon\is \frac{\ln\varrho}{\ln\lambda}$. Therefore, we choose to neglect the $\varrho$\dash dependent $\epsilon$\dash corrections and keep only the ``constant'', i.e., $\varrho$\dash independent, term $\ln\lambda$ when calculating $\sus$. We verify this simplification for our system with a heavy $b$ quark in \hyperref[fig:density_ratios]{figure \ref{fig:density_ratios}}.

These small caloron sizes allows us to use the \textsl{small\dash constituent approximation} at high enough temperatures, i.e., to describe an $n$\dash caloron configurations as superpositions of spatially well\dash separated and thus non\dash interacting\footnote{Instanton and caloron interactions are short\dash ranged; e.g., for well\dash separated instantons at locations $z_i$ with typical separation scale $d$, the interaction corrections compared to an exact solution are $\lesssim \frac{\varrho^2}{d^3}$ in the ``near region'' $|x-z_i|\lesssim \varrho$ (for some $i$) and $\lesssim \frac{\varrho^4}{d^5}$ in the ``far region'' $|x-z_i|\gtrsim\varrho$ $\forall i$ \cite{inst_interactions}.} single HS (anti$\,$-)calorons.
At leading order, all higher charge\dash calorons can then be described this way. A general caloron background is then populated by calorons of all $n\in\mathbb{Z}$ and one describes $n$\dash caloron configurations using $n_+$ HS calorons and $n_-$ HS anti\dash calorons with $n_+-n_-\is n$. This is the \textsl{dilute gas approximation} (DGA) \cite{callan_dashen_gross, higher_charge}. In this approximation, the partition function and thus the topological susceptibility take a simple form (in terms of (\ref{eq:partitioN_func_cal_density})):
\begin{align}
& Z_\text{DGA}(\theta)\is \exp\!\big(2 V_4\,\mathpzc{D}(T,m_f)\,\cos(\theta)\big) \label{eq:partitioN_funct_dga} \\
& \Rightarrow\quad \sus(T>T_\text{c})\stackrel{\text{DGA}}{\is}2\mathpzc{D}(T,m_f) \is 2\int_0^\infty\!\!\ab\varrho\,\mathpzc{d}(m_f,\varrho,\lambda)\,, \label{eq:top_suscep_dga}
\end{align}
where $T_{c} \simeq 155\,\text{MeV}$ is the critical temperature of chiral perturbation theory \cite{Borsanyi:2013bia,HotQCD:2014kol}.

For zero temperature but non\dash vanishing fermion masses, the fermion determinant relation was evaluated and one has analytical results for small\dash mass and large\dash mass expansions
\begin{equation}
\label{eq:correction_factor_inst_small_large_mass}
\left.\fferm(m_f,\varrho)\right|_{T^{\,}=^{\,} 0} \is \left\lbrace\begin{aligned} & \prod_f\exp\!\Big( m^2_f\varrho^2 \ln(m_f\varrho) + (\gamma_\text{E}-\ln 2)m^2_f\varrho^2 \Big) && \text{: }m_f\varrho\lesssim 0.5 \\ & \begin{aligned} \prod_f \frac{e^{-2\alpha\left(\frac{1}{2}\right)}}{(m_f\varrho)^\frac{1}{3}}\exp\!\Bigg(\! &-\frac{2}{75\,(m_f\varrho)^2} - \frac{34}{735\,(m_f\varrho)^4}\, + \\ & + \frac{464}{2835\,(m_f\varrho)^6} - \frac{15832}{148225\,(m_f\varrho)^8}\Bigg)\end{aligned} && \text{: }m_f\varrho\gtrsim 1.8 \end{aligned}\right.
\end{equation}
as well as an interpolation between them, covering arbitrary masses, in \cite{high_low_m_instantons}:
\begin{equation}
\label{eq:pade_T=0}
\begin{aligned}
& \left.\fferm(m_f,\varrho)\right|_{T^{\,}=^{\,} 0} \is \\
& \is \prod_f \frac{e^{-2\alpha\left(\frac{1}{2}\right)}}{(m_f\varrho)^\frac{1}{3}}\exp\!\left(\frac{\frac{1}{3}\ln(m_f\varrho)+2\alpha\!\left(\frac{1}{2}\right) - \left(6\alpha\!\left(\frac{1}{2}\right)-\gamma_\text{E}+\ln 2\right)(m_f\varrho^2) - \frac{2}{5}(m_f\varrho)^4}{1-3(m_f\varrho)^2 + 20(m_f\varrho)^4+15(m_f\varrho)^6}\right)\,.
\end{aligned}
\end{equation}
Furthermore, an explicit numerical solution covering arbitrary masses was found \cite{arbitrary_mass}:
\begin{equation}
\label{eq:correction_factor_inst_arbitrary_mass}
\begin{aligned}
& \left.\fferm(m_f,\varrho)\right|_{T^{\,}=^{\,} 0} \is \\
& \is \prod_f \exp\!\Bigg(\!-2\alpha\!\left(\frac{1}{2}\right)-2\lim_{L^{\,}\rightarrow^{\,}\infty}\Bigg(\sum_{l^{\,}=^{\,}0,\frac{1}{2},...}^{L}\!(2l+1)(2l+2)P_{M_{f\!},{}_{\,}\varrho}(l)+ 2L^2+4L \,- \\
& \phantom{\is \prod_f \exp\!\Bigg(} - \left(\frac{1}{6} + \frac{m_f^2\varrho^2}{2}\right)\ln(L) + \frac{m_f^2\varrho^2}{2}\big(\ln(m_f\varrho) + 1 -2\ln 2\big)+\frac{127}{72}-\frac{\ln 2}{3}\Bigg)\Bigg)\,,
\end{aligned}
\end{equation}
where $P_{M_{f\!},{}_{\,}\varrho}(l)\is S_{M_{f\!},{}_{\,}\varrho}^{l,l+\frac{1}{2}}(\ncoverline{r}\rightarrow\infty)+S_{M_{f\!},{}_{\,}\varrho}^{l+\frac{1}{2},l}(\ncoverline{r}\rightarrow\infty)$ and $S_{M_{f\!},{}_{\,}\varrho}^{l,j}(\ncoverline{r})$ is the numerical solution to the ordinary differential equation
\begin{equation}
\label{eq:ode_instantons}
\ab^2_{\ncoverline{r}}S^{l,j} + \big(\ab_{\ncoverline{r}} S^{l,j}\big)^2 + \left(\frac{1}{\ncoverline{r}} + 2\frac{\ab_{\ncoverline{r}}I_{2l+1}(m_f\ncoverline{r})}{I_{2l+1}(m_f\ncoverline{r})}\right)\ab_{\ncoverline{r}}S^{l,j}\is \frac{4(j-l)(j+l+1)}{\ncoverline{r}^2+\varrho^2}-\frac{3\varrho^2}{\big(\ncoverline{r}^2+\varrho^2\big)^2}\,,
\end{equation}
with $I_{\alpha}(x)$ the modified Bessel function of the first kind.

We aim to calculate the correction factor $\left.\mathpzc{f}(m_f,\varrho)\right|_{T\,>\,0}$ for the general case of massive quarks at finite temperature. For that, we calculate the regularized, vacuum\dash normalized Klein\dash Gordon operator determinant with anti\dash periodic temporal boundaries from (\ref{eq:cal_density_general}):
\begin{align}
\label{eq:log_fermion_cal_density}
 -\ln(\,\fferm) \is \gamma_\text{ferm}
 & = \ln \frac{m}{\Lambda}
 -2\ln\!\left(\frac{\det(-D_-^2+m^2)\,\det(-\partial_-^2+\lambda^2)}{\det(-D_-^2+\lambda^2)\,\det(-\partial_-^2+m^2)}\right)\is -2\gammas \; ,
\end{align}
where $\ln(m/\Lambda)$ arises from the contribution of the zero-mode of $\slashed{D}$.
As in (\ref{eq:correction_factor_inst_small_large_mass}) - (\ref{eq:correction_factor_inst_arbitrary_mass}), which only consider the $\mathfrak{su}(2)$\dash caloron, we can also limit ourselves to $D(A_\text{HS})$ in (\ref{eq:log_fermion_cal_density}). This is justified by (\ref{eq:caloron_embedding_SU(N)}) and the connection to $m\is 0$, which contains the right $SU(N)$\dash dependent factors, cf. (\ref{eq:cal_density_T=m=0}) and (\ref{eq:correction_factor_massless_quarks}).

The finite\dash $T$ spacetime $\spt$ has a distinct, bounded time direction compared to the unbounded space directions.
This breaks the $SO(4)$\dash symmetry of $T\is 0$\dash physics down to $SO(3)$. From the physical point of view, the presence of the external heat bath implies a preferred Lorentz frame, the heat bath's rest frame. Therefore, Lorentz invariance is broken: rotations and translations are still symmetries, whereas boosts are not, and the Lorentz group $SO^+(1,3)$ is broken to $SO(3)$ \cite{thermal_sym_1, thermal_sym_2}.
As the HS caloron, compared to the BPST instanton, is only radially symmetric in the three space dimensions (and not explicitly so, cf. (\ref{eq:HS_caloron}), (\ref{eq:phi_function})), calculating $\det(-D^2_-+m_f^2)$ requires solving a complicated \mbox{2\dash dimensional} (partial) differential equation - we obtain this equation \hyperref[appendix:pde]{appendix \ref{appendix:pde}} -, compared the 1\dash dimensional ordinary differential equation at $T\is 0$ (\ref{eq:ode_instantons}).
We instead follow an alternative approach which was first used for $T\is 0$ in \cite{high_low_m_instantons} and that we adapt and generalize to finite $T$. Our approach is discussed in detail in the following \hyperref[sec:strategy]{section \ref{sec:strategy}}.

\section{Strategy of our Approach}
\label{sec:strategy}

As we stated in \hyperref[sec:intro]{section \ref{sec:intro}}, we want to compare topological susceptibilities for theories with $2+1+1+1$ (including a dynamical $b$ quark) and $2+1+1$ dynamical quarks (with $b$ asymptotically heavy, accessible via lattice QCD).
For that, we calculate $\sus(m_f,T)$ using (\ref{eq:top_suscep_dga}), once with physical $m_b$ and once with an asymptotic $m_\text{asy}$, keeping the 4\dash flavor effective theories in the IR fixed, i.e., equal for both cases.
Then we compute the ratio $\frac{\sus(m_b,T)}{\sus(m_\text{asy},T)}$ which, together with the 4\dash flavor lattice result, gives the full 5\dash flavor topological susceptibility at high temperatures.

In order to obtain the caloron density, as required in (\ref{eq:top_suscep_dga}), we to obtain the fermionic part of the caloron density correction factor (cf. (\ref{eq:caloron_density_with_correction_factor})) in the general case of heavy quarks at finite temperatures. For this, we proceed as follows:\newline

\noindent \textbf{\underline{\textsl{Step 1)}}}

We calculate $\gammas$ (\ref{eq:log_fermion_cal_density}) for small (but non\dash vanishing) and large fermion masses separately and obtain two expansions, analogous to (\ref{eq:correction_factor_inst_small_large_mass}):
\begin{equation}
\label{eq:correction_factor_general}
\left.\fferm(m_f,\varrho)\right|_{T\,>\,0}\is\left\lbrace
    \begin{aligned}
    & \mathpzc{f}_\text{small}\is \exp\!\big(2\gammas(m,\varrho)\big) && \text{: small }m \\
    & \mathpzc{f}_{\,\text{large}}\is \frac{e^{-2\alpha\left(\frac{1}{2}\right)}}{(\lambda\rho)^\frac{1}{3}}\,\exp\!\big(2\gammas(m,\varrho,\lambda)\big)  && \text{: large }m
    \end{aligned}\right. .
\end{equation}
The large\dash $m$ factor $e^{-2\alpha(1/2)}(\lambda\varrho)^{-1/3}$ cancels $m\ll 1$\dash terms in (\ref{eq:cal_density_T=m=0}).

For small fermion masses $m\ll 1$ we Taylor expand (\ref{eq:log_fermion_cal_density}) up to first order in $m^2$. Such a Taylor expansion is not possible at zero temperature, where the known result for $\gammas$ contains a term $(m\varrho)^2\ln(m\varrho)$ \cite{high_low_m_instantons} which is non\dash analytical at $m\varrho\is 0$.
Noting that $\ln\det(-\ncoverline{D}^2+m^2) \is \Tr\ln(-\ncoverline{D}^2+m^2)$, this IR non\dash analyticity can be traced back to eigenmodes of $-\ncoverline{D}^2$ with arbitrarily low momenta that get affected arbitrarily strongly by the introduction of even an infinitesimal mass.
At finite temperature, however, fermions have anti\dash periodic boundary conditions which raise the lowest fluctuation frequencies in $D_-^2$ to $\pi T$, the lowest fermionic Matsubara frequency, or $\pi$ in our dimensionless units.
Therefore, as long as $m \ll 1$, the introduction of a small mass has only an infinitesimal effect the determinant.
Indeed, one expects the small\dash mass expansion to be a Taylor series with a radius of convergence $\sim \pi$.
In \hyperref[sec:smallmass]{section \ref{sec:smallmass}} we thus obtain a Taylor expansion of the general structure
\begin{equation}
\label{eq:small_mass_exp_general}
\gamma_\text{ferm}(m\text{ small},\varrho) \is -2\gammas(m\text{ small},\varrho)\is  \frac{(\pi\varrho)^2}{3} -2 A(\pi\varrho) -2 m^2\,\gammas^\text{small}(\varrho) + \mathcal{O}(m^4) \,,
\end{equation}
where the purely $\varrho$\dash dependent first terms are given by the fermionic part of the known result (\ref{eq:correction_factor_massless_quarks}).
Computing the $m^4$\dash coefficient is possible in principle but much more challenging than the $m^2$\dash coefficient and we will not attempt it here.

For large quark masses $m\gg 1$ we employ the asymptotic \textsl{heat kernel expansion} of  (\ref{eq:log_fermion_cal_density}). In \hyperref[sec:largemass]{section \ref{sec:largemass}} we find the resulting expansion
\begin{equation}
\label{eq:large_mass_exp_general}
\gammas(m\text{ large},\varrho,\lambda) \is \frac{1}{6}\ln\!\left(\frac{\lambda}{m}\right) +\sum_{k^{\,}=^{\,} 1}^{k_\text{max}}\frac{\gammas^{\text{large, }k}(\varrho)}{m^{2k}} .
\end{equation}
The series is asymptotic and its Borel resummation contains an ambiguity $\sim m^b e^{-m}$, $b\sim 1$, which is sensitive to the anti\dash periodic boundary conditions (we discuss this in \hyperref[appendix:finite_T_corrections]{appendix \ref{appendix:finite_T_corrections}}).\newline

\noindent \textbf{\underline{\textsl{Step 2)}}}

We interpolate between the small\dash mass result (\ref{eq:small_mass_exp_general}) and the large\dash mass result (\ref{eq:large_mass_exp_general}) in order to obtain the full correction factor following (\ref{eq:correction_factor_general}).
For this we seek an approximate fitting function for $\gamma_{\text{ferm}}$ which matches both limiting behaviors.
Specifically, we seek a function $\mathpzc{p}(m,\varrho)$ defined as
\begin{equation}
-\ln\!\big(\,\fferm(m,\varrho)\big) \is \gamma_\text{ferm}\is 2\alpha\!\left(\frac{1}{2}\right)+\mathpzc{p}(m,\varrho)
\end{equation}
obeying
\begin{equation}
\label{eq:correction_factor_small_and_large_m}
\mathpzc{p}(m,\varrho) \to 
\left\lbrace \begin{aligned}
& -2\alpha\!\left(\frac{1}{2}\right)+\frac{(\pi\varrho)^2}{3}-2A(\pi\varrho) - 2m^2 \,\gammas^\text{small}(\varrho) \quad &\text{: small }m \\
& \phantom{-} \frac{1}{3}\ln(m\varrho)-2\sum_{k^{\,}=^{\,} 1}^{k_\text{max}}\frac{\gammas^{\text{large, }k}(\varrho)}{m^{2k}} &\text{: large }m \end{aligned}\right.\, .
\end{equation}
Because the large\dash mass behavior features a logarithm $\ln(m\varrho)$ while the small\dash mass behavior does not, a regular Pad\'e approximant cannot work for $\mathpzc{p}$ in (\ref{eq:correction_factor_small_and_large_m}); instead we make the ``Pad\'e\dash like'' approximant \textsl{Ansatz} 
\begin{align}
& \mathpzc{p}(m,\varrho) \is \frac{\sum_{i^{\,}=^{\,}0}^K \mathpzc{p}_{\,{}^{\!}i}(\varrho)\,m^{2i}}{\prod_{j^{\,}=^{\,}1}^{K+1}\big(1+\mathpzc{P}_j(\varrho)\,m^2\big)} + \frac{1}{6}\ln\!\left(m^2\varrho^2 + \xi^2(\varrho)\right),\;\,\mathpzc{P}_j>0\,\forall j,\varrho\,, \label{eq:pade_ansatz_general} \\
& K \is \left\lbrace\begin{aligned} &\frac{k_\text{max}-1}{2} && \text{: $k_\text{max}$ odd} \\ & \frac{k_\text{max}-2}{2} && \text{: $k_\text{max}$ even} \end{aligned}\right.,\label{eq:K_pade_ansatz_general}
\end{align}
which contains a regulated logarithmic function that becomes a simple log at large mass and a (purely $\varrho$\dash dependent) ``constant'' at small mass.
The coefficients of the polynomial in the numerator, the roots of the polynomial in the denominator, and the constant in the log then represent $2K+3$ $\varrho$\dash dependent coefficient functions.
For $m\ll 1$ this \textsl{Ansatz} approaches an $m$\dash independent function of $\varrho$ with $\mathcal{O}(m^2)$\dash corrections. For $m\gg 1$ the logarithm correctly reproduces the corresponding term in (\ref{eq:correction_factor_small_and_large_m}), while the rational part falls off as $m^{-2}$.
To fix the coefficients, we perform a Taylor expansion of (\ref{eq:pade_ansatz_general}) up to $\mathcal{O}(m^2)$ for small masses as well as a Laurent expansion up to $\mathcal{O}(m^{-2k_\text{max}})$ for large masses and identify these expansions with the corresponding ones in (\ref{eq:correction_factor_small_and_large_m}) by equating the $\varrho$\dash dependent coefficients.
This yields a non\dash linear system of equations.
Note that, if the number of known Taylor and Laurent expansion coefficients is even, then the \textsl{Ansatz} has the wrong number of free parameters; but in this case one can fix the value of $\xi$, for instance $\xi\is 1$, and fit only the numerator and denominator coefficients.

Our approach is motivated by the excellent agreement of $-\frac{1}{2}\ln\!\big(\left.\fferm(m_f,\varrho)\right|_{T^{\,}=^{\,}0}\big)$ given by the interpolation (\ref{eq:pade_T=0}) in \cite{high_low_m_instantons} with the full numerical result (\ref{eq:correction_factor_inst_arbitrary_mass}) from \cite{arbitrary_mass} as shown in figure~6 of \cite{arbitrary_mass}.

Having determined $\mathpzc{p}(m,\varrho)$, we can describe the full caloron density (\ref{eq:caloron_density_with_correction_factor}) using the $T\is 0$\dash caloron density (\ref{eq:cal_density_T=m=0}) and the correction factors (\ref{eq:correction_factor_massless_quarks}) for $N_{\!f_l}$ (vanishingly) light quarks as well as our result ((\ref{eq:correction_factor_small_and_large_m}), (\ref{eq:pade_ansatz_general})) for $\left.\fferm(m_{f_\text{h}},\varrho)\right|_{T^{\,}>^{\,}0}\is \exp\!\big(-2\alpha\!\left(\frac{1}{2}\right)- \mathpzc{p}(m_{f_\text{h}},\varrho)\big)$ describing $N_{\!f_\text{h}}$ heavy quarks, $N_{\! f}\is N_{\!f_l}+N_{\!f_\text{h}}$:
\begin{equation}
\label{eq:cal_density_with_pade_general}
\begin{aligned}
\mathpzc{d}(m_{f_l}, m_{f_\text{h}} ,\varrho, \lambda)\is & \frac{2 e^{-\alpha(1) + 4\alpha\left(\frac{1}{2}\right) + \ln 2 - N\left(2\alpha\left(\frac{1}{2}\right) + 2\ln 2 \right) + 2N_{\!f_l} \alpha\left(\frac{1}{2}\right)}}{\pi^2(N-1)!(N-2)!} \left(\frac{\ln\lambda(11N-2N_{\! f})}{3}\right)^{\!2N}\!\times \\
& \times e^{-\frac{8\pi^2}{g^2(N_{\! f},N,1/\varrho)}}\, \frac{\prod_f m_f\varrho}{\varrho^5}\, \left.\mathpzc{f}(0,\varrho)\right|_{N_{\! f_l},\,T^{\,}>^{\,}0}\prod_{f_\text{h}}e^{-\mathpzc{p}(m_{f_\text{h}},\varrho)}
\end{aligned}
\end{equation}
with $g^2\big(N_{\! f},N,\frac{1}{\varrho}\big)$ given in (\ref{eq:coupling}) and $g^{-2N}$ replaced by the purely $\lambda$\dash dependent term as discussed below (\ref{eq:correction_factor_massless_quarks}); the light quark masses appear only as factors. Generally, we keep the light and heavy flavors $N_{\!f_l}$ and $N_{\!f_\text{h}}$ unspecified and only fix them to the physical case of $N_{\!f_l}\is 4$, $N_{\!f_\text{h}}\is 1$, $N_{\! f} \is N_{\!f_l}+N_{\!f_\text{h}} \is 5$ when presenting explicit results.\newline

\noindent \textbf{\underline{\textsl{Step 3)}}}

We compute the topological susceptibility in the DGA by integration over all HS caloron sizes following (\ref{eq:top_suscep_dga}) with our result (\ref{eq:cal_density_with_pade_general}) plugged in.
Then we take the ratio $\kappa\is\frac{\sus(m_{f_\text{h}},{}^{\,}T)}{\sus(m_{f_\text{h, asy}},{}^{\,}T)}$, comparing a theory with physically heavy quarks to a theory with asymptotically heavy quark masses $m_{f_\text{h(, asy)}}$. Note that $\kappa\neq\kappa(\lambda)$.

\begin{figure}
    \centering
    \includegraphics[width=0.45\textwidth]{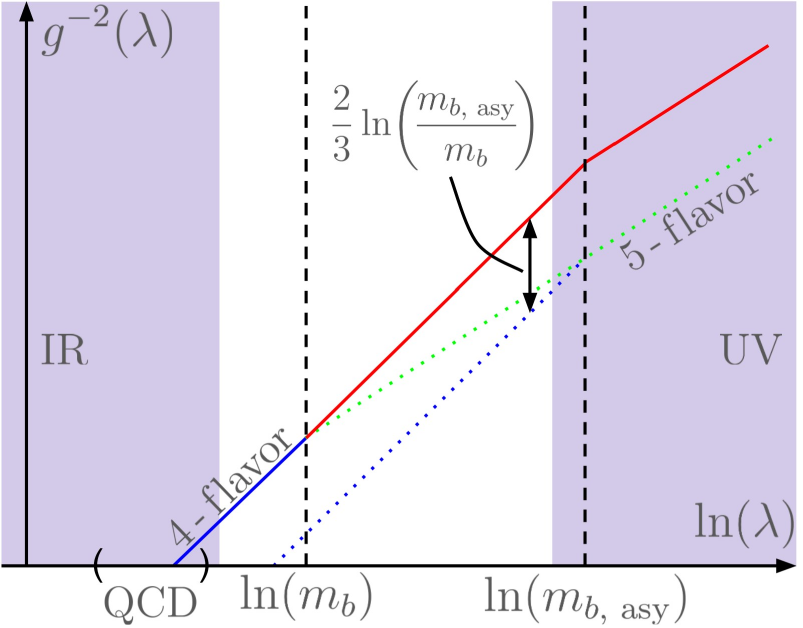}
    \caption{The running coupling $g^{-2}(\lambda)$ for a theory with four light and a heavy $b$ quark. At large energy scales $\gg m_b$ one has a 5\dash flavor running (\textcolor{green}{\rule[2pt]{3pt}{2pt} \rule[2pt]{3pt}{2pt} \rule[2pt]{3pt}{2pt}}) which switches to 4\dash flavor running at the energy scale $m_b$ (\textcolor{blue}{\rule[2pt]{18pt}{1.5pt}}).
    For a theory with an asymptotically heavy $b$ quark, the switch occurs at the \textcolor{violet}{UV} scale $m_{b\text{, asy}}$ (\textcolor{blue}{\rule[2pt]{3pt}{2pt} \rule[2pt]{3pt}{2pt} \rule[2pt]{3pt}{2pt}}) and the two theories disagree in the \textcolor{violet}{IR}, with the asymptotic $b$\dash theory failing to describe known 4\dash flavor QCD/IR theory.\newline
    In order to compare the $m_b\,$- and $m_{b\text{, asy}}$\dash theories with matching IR physics, we modify the coupling $g_\text{asy}$ and describe it in terms of $g_\text{phys}$ (\ref{eq:coupling_modified}) for scales $>m_b$ (\textcolor{red}{\rule[2pt]{18pt}{1.5pt}}).
    Overall, $g_\text{asy}$ is thus given by (\textcolor{blue}{\rule[2pt]{9pt}{1.5pt}}\textcolor{red}{\rule[2pt]{9pt}{1.5pt}}).
    While this description disagrees with the physical description in the UV, it agrees in the IR and thus corresponds better to what happens in a $2+1+1$\dash flavor lattice calculation.}
    \label{fig:coupling_modified}
\end{figure}

In order to ensure equal $N_{\!f_l}$\dash flavor IR theories, we define the running coupling constant (\ref{eq:coupling}) for the theory containing asymptotically heavy quarks by
\begin{equation}
\label{eq:coupling_modified}
\frac{8\pi^2}{g^2_\text{asy}(N_{\!f},N,1/\varrho,\lambda)} \is \underbrace{\frac{8\pi^2}{g^2_\text{phys}(N_{\!f},N,1/\varrho,\lambda)}}_{-\frac{1}{3}\ln(\lambda\varrho)(11N-3N_{\! f})}\, +\, \frac{2}{3} \sum_{f_\text{h}}\ln\!\left(\frac{m_{f_\text{h, asy}}}{m_{f_\text{h}}}\right).
\end{equation}
This is illustrated and described in \hyperref[fig:coupling_modified]{figure~\ref{fig:coupling_modified}}.

Together with $\lim_{{}^{\,}m_{f_\text{h, asy}}\rightarrow^{\,}\infty}\mathpzc{p}(m_{f_\text{h, asy}},\varrho) \is \frac{1}{3}\ln(m_{f_\text{h, asy}}\varrho)$, the modifier term in (\ref{eq:coupling_modified}) cancels $m_{f_\text{h, asy}}$ in the caloron density, so that $\kappa\neq \kappa(m_{f_\text{h, asy}})$. We therefore obtain the $\sus$\dash ratio
\begin{equation}
\label{eq:top_suscep_ratio_general}
\kappa(m_{f_\text{h}},N_{\! f_l},N_{\! f_\text{h}},N ) \is \frac{\sus\big( N, m_{f_l}, m_{f_\text{h}} , g_\text{phys}\big)}{\sus\big( N, m_{f_l}, m_{f_\text{h, asy}}, g_\text{asy}\big)}\,.
\end{equation}

\section{Small Mass -- Taylor Expansion}
\label{sec:smallmass}

\subsection{Structure of the Expansion}
\label{subsec:small_m_exp_structure}

We now expand $\fferm$ for quarks with a non\dash vanishing (as opposed to (\ref{eq:cal_density_T=m=0}) and (\ref{eq:correction_factor_massless_quarks})) but small mass, in order to obtain $\mathpzc{f}_\text{small}$ in (\ref{eq:correction_factor_general}).
As we discussed in \hyperref[sec:strategy]{section~\ref{sec:strategy}}, we can perform a Taylor expansion of the logarithmic determinant ratio (\ref{eq:log_fermion_cal_density}) at finite $T$.
We use
\begin{equation}
\begin{aligned}
\frac{\ab}{\ab m^2}\ln\det(-D_-^2+m^2) & \is \frac{\ab}{\ab m^2}\Tr \ln(-D_-^2+m^2) \is \\
& \is \Tr\left(\frac{1}{-D_-^2+m^2}\right) \is \int^1\!\!\ab^4 x\,\tr\Braket{x|\frac{1}{-D_-^2+m^2}|x}
\end{aligned}
\end{equation}
with the anti\dash periodic closed loop or coincident propagator $\Delta^-(x,x,m^2)\is\big\langle x|\frac{1}{-D_-^2+m^2}|x\big\rangle$ for the $\mathcal{O}(m^2)$\dash coefficient.
Including higher orders in the Taylor expansion would require convolutions of such propagators and we thus avoid them.

\begin{figure}[htbp]
\centering
\includegraphics[width=0.4\textwidth]{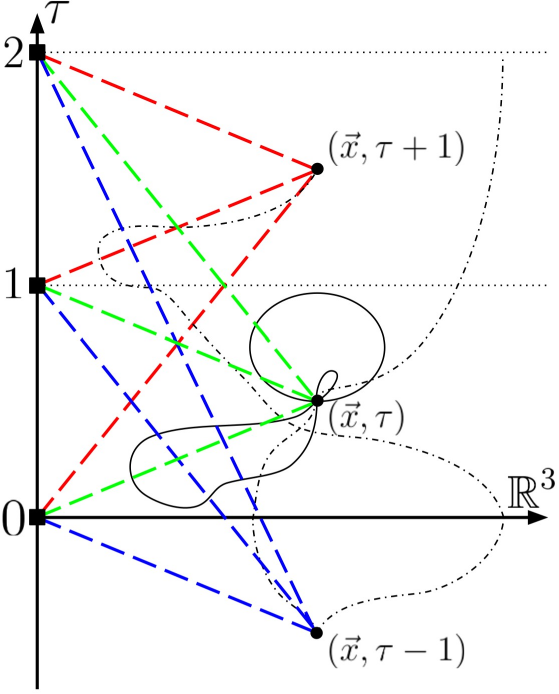}
\caption[Coincident propagators in caloron background]{Some closed loop\dash propagators in the periodic spacetime $\spt$ as they appear in (\ref{eq:eff_act_Tay_ansatz}). The periodicity of the spacetime is made explicit by showing all the time copies of the boson and BPST instanton making up the thermal boson and HS caloron, respectively.\newline
The anti\dash periodic boson copies are located at ($\bullet$) $x+j\e{4}$, $j\in\mathbb{Z}$ and are connected by closed loops; the solid lines (\rule[2.5pt]{18pt}{1.5pt}) show ``aperiodically closed loops'' which do not encounter the spacetime periodicity, the dash\dash dotted lines (\rule[2.5pt]{8pt}{1.5pt}$\,\boldsymbol{\cdot}\,$\rule[2.5pt]{8pt}{1.5pt}$\,\boldsymbol{\cdot}\,$\rule[2.5pt]{8pt}{1.5pt}) show loops which encounter the periodicity $j$ times and close (anti$\,$-)periodically for $j$ even (odd). The caloron is made up of periodic instanton copies located at ($\blacksquare$) $0+j\e{4}$.\newline
All boson copies and all connecting, closed loop\dash propagators are affected by all periodic instanton copies; this is symbolized by the dashed red, green, and blue lines (\textcolor{red}{\rule[2.5pt]{8pt}{1.5pt}} \textcolor{green}{\rule[2.5pt]{8pt}{1.5pt}} \textcolor{blue}{\rule[2.5pt]{8pt}{1.5pt}}) connecting the instanton and boson copies.}
\label{fig:closed_loops}
\end{figure}

Coincident propagators are naturally divergent and we achieve regularization via point splitting, i.e., by considering ``almost closed loop'' propagators $\Delta^-(x',x,m^2)$ from $x$ to $x'\is x + \varepsilon$, $\varepsilon\rightarrow 0$. To retain the gauge invariance of $\gammas$, we insert an appropriate Wilson line $\,\text{P}\exp\Big(i\!\int_{x}^{x'}\!\ab z^\mu A^\mu_\text{HS}(z)\Big) \is 1 + i A^\mu_\text{HS}(x)\,\varepsilon^\mu + \frac{1}{2}\Big(i\!\left.\partial^{\mu\!} A^\nu_\text{HS}\right|_x - (A^\mu_\text{HS}A^\nu_\text{HS})(x)\Big)\varepsilon^\mu\varepsilon^\nu + \mathcal{O}\big(\varepsilon^3\big)$ into the propagator.
Using also (\ref{eq:correction_factor_massless_quarks}), we find the Taylor expansion up to $\mathcal{O}(m^2)$:
\begin{equation}
\label{eq:eff_act_Tay_ansatz}
\begin{aligned}
\gamma_\text{ferm}\is & -\frac{1}{3}\ln(\lambda\varrho)-2\alpha\!\left(\frac{1}{2}\right)+\frac{1}{3}(\pi\varrho)^2-2A(\pi\varrho) - \\
& -2m^2\lim_{\varepsilon^{\,}\rightarrow^{\,} 0}\Tr\!\left(\Braket{x'|\frac{1}{-D_-^2}|x}\,\text{P}\exp\Big(i\!\int_{x}^{x'}\!\!\ab z^\mu A^\mu_\text{HS}(z)\Big) -\Braket{x'|\frac{1}{-\partial_-^2}|x} \right).
\end{aligned}
\end{equation}
This expansion contains the spacetime integral over all massless, traced, closed loop propagators of the anti\dash periodic scalar boson in a periodic caloron background. We try to sketch the intricacy of this system in \hyperref[fig:closed_loops]{figure \ref{fig:closed_loops}}.

We therefore require the massless anti\dash periodic scalar propagator in a caloron and the vacuum background. In general, any (anti$\,$-)periodic propagator $\Delta^\pm_{(0)}(x,y,m^2)$ can be obtained from the corresponding aperiodic propagator $\ncoverline{\Delta}_{(0)}(\ncoverline{x},\ncoverline{y},m^2)$ in $\mathbb{R}^4$ by adding up time copies \cite{qcd_at_finite_T}
\begin{equation}
\label{eq:finite_T_propagator}
\Delta^\pm_{(0)}(x,y,m^2) \is \sum_{j^{\,}\in^{\,}\mathbb{Z}}(\pm 1)^{j\,} \ncoverline{\Delta}_{(0)}(\ncoverline{x},\ncoverline{y}+j\e{4},m^2)
\end{equation}
(compare the construction of the caloron from instantons). Performing the time\dash copy sums amounts to compactifying the spacetime in the temporal direction $\mathbb{R}^4\rightarrow\spt$, therefore the bars are dropped: $\ncoverline{\Delta}\rightarrow\Delta$, $|\vec{\ncoverline{x}}|,|\vec{\ncoverline{y}}|\rightarrow r_x,r_y$, $\ncoverline{t}, \ncoverline{t}_y\rightarrow\tau_x,\tau_y$, etc.

In the following \hyperref[subsec:propagator]{section \ref{subsec:propagator}} we obtain this massless closed loop propagator $\Delta^\pm(x,x')$ in point splitting regularization. In \hyperref[appendix:propagators]{appendix \ref{appendix:propagators}} we derive the traceful parts for general massless propagators $\Delta^\pm(x,y)$ (the traceless parts can be obtained analogously).

\subsection{Massless closed loop scalar Propagator}
\label{subsec:propagator}

First, we require the aperiodic $\mathbb{R}^4$\dash propagator for a scalar field in a HS caloron background. An \textsl{Ansatz} for this propagator can be found by employing the results of \cite{propagators_pseudoparticle_fields}:
\begin{align}
& \ncoverline{\Delta}(\ncoverline{x},\ncoverline{y})\is \frac{1}{\sqrt{\phi(x)}}\frac{F(\ncoverline{x},\ncoverline{y})}{4\pi^2(\ncoverline{x}-\ncoverline{y})^2}\frac{1}{\sqrt{\phi(y)}}\;\, , \label{eq:aperiodic_propagator_via_F} \\
& \begin{aligned}
& F(\ncoverline{x},\ncoverline{y})  \is 1 + \varrho^2\sum_{k^{\,}\in^{\,}\mathbb{Z}}\frac{\sigma^\mu (\ncoverline{x}-k^{\,}\e{4})^\mu}{(\ncoverline{x}-k^{\,}\e{4})^2} \frac{\sigma^{\dagger\,\nu}(\ncoverline{y}-k^{\,}\e{4})^\nu}{(\ncoverline{y}-k^{\,}\e{4})^2} \is \\
& \is 1+ \varrho^{2\!} \sum_{k^{\,}\in^{\,}\mathbb{Z}} \frac{\vec{\ncoverline{x}}\cdot\vec{\ncoverline{y}} + (\ncoverline{t}_x-k)(\ncoverline{t}_y-k)+i(\ncoverline{t}_x-k)\vec{\ncoverline{y}}\cdot\vec{\sigma} - i(\ncoverline{t}_y-k)\vec{\ncoverline{x}}\cdot\vec{\sigma} + i(\vec{\ncoverline{x}}\times\vec{\ncoverline{y}}{}^{\,})\cdot\vec{\sigma}}{\vec{\ncoverline{x}}{}^{\,2}\, \vec{\ncoverline{y}}{}^{\,2} + \vec{\ncoverline{x}}{}^{\,2}(\ncoverline{t}_y-k)^2 + \vec{\ncoverline{y}}{}^{\,2}(\ncoverline{t}_x-k)^2+(\ncoverline{t}_x-k)^2(\ncoverline{t}_y-k)^2}\,.
\end{aligned}\label{eq:F_function}
\end{align}
Since $\phi$ (\ref{eq:phi_function}) is periodic in the $\ncoverline{t}$\dash direction, the $\ncoverline{x}\,$- and $x$\dash coordinates are equivalent. We then plug (\ref{eq:aperiodic_propagator_via_F}) and (\ref{eq:F_function}) into (\ref{eq:finite_T_propagator}).

In order to obtain closed loop\dash propagators $\Delta^\pm(x',x)$ from (\ref{eq:finite_T_propagator}), we set $\ncoverline{y}\rightarrow \ncoverline{x}$ and $\ncoverline{x}\rightarrow\ncoverline{x}' \is \ncoverline{x}+\ncoverline{\varepsilon}$ in (\ref{eq:aperiodic_propagator_via_F}) and (\ref{eq:F_function}).
The $j\is 0$\dash term in (\ref{eq:finite_T_propagator}) then corresponds to the aperiodically closed loops in \hyperref[fig:closed_loops]{figure \ref{fig:closed_loops}}, while the $j\neq 0$\dash modes correspond to the anti\dash periodically closed loops.
Thus we can identify the diverging and finite parts of $\Delta^\pm(x,x')$ corresponding to $j\is 0$\dash and $j\neq 0$\dash modes, respectively. 

For the finite part of $\Delta^\pm(x',x)$, we can safely take the limit $\ncoverline{\varepsilon}\rightarrow 0$ as $j\neq 0$. First, we focus on the traceful part of (\ref{eq:aperiodic_propagator_via_F}), denoted as $\istr$, and calculate (\ref{eq:F_function}):
\begin{align}
\label{eq:F_function_coinc_traceful_finite}
 F(\ncoverline{x},\ncoverline{x}+j^{\,}\e{4}) & \istr
 1 + \sum_{k{}^{\,}\in{}^{\,}\mathbb{Z}}
 \frac{\varrho^2\big(\vec{\ncoverline{x}}{}^{\,2} + (\ncoverline{t}-k)(\ncoverline{t}+j-k)\big)}{\vec{\ncoverline{x}}{}^{\,4} + \vec{\ncoverline{x}}{}^{\,2}(\ncoverline{t}+j-k)^2 + \vec{\ncoverline{x}}{}^{\,2}(\ncoverline{t}-k)^2+(\ncoverline{t}-k)^2(\ncoverline{t}+j-k)^2} \is  \nonumber \\  
&\is 1 + \frac{4\pi\varrho^2|\vec{\ncoverline{x}}| \sinh(2\pi |\vec{\ncoverline{x}}|)}{(4\,\vec{\ncoverline{x}}{}^{\,2}+j^2)\big(\cosh(2\pi|\vec{\ncoverline{x}}|) - \cos(2\pi\ncoverline{t})\big)}\,.
\end{align}
We can now perform the $j$\dash summation according to (\ref{eq:finite_T_propagator}) and find the traceful finite part of the coincident (anti$\,$-)periodic propagator:
\begin{equation}
\label{eq:finite_T_propagator_traceful_finite}
\begin{aligned}
& \sum_{j{}^{\,}\in{}^{\,}\mathbb{Z}\setminus\lbrace 0\rbrace}\!\!(\pm 1)^j\ncoverline{\Delta}(\ncoverline{x},\ncoverline{x}+j\e{4}) \is \!\sum_{j{}^{\,}\in{}^{\,}\mathbb{Z}\setminus\lbrace 0\rbrace}\!\!\frac{(\pm 1)^j F(\ncoverline{x},\ncoverline{x}+j\e{4})}{4\pi^2 j^2\,\phi(x)} \istr \\
&\istr \!\sum_{j{}^{\,}\in{}^{\,}\mathbb{Z}\setminus\lbrace 0\rbrace}\frac{(\pm 1)^j}{4\pi^2\,\phi(x)}\Bigg(\frac{1}{j^2} + \left(\frac{1}{j^2} - \frac{1}{4\,\vec{\ncoverline{x}}{}^{\,2}+j^2}\right)\frac{\pi\varrho^2 \sinh(2\pi |\vec{\ncoverline{x}}|)}{|\vec{\ncoverline{x}}|\big(\cosh(2\pi|\vec{\ncoverline{x}}|) - \cos(2\pi\ncoverline{t})\big)}\Bigg) \is \\
& \is \left\lbrace\begin{aligned} & + \frac{1}{12} \\ & -\frac{1}{24} \end{aligned}\right\rbrace +\frac{\varrho^2\sinh(2\pi r)}{16\pi r^3\big(\cosh(2\pi r) - \cos(2\pi\tau)\big)\phi(x)}\left(1-2\pi r\cdot\left\lbrace \begin{aligned} & \coth(2\pi r) \\ & {}^{\,}\mathrm{csch}(2\pi r) \end{aligned}\right\rbrace\right),
\end{aligned}
\end{equation}
where we used that $\frac{1}{j^2(4\vec{\ncoverline{x}}{}^{\,2}+j^2)}\is \frac{1}{4\,\vec{\ncoverline{x}}{}^{\,2} j^2} - \frac{1}{4\,\vec{\ncoverline{x}}{}^{\,2}(4\,\vec{\ncoverline{x}}{}^{\,2}+j^2)}$. We abbreviate $\lbrace +\frac{1}{12},-\frac{1}{24} \rbrace \is C^\pm$. This finite contribution contains the corresponding finite contribution to the free propagator
\begin{equation}
\label{eq:finite_T_propagator_traceful_finite_free}
\sum_{j{}^{\,}\in{}^{\,}\mathbb{Z}\setminus\lbrace 0\rbrace}\!\!(\pm 1)^j\ncoverline{\Delta}_0(\ncoverline{x},\ncoverline{x}+j\e{4})\is \sum_{j^{\,}\in^{\,}\mathbb{Z}}\frac{(\pm 1)^j}{4\pi^2 j^2} \is C^\pm \,,
\end{equation}
which physically represents the well known effect of a mass on the thermal part of the pressure.

Second, we find the finite traceless part of (\ref{eq:finite_T_propagator}) to vanish. To see that, we calculate the traceless part of (\ref{eq:F_function}), denoted as $\isnotr$,
\begin{equation}
\label{eq:F_function_coinc_traceless_finite}
\begin{aligned}
F(\ncoverline{x},\ncoverline{x} +j\e{4}) & \isnotr \sum_{k{}^{\,}\in{}^{\,}\mathbb{Z}}\frac{-i\varrho^2 j\,\vec{\ncoverline{x}}\cdot\vec{\sigma}}{\vec{\ncoverline{x}}{}^{\,4} + \vec{\ncoverline{x}}{}^{\,2}(\ncoverline{t}-k)^2 + \vec{\ncoverline{x}}{}^{\,2}(\ncoverline{t}+j-k)^2+(\ncoverline{t}-k)^2(\ncoverline{t}+j-k)^2} \is \\
& \is -\frac{2i\pi\varrho^2\sinh(2\pi |\vec{\ncoverline{x}}|)\,\e{\vec{\ncoverline{x}}}\cdot\vec{\sigma}}{\cosh(2\pi|\vec{\ncoverline{x}}|) - \cos(2\pi\ncoverline{t})}\frac{j}{4\,\vec{\ncoverline{x}}{}^{\,2} + j^2}\
\end{aligned}
\end{equation}
and note that it is odd in $j$, i.e., the time copy sums vanish:
\begin{equation}
\label{eq:finite_T_propagator_traceless_finite}
\sum_{j{}^{\,}\in{}^{\,}\mathbb{Z}\setminus\lbrace 0\rbrace}\!\!(\pm 1)^j\ncoverline{\Delta}\big(\ncoverline{x},\ncoverline{x}+j\e{4})\big) \propto_{\setminus\tr} \!\sum_{j{}^{\,}\in{}^{\,}\mathbb{Z}\setminus\lbrace 0\rbrace}\!\!\frac{(\pm 1)^j j}{(4\,\vec{\ncoverline{x}}{}^{\,2}+j^2)j^2} \is 0 \,.
\end{equation}

Turning to the infinite (or rather, regularized) part of $\Delta^\pm(x',x)$, we set $j\is 0$, but have to keep $\ncoverline{\varepsilon}$. We choose a temporal splitting $\ncoverline{x}'\is \ncoverline{x} + \ncoverline{\varepsilon}_\tau{}^{\,}\e{4}$.
Note that this procedure for the $j\is 0$\dash mode is equivalent for the periodic and anti\dash periodic case.
In temporal point splitting regularization the traceful part of (\ref{eq:F_function}) reads
\begin{equation}
\label{eq:F_function_coinc_traceful_infinite}
\begin{aligned}
& F(\ncoverline{x}',\ncoverline{x}) \istr\\
& \istr 1 + \sum_{k{}^{\,}\in{}^{\,}\mathbb{Z}}\frac{\varrho^2\left(\vec{\ncoverline{x}}{}^{\,2} + (\ncoverline{t}+\ncoverline{\varepsilon}_\tau-k)(\ncoverline{t}-k)\right)}{\vec{\ncoverline{x}}{}^{\,4} + \vec{\ncoverline{x}}{}^{\,2}(\ncoverline{t}-k)^2 + \vec{\ncoverline{x}}{}^{\,2}(\ncoverline{t}+\ncoverline{\varepsilon}_\tau-k)^2+(\ncoverline{t}+\ncoverline{\varepsilon}_\tau-k)^2(\ncoverline{t}-k)^2} \is \\
& \is 1+\frac{\pi\varrho^2}{4\,\vec{\ncoverline{x}}{}^{\,2} + \ncoverline{\varepsilon}_\tau^{\,2}} \Big(2\,|\vec{\ncoverline{x}}|\sinh(4\pi|\vec{\ncoverline{x}}|) - 2\,|\vec{\ncoverline{x}}|\sinh(2\pi|\vec{\ncoverline{x}}|)\big(\cos(2\pi\ncoverline{t}) + \cos(2\pi(\ncoverline{t}+\ncoverline{\varepsilon}_\tau))\big) - \\
&\phantom{\is 1+\frac{1\pi\varrho^2}{4\,\vec{\ncoverline{x}}{}^{\,2} + \ncoverline{\varepsilon}_\tau^{\,2}} \Big(} - \ncoverline{\varepsilon}_\tau\cosh(2\pi|\vec{\ncoverline{x}}|)\big(\sin(2\pi(\ncoverline{t}+\ncoverline{\varepsilon}_\tau)) - \sin(2\pi\ncoverline{t})\big) + \ncoverline{\varepsilon}_\tau\sin(2\pi\ncoverline{\varepsilon}_\tau)\Big)\times \\
& \phantom{\is 1+\frac{1\pi\varrho^2}{4\,\vec{\ncoverline{x}}{}^{\,2} + \ncoverline{\varepsilon}_\tau^{\,2}}} \times \frac{1}{\big(\cosh(2\pi|\vec{\ncoverline{x}}|) - \cos(2\pi\ncoverline{t})\big)\big(\cosh(2\pi|\vec{\ncoverline{x}}|) - \cos(2\pi(\ncoverline{t}+\ncoverline{\varepsilon}_\tau))\big)}
\end{aligned}
\end{equation}
and for the traceless part we find
\begin{equation}
\label{eq:F_function_coinc_traceless_infinite}
\begin{aligned}
& F(\ncoverline{x}',\ncoverline{x}) \isnotr \sum_{k{}^{\,}\in{}^{\,}\mathbb{Z}} \frac{i\varrho^{2\,}\ncoverline{\varepsilon}_\tau\,\vec{\ncoverline{x}}\cdot\vec{\sigma}}{\vec{\ncoverline{x}}{}^{\,4} + \vec{\ncoverline{x}}{}^{\,2}(\ncoverline{t}-k)^2 + \vec{\ncoverline{x}}{}^{\,2}(\ncoverline{t}+\ncoverline{\varepsilon}_\tau-k)^2+(\ncoverline{t}+\ncoverline{\varepsilon}_\tau-k)^2(\ncoverline{t}-k)^2} \is \\
& \is \frac{i\pi\varrho^2 \,\e{\ncoverline{x}}\cdot\vec{\sigma}}{4\,\vec{\ncoverline{x}}{}^{\,2}+\ncoverline{\varepsilon}_\tau^{\,2}}\Big(\ncoverline{\varepsilon}_\tau\sinh(4\pi|\vec{\ncoverline{x}}|) + 2\,|\vec{\ncoverline{x}}|\cosh(2\pi|\vec{\ncoverline{x}}|)\big(\sin(2\pi(\ncoverline{t}+\ncoverline{\varepsilon}_\tau))-\sin(2\pi\ncoverline{t})\big) - \\
& \phantom{\is\frac{i\pi\varrho^2}{4\,\vec{\ncoverline{x}}{}^{\,2}+\ncoverline{\varepsilon}_\tau^{\,2}}\Big(} - \ncoverline{\varepsilon}_\tau\sinh(2\pi|\vec{\ncoverline{x}}|)\big(\cos(2\pi\ncoverline{t}) + \cos(2\pi(\ncoverline{t}+\ncoverline{\varepsilon}_\tau))\big) - 2\,|\vec{\ncoverline{x}}|\sin(2\pi\ncoverline{\varepsilon}_\tau)\Big)\times \\
&\phantom{\is \frac{i\pi\varrho^2}{4\,\vec{\ncoverline{x}}{}^{\,2}+\ncoverline{\varepsilon}_\tau^{\,2}}}\,\times\frac{1}{\big(\cosh(2\pi|\vec{\ncoverline{x}}|) - \cos(2\pi\ncoverline{t})\big)\big(\cosh(2\pi|\vec{\ncoverline{x}}|) - \cos(2\pi(\ncoverline{t}+\ncoverline{\varepsilon}_\tau))\big)}\,.
\end{aligned}
\end{equation}
Multiplied by $(\phi(x')\phi(x))^{-1/2}$, (\ref{eq:F_function_coinc_traceful_infinite}) and (\ref{eq:F_function_coinc_traceless_infinite}) give the traceful and traceless $j\is 0$\dash contributions to (\ref{eq:finite_T_propagator}), respectively.

By plugging (\ref{eq:finite_T_propagator_traceful_finite}) - (\ref{eq:F_function_coinc_traceless_infinite}) into (\ref{eq:finite_T_propagator}) we find the massless periodic and anti\dash periodic coincident propagator at finite temperature, regularized via point splitting in the temporal direction. Finally, we perform a Taylor expansion in $\ncoverline{\varepsilon}_\tau$:
\begingroup
\allowdisplaybreaks
\begin{align}
& \Delta^\pm(x'\is x + \varepsilon_\tau{}^{\,}\e{4},x)\is \frac{F(\ncoverline{x}',\ncoverline{x})}{4\pi^2\, \ncoverline{\varepsilon}_\tau^{\,2}\sqrt{\phi(x')\phi(x)}}\, + \!\sum_{j{}^{\,}\in{}^{\,}\mathbb{Z}\setminus\lbrace 0\rbrace}\!\!\frac{(\pm 1)^j F(\ncoverline{x},\ncoverline{x}+j\e{4})}{4\pi^2 j^2\,\phi(x)} \is \nonumber\\
& \is \frac{1}{4\pi^2\,\varepsilon_\tau^2} + C^\pm + \frac{\varrho^2\sinh(2\pi r)\left(1-2\pi r\cdot\left\lbrace \begin{aligned} & \coth(2\pi r) \\ & {}^{\,}\mathrm{csch}(2\pi r) \end{aligned}\right\rbrace\right)}{16\pi r^3\big(\cosh(2\pi r)-\cos(2\pi\tau) \big)\phi(x)} - \nonumber\\
&\phantom{\is} - \frac{\varrho^2}{64\pi r^3} \Bigg[\sinh(6\pi r) + 4\pi r \cosh(4\pi r) \cos(2\pi\tau) - \nonumber\\
& \phantom{\is} - \sinh(2\pi r) \left(\frac{8 \pi^3 \varrho^2 r \sinh(2\pi r) \sin^2(2\pi\tau)}{\big( \cosh(2\pi r) - \cos(2\pi\tau)\big)\phi(x)} - 2\cos(2\pi\tau) - 3\right) - \nonumber\\
&\phantom{\is} - 4 (\sinh(4\pi r) - 3\pi r) \cos(2\pi\tau)-4 \pi  r \cosh (2 \pi  r) (\cos (4\pi\tau) +3)\Bigg]\times \nonumber\\
&\phantom{\is}\times\Big(\big(\cosh (2\pi r) - \cos(2\pi\tau)\big)^3\, \phi(x) \Big)^{-1} + \nonumber\\
&\phantom{\is} + \frac{i\varrho^2\,\sigma^r}{64\pi r^3\, \varepsilon_\tau}\Bigg[ r \sinh(8\pi r) + \pi\varrho^2 \cosh(8\pi r) + \big((4\pi^2\varrho^2 - 6) r \sinh(6\pi r) + \nonumber\\
& \phantom{\is} + 4\pi(r^2 - \varrho^2)\cosh(6\pi r)\big)\cos(2\pi\tau) - 2\pi\Big((4r^2 - \varrho^2)\big(2+\cos(4\pi\tau)\big) + \varrho^2\Big)\cosh(4\pi r)  - \nonumber\\
&\phantom{\is} - 12\pi^2\varrho^2 r \sinh(4\pi r) + 8r \sinh(4\pi r) - 2(2\pi^2\varrho^2-3) r \sinh(4\pi r) \cos(4\pi\tau) + \nonumber\\
&\phantom{\is} + 4\pi\cosh(2\pi r) \big((14r^2 +\varrho^2)\cos(2\pi\tau) + r^2\cos(6\pi\tau)\big) - 2r \sinh(2\pi r) \cos(6\pi\tau) + \nonumber\\
&\phantom{\is}  + 4(5\pi^2\varrho^2 - 3) r \sinh(2\pi r)\cos(2\pi\tau) -\pi(8r^2+\varrho^2)(2\cos(4\pi\tau)+3)\Bigg]\times \nonumber\\
&\phantom{\is}\times\Big(\big(\cosh(2\pi r) - \cos(2\pi\tau)\big)^4\, \phi^2(x)\Big)^{-1} - \nonumber\\
&\phantom{\is} - \frac{i\varrho^2\,\sigma^r}{64\pi r^3}\Bigg[ 2\pi r \big(2\pi r \cosh(6 \pi r) + 8\pi^2\varrho^2 \sinh^3(2\pi r) + \sinh(6\pi r) \big)\sin(2\pi\tau)- \nonumber\\
&\phantom{\is} - 4\pi r \sinh(4\pi r) -  4\pi^2r^2 \cosh(2\pi r) \big(6 \sin(2\pi\tau) + \sin(6\pi\tau)\big) + \nonumber\\
&\phantom{\is} + 2\pi r \sinh(2\pi r)\big(2\sin(4\pi\tau) + \sin(6\pi\tau)\big) + 16\pi^2r^2\sin(4\pi\tau)\Bigg]\times \nonumber\\
&\phantom{\is}\times\Big(\big(\cosh(2\pi r) - \cos(2\pi\tau)\big)^4\, \phi^2(x)\Big)^{-1} + \mathcal{O}(\varepsilon_\tau) \label{eq:coincident_propagator}
\end{align}
\endgroup
with $C^\pm \is \lbrace \frac{1}{12},-\frac{1}{24} \rbrace$ and $\sigma^r \is \begin{pmatrix} \cos(\theta) & \sin(\theta)e^{-i\varphi} \\ \sin(\theta)e^{i\varphi} & -\cos(\theta) \end{pmatrix}$ being a function of only the polar and azimuthal angles $\theta$, $\varphi$. Note also: $\tr(\sigma^r)\is 0$ and $(\sigma^r)^2\is 1$.

The first two (constant) diagonal terms in $\Delta^\pm(x',x)$ correspond to the periodic/anti\dash periodic free field coincident propagator $\Delta^\pm_0(x',x)\is \left.\Delta^\pm(x',x)\right|_{\varrho{\,} =^{\,} 0}\is \frac{1}{4\pi^2\,\varepsilon_\tau^2} + C^\pm$.
The spacetime\dash dependent diagonal terms (third and fourth term) we write as $\Delta^\pm_\text{diag, finite}(x)$. In the periodic case, this finite contribution vanishes polynomially as $\Delta^+_\text{diag, finite}\stackrel{r^{\,}\rightarrow^{\,}\infty}{\propto} - \frac{\varrho^2}{8 r^2}$ for large distances from the caloron, while for the anti\dash periodic scalar of interest this term falls off exponentially as \mbox{$\Delta^-_\text{diag, finite} \stackrel{r^{\,}\rightarrow^{\,}\infty}{\propto} -\frac{\varrho^2}{2 r^2}\,e^{-2\pi r}\cos^2(\pi\tau)$}.
The off\dash diagonal part of the coincident propagator (fifth and sixth term) contains an ``$\varepsilon_{\tau}\rightarrow 0$\dash diverging'' term and a finite one, which we denote as $i\sigma^r(x)\,\varepsilon_\tau^{-1}\Delta_\text{off-diag, infinite}(x)$ and $i\sigma^r(x)\,\Delta_\text{off-diag, finite}(x)$. They fall off as $\Delta_\text{off-diag, infinite} \stackrel{r^{\,}\rightarrow^{\,}\infty}{\propto} \frac{\varrho^2}{8\pi r^2}$ and $\Delta_\text{off-diag, finite}\stackrel{r^{\,}\rightarrow^{\,}\infty}{\propto} -\frac{\pi\varrho^2}{2 r}\,e^{-2\pi r}\sin(2\pi\tau)$.

The fact that the anti\dash periodic propagator falls off exponentially at large separation, but the periodic one does not, explains why it is possible to perform an $m^2$\dash expansion in the anti\dash periodic but not the periodic case.
For periodic boundary conditions, the lowest Matsubara frequency is zero, and the logarithmic IR effects present in vacuum become more severe, appearing as a linear divergence in the $m^2$\dash coefficient.
We expect that the true $m^2$\dash dependence will be non\dash analytic $\propto m$, similar to what happens in the finite\dash $m$ expansion of the pressure \cite{Kirzhnits:1976ts}.

For small distances from the caloron (center) the coincident propagator\dash terms scale as $\Delta^+_\text{diag, finite}\stackrel{r^{\,} \rightarrow^{\,} 0}{\propto}  -\frac{\pi^2\varrho^2(1+\pi^2\varrho^2)}{2\big(2\pi^2\varrho^2 + 1 -\cos(2\pi\tau)\big)^2}\,$, $\,\Delta^-_\text{diag, finite}\stackrel{r^{\,} \rightarrow^{\,} 0}{\propto}  -\frac{\pi^2\varrho^2\cos^2(\pi\tau)}{2\big(2\pi^2\varrho^2 + 1 -\cos(2\pi\tau)\big)^2}$ for the periodic and anti\dash periodic traceful terms and $\Delta_\text{off-diag, infinite} \stackrel{r^{\,} \rightarrow^{\,} 0}{\propto} r\,\frac{\pi^2\varrho^2\big(2+\cos(2\pi\tau)\big)\csc^2(\pi\tau)}{6 \big(2\pi^2\varrho^2 + 1 -\cos(2\pi\tau)\big)}$ and \mbox{$\Delta_\text{off-diag, finite} \stackrel{r^{\,} \rightarrow^{\,} 0}{\propto}  -r\, \frac{\pi^3\varrho^2 \big(12\pi^2\varrho^2 + 9 - 8\cos(2\pi\tau) - \cos(4\pi\tau)\big)\cot(\pi\tau)\,\csc^2(\pi\tau)}{12\big(2\pi^2\varrho^2+1-\cos(2\pi\tau)\big)^2}$} for the traceless parts.

Using all of the above abbreviations, we write the closed loop propagator (\ref{eq:coincident_propagator}) as
\begin{equation}
\label{eq:coincident_propagator_short}
\begin{aligned}
\Delta^\pm(x',x) \is &\, \frac{1}{4\pi^2\,\varepsilon_\tau^2} + C^\pm + \Delta^\pm_\text{diag, finite}(r,\tau) + \frac{i\sigma^r}{\varepsilon_\tau}\,\Delta_\text{off-diag, infinite}(r,\tau)\,+ \\
& \,+ i\sigma^r\,\Delta_\text{off-diag, finite}(r,\tau) + \mathcal{O}(\varepsilon_\tau)\,.
\end{aligned}
\end{equation}

\subsection{Taylor Expansion -- Numerical Results}
\label{subsec:small_m_results}

We can now plug (\ref{eq:coincident_propagator_short}) into (\ref{eq:eff_act_Tay_ansatz}). According to the temporal point splitting employed for the coincident propagator, we also have to include a temporal Wilson line from $x$ to $x'$ in (\ref{eq:eff_act_Tay_ansatz}). Using $A^4_\text{HS}\is -\ncoverline{\eta}{}^{a4\nu}\partial^\nu\ln(\phi(r,\tau))\frac{\sigma^a}{2} \is -\vec{\partial}\ln\phi\,\cdot\frac{\vec{\sigma}}{2} \is - \frac{\partial_r \phi}{\phi}\,\frac{\sigma^r}{2}$ we find the Wilson line
\begin{equation}
\label{eq:Wilson_line_temp_splitting}
\begin{aligned}
& e^{i\!\int_{\tau}^{\tau+\varepsilon_\tau}\!\ab\tau\, A^4_\text{HS}(\vec{x},\tau)} \is 1 + i A^4_\text{HS}(x)\,\varepsilon_\tau + \frac{1}{2}\Big(i\!\left.\partial_\tau A^4_\text{HS}\right|_{x} - \big(A^4_\text{HS}(x)\big)^2\Big)\varepsilon_\tau^2 + \mathcal{O}\big(\varepsilon_\tau^3\big) \is \\
& \is 1-\frac{1}{8}\left(\left.\frac{\partial_r \phi}{\phi}\right|_{x}\right)^2 \varepsilon_\tau^2 - i\,\frac{\sigma^r}{2}\left.\left(\frac{\partial_r \phi}{\phi}\,\varepsilon_\tau + \frac{\partial_\tau\partial_r\phi}{2\phi}\,\varepsilon_\tau^2 - \frac{\partial_\tau\phi\,\partial_r\phi}{2\phi^2}\,\varepsilon_\tau^2\right)\right|_{x} + \mathcal{O}(\varepsilon_\tau^3)\,.
\end{aligned}
\end{equation}
We note the large$\,$- and small\dash distance behavior of the caloron field $A^4_\text{HS}\propto\frac{\partial_r\phi}{\phi}\stackrel{r^{\,}\rightarrow^{\,}\infty}{\propto}-\frac{\pi\varrho^2}{r^2}$ and $\frac{\partial_r\phi}{\phi}\stackrel{r^{\,}\rightarrow^{\,}0}{\propto} -r\,\frac{8\pi^4\varrho^2\big(2+\cos(2\pi\tau)\big)}{3\big(2\pi^2\varrho^2 +1 -\cos(2\pi\tau)\big)\big(1-\cos(2\pi\tau)\big)}$.

Plugging now also (\ref{eq:Wilson_line_temp_splitting}) into (\ref{eq:eff_act_Tay_ansatz}), we find the $m^2$\dash coefficient $- 2\gammas^\text{small}$ of the $\gamma_\text{ferm}$\dash expansion (\ref{eq:small_mass_exp_general}):
\begin{equation}
\label{eq:gamma_small}
\begin{aligned}
- 2\gammas^\text{small}(\varrho) \is & -2\lim_{\varepsilon_\tau{}^{\,}\rightarrow^{\,} 0}\Tr\!\left( \frac{1}{4\pi^{2\,}\varepsilon_\tau^2} +C^- - \frac{1}{32\pi^2}\Big(\left.\frac{\partial_r \phi}{\phi}\right|_{x}\Big)^2 + \Delta^-_\text{diag, finite}(r,\tau) \,+ \right. \\
&\phantom{-2\lim_{\varepsilon_\tau{}^{\,}\rightarrow^{\,} 0}\Tr\Big(}\; \left.+ \,\frac{1}{2}\, \Delta_\text{off-diag, infinite}(r,\tau)\left.\frac{\partial_r\phi}{\phi}\right|_x + \mathcal{O}(\varepsilon_\tau) - \Delta^-_0(x',x) \right) \is \\
& \is - 2\int_0^1\!\!\ab\tau\int_0^\infty\!\!\ab r\,r^2\left(-\frac{1}{4\pi}\Big(\left.\frac{\partial_r \phi}{\phi}\right|_{x}\Big)^2 + 8\pi\,\Delta^-_\text{diag, finite}(r,\tau) \,+\right. \\
&\phantom{\is - \int_0^1\!\!\ab\tau\int_0^\infty\!\!\ab r\,r^2\,\Big(}\;\, \left.+ \,4\pi\,\Delta_\text{off-diag, infinite}(r,\tau)\left.\frac{\partial_r\phi}{\phi}\right|_{x\,} \right),
\end{aligned}
\end{equation}
where the factor $8\pi$ in the second step is due to the solid angle integration and taking the $\mathfrak{su}(2)$\dash trace.

Due to the aforementioned large$\,$- and small\dash distance behavior of $\frac{\partial_r\phi}{\phi}$, $\Delta^+_\text{diag, finite}$, and $\Delta^+_\text{off-diag, infinite}$, the integral $\gammas^\text{small}$ is finite and we can calculate it numerically for different values of the parameter $\varrho$.
It is of interest to note that in the periodic case the corresponding integral $\gamma_{\text{s, }+}$ is linearly divergent. This is due to $\Delta^+_\text{diag, finite}(r,\tau)$ scaling as $r^{-2}$ for large distances as we stated above.

\begin{figure}[htbp]
\includegraphics[width=0.47\textwidth]{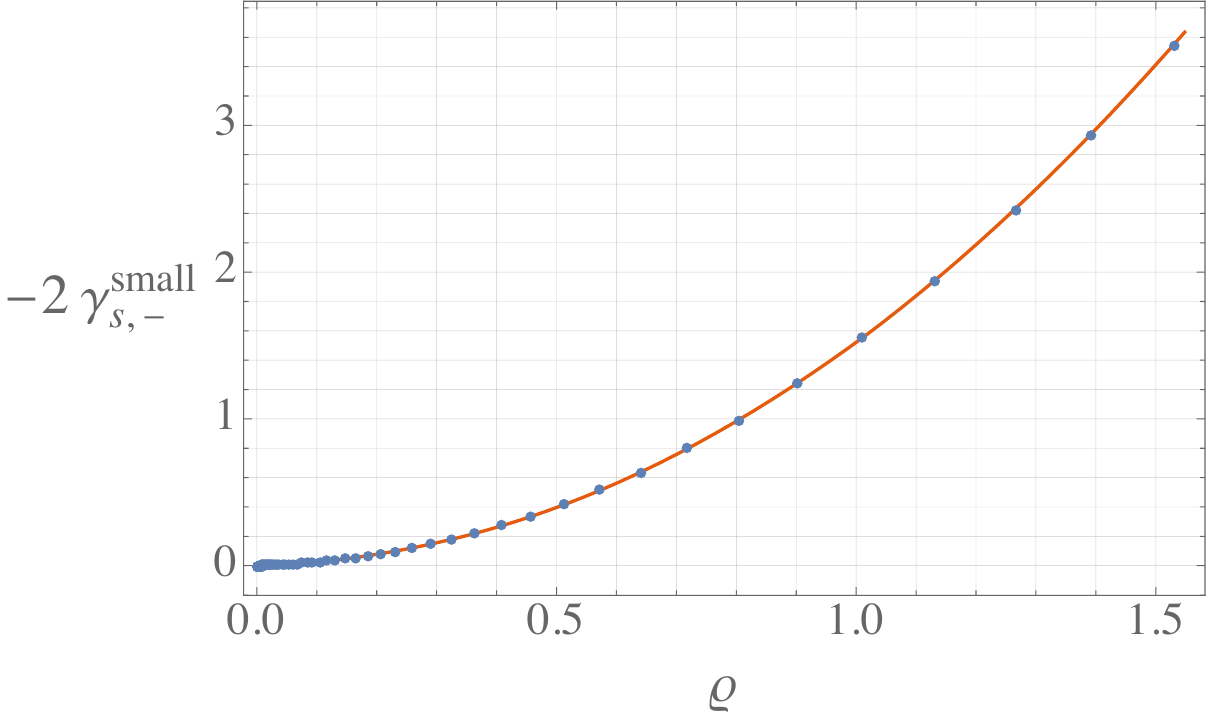}
\hfill
\includegraphics[width=0.47\textwidth]{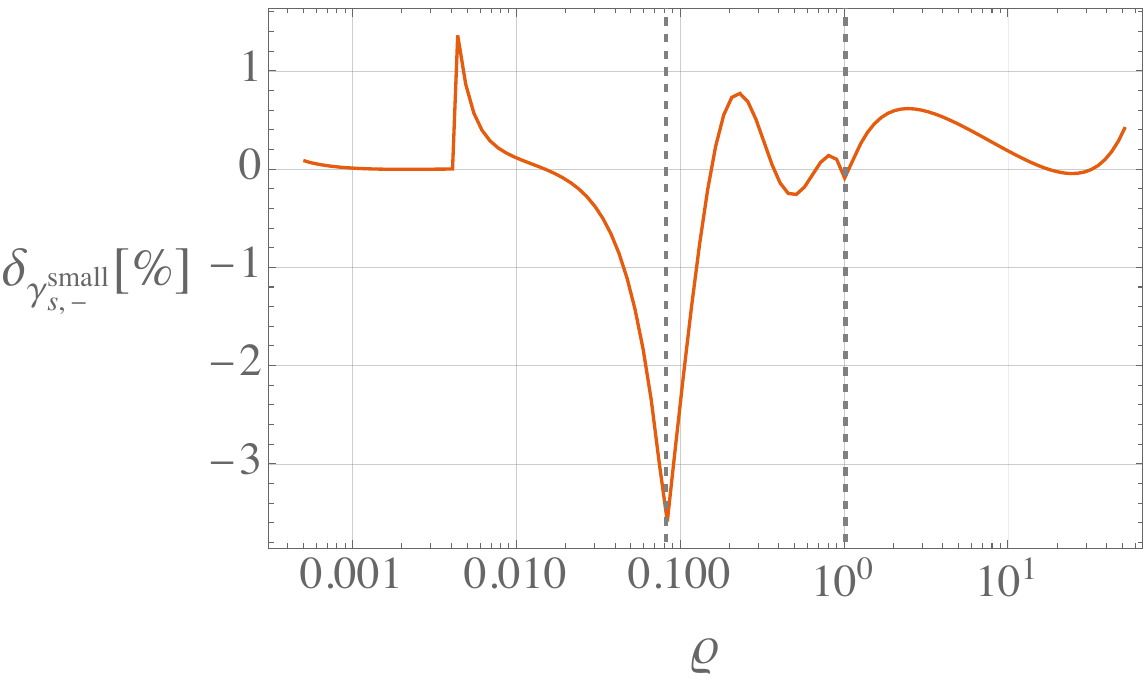}
\caption{Left:  $m^2$\dash coefficient $-2 \gammas^{\text{small}}$ as a function of the caloron size $\varrho$.
Right:  percent relative error of our fitting function, (\ref{eq:m^2_coeff}).
For the physically most important region $0.2 < \varrho < 0.6$, the fit is accurate to better than $1\%$.
}
\label{fig:gamma_small}
\end{figure}

We obtain $\gammas^\text{small}$ by numerical integration for a range of logarithmically spaced caloron sizes $5\cdot 10^{-4} \leq \varrho < 135$; it is well described by the following approximate form (we include the values and error of the fitting function in the ancillary files):
\begin{equation}
\label{eq:m^2_coeff}
\gammas^\text{small}(\varrho)\is \left\lbrace \begin{aligned}
& -0.5\tilde{\varrho}^2\ln(0.946\tilde{\varrho}) && 
0\leq \varrho\leq 0.082 \\
& -0.85\varrho^{1.80} + 0.59\tilde{\varrho}^{1.28} \quad && 
0.082 < \varrho \leq 1.045 \\
& -0.76\varrho^{2} && 
1.045<\varrho \end{aligned}\right. .
\end{equation}
We have divided $\varrho$ into three regions: a small region where the heavy quark effect is approximately the same as for a zero\dash temperature instanton (described by $\tilde{\varrho}$), an intermediate region which is of the most physical interest, and a large region where thermal effects grow with caloron size as $\varrho^2$.
We display the numerical values for $\gammas^{\text{small}}(\varrho)$ in
\hyperref[fig:gamma_small]{figure \ref{fig:gamma_small}}.

Our final small\dash mass result for the fermionic correction factor (\ref{eq:correction_factor_general}), using  (\ref{eq:eff_act_Tay_ansatz}) together with (\ref{eq:m^2_coeff}),
reads:
\begin{equation}
\label{eq:correction factor_small_m}
\mathpzc{f}_\text{small}(m,\varrho) \is \left.\mathpzc{f}(0,\varrho)\right|_{T^{\,}>^{\,}0\text{, ferm}}\,e^{2m^2\gammas^\text{small}(\varrho)} \is
\exp\!\left(-\frac{(\pi\varrho)^2}{3} + 2A(\pi\varrho) + 2m^2\gammas^\text{small}(\varrho)\right).
\end{equation}

The small\dash mass expansion is reasonable where $\gammas^{\text{small}}(\varrho)$ itself is small, which unsurprisingly is when $m < \varrho^{-1}$.
We add the condition that $m < 1$, as otherwise the small\dash mass expansion of the thermal part of the pressure is also not under control.
For the physically most important region around $\varrho \sim 0.4$, this latter condition is stronger.

\section{Large Mass -- Heat Kernel Expansion}
\label{sec:largemass}

\subsection{Structure of the Expansion}
\label{subsec:large_m_exp_structure}

In order to perform the large\dash mass ($m\gg 1$) expansion, we proceed analogously to \cite{high_low_m_instantons} and employ the \textsl{Schwinger proper time} ($s$) \textsl{representation} of the logarithmic correction factor term (\ref{eq:log_fermion_cal_density}). In this representation the mass is again (like for the Taylor expansion) separated from the purely caloron\dash dependent differential operator $-D^2_-$ and from $-\partial^2_-$:
\begin{equation}
\label{eq:proper_time_rep}
\gammas \is -\int_0^\infty\frac{\ab s}{s}\,\big(e^{-m^2 s}-e^{-\lambda^2 s}\big)\,\Tr\!\left(\Braket{x|\big(e^{-(-D_-^2)s}-e^{-(-\partial_-^2)s})|x}\right) .
\end{equation}
The proper time $s$ is also $\beta^{-2}$\dash rescaled with $[s]\is\text{mass}^{-2}$.
$\braket{x|e^{-(-D_-^2)s}|y}\is \Braket{xs|y}^-$ and $\braket{x|e^{-(-\partial_-^2)s}|y} \is\Braket{xs|y}^-_0$ are the anti\dash periodic proper time\dash Green's functions, i.e., they satisfy proper time\dash Schr\"odinger equations: $-\partial_s \Braket{xs|y}^-\is -D_{x,\,-}^2\Braket{xs|y}^-$ and analogously for $\Braket{xs|y}^-_0$. These Green's functions describe a propagation from $y$ to $x$ in proper time $s$.
The ordinary anti\dash periodic propagators from $y$ to $x$ in Euclidean time are then reproduced by $s$\dash integration: 
\begin{equation}
\label{eq:propagator_from_heat_kernel}
\Delta^-(x,y,m^2)\is\Braket{x|\frac{1}{-D_-^2+m^2}|y} \is \int_0^\infty\!\!\ab s\,\Braket{xs|y}^- e^{-m^2 s}
\end{equation}
and analogously for $\Delta^-_0(x,y,m^2)$ \cite{det_with_spacetime_mass}.
Since the ordinary propagator $[\Delta^\pm_{(0)}] \is \text{mass}^2$ and the $s$\dash integration adds $[\ab s]\is \text{mass}^{-2}$, the proper time\dash Green's functions must be $[\Braket{xs|y}^\pm_{(0)}]\is\text{mass}^{4}$. Their integration with $[\ab^4 x] \is \text{mass}^{-4}$ then yields a dimensionless quantity.
As the proper time representation splits off the mass, we can consider the proper time Green's functions of massless, anti\dash periodic bosons in a caloron background.
Thus, the logarithm (\ref{eq:proper_time_rep}) is again given by the spacetime integral over all massless and traced closed loops with a caloron background as shown in \hyperref[fig:closed_loops]{figure \ref{fig:closed_loops}}.

The proper time Green's functions are also called the \textsl{heat kernels} of their respective operators, here of $-D^2_-$ and $-\partial^2_-$, with respect to proper time and the coincident heat kernels in (\ref{eq:proper_time_rep}) naturally diverge.
Regularization is achieved by performing an asymptotic expansion (as mentioned in \hyperref[sec:strategy]{section \ref{sec:strategy}}), the so\dash called \textsl{heat kernel expansion} \cite{deWitt, heat_kernel_user_manual, thermal_heat_kernel, thermal_heat_kernel_short, thermal_heat_kernel_alt}. This expansion is achieved by expanding (\ref{eq:proper_time_rep}) for $s\lesssim m^{-2}\ll 1$, which is enforced by the exponential damping $e^{-m^2 s}$, $m\gg 1$ for non\dash infinitesimal $s$:
\begin{equation}
\label{eq:heat_kernel_exp_finite_t}
\Tr\!\left(\Braket{xs|x}^-\right)\is\Tr\!\left(\Braket{x|e^{-(-D^2_-)s}|x}\right)\,\stackrel{s\,\searrow\, 0}{\cong}\!\sum_{k^{\,}\in^{\,}\mathbb{N}{}^{\,}\cup{}^{\,}\mathbb{N}+\frac{1}{2}}\!\frac{s^{k-2}}{(4\pi)^2}\int^1\!\!\ab^4 x\,\tr\big( b_{2k}(A_\text{HS})\big)
\end{equation}
with the finite\dash $T$ \textsl{heat kernel} or \textsl{Seeley\dash DeWitt coefficients} $[b_{2k}]\is \text{mass}^{2k}$ given in \cite{thermal_heat_kernel, thermal_heat_kernel_short, thermal_heat_kernel_alt}.

The large\dash mass expansion corresponds to inserting the series expansion (\ref{eq:heat_kernel_exp_finite_t}) in (\ref{eq:proper_time_rep}), switching the order of the $k$\dash sum and $s$\dash integral, and performing the $s$\dash integral followed by the trace including a spacetime integral.
The uniform convergence properties needed to exchange sum and integral are not generally fulfilled, so the large\dash mass $k$\dash summation is generally only asymptotic.
This is actually expected; the identical sum arises for the case of an anti\dash periodic Klein\dash Gordon operator $D_-^2$ as for a periodic one $D_+^2$, but the complete result should differ for the two cases.
The boundary condition is only expected to matter through terms of order $m^b e^{-m}$, $b \sim 1$, which is the typical level of ambiguity associated with summing such an asymptotic series.
In using this large\dash mass series, we must either take care that $m$ is large enough to allow us to neglect such exponentially small effects, or we must incorporate some other boundary condition\dash dependent effects.
Later we will do so by combining this series with the small\dash $m^2$ series, which knows explicitly about the boundary conditions.
We will also return to the explicit treatment of boundary condition\dash dependent $m^b e^{-m}$\dash effects in 
\hyperref[appendix:finite_T_corrections]{appendix \ref{appendix:finite_T_corrections}}.

\begin{wrapfigure}{r}{0.22\textwidth}
\centering
\includegraphics[width=0.17\textwidth]{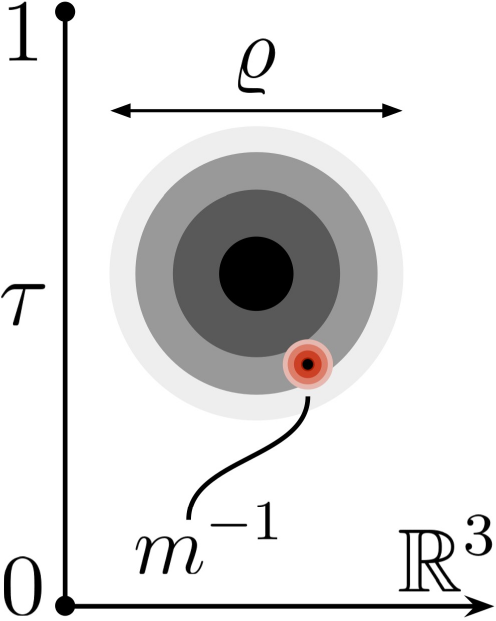}
\caption{}
\label{fig:large_m_exp}
\end{wrapfigure}
The exponential damping of boundary\dash condition dependence is due to heavy quarks $m\gg 1$ exploring $\spt$ on length scales $m^{-1}\ll 1$, i.e., essentially as they would $\mathbb{R}^4$. \hyperref[fig:large_m_exp]{Figure~\ref{fig:large_m_exp}} sketches this: a typical caloron $\varrho\approx 0.5$ (gray) and the heavy quark\dash propagation range (red).
Temperature only enters in the way that it modifies the caloron fields; the short\dash range propagation only feels the boundary conditions in an exponentially suppressed fashion because propagation of a heavy field over a distance $\beta^{\,}\,\widehat{=}^{\,}\,1$ is exponentially small.

The $m^{-(2k-4)}$\dash expansion at finite temperature is due to the $T\is 0$\dash heat kernel coefficients $\ncoverline{b}_{2k^{\,}=^{\,}\text{even}}(A_\text{HS})\subset b_{2k}$ (we discuss the full $b_{2k}$\dash coefficients for $T>0$ in \hyperref[appendix:finite_T_corrections]{appendix \ref{appendix:finite_T_corrections}}). The $T\is 0$\dash heat kernel coefficients are given in \cite{heat_kernel_user_manual, heat_kernel_higher_order}:
\begin{equation}
\label{eq:heat_kernel_coeff_T=0}
\begin{aligned}
& \ncoverline{b}_0 \is 1\,, \quad \ncoverline{b}_2(\ncoverline{x}) \is 0\,,\quad \ncoverline{b}_4(\ncoverline{x}) \is -\frac{1}{12}G^{\mu\nu}G^{\mu\nu}\,,\quad \ncoverline{b}_6\is \frac{i}{90}G^{\mu\nu}G^{\mu\kappa}G^{\nu\kappa}\,, \\
& \begin{aligned} \ncoverline{b}_8\is & \frac{1}{24}\left(\frac{17}{210}G^{\mu\nu}G^{\mu\nu}G^{\kappa\lambda}G^{\kappa\lambda}+\frac{2}{35}G^{\mu\nu}G^{\mu\kappa}G^{\nu\lambda}G^{\kappa\lambda}+\frac{1}{105}G^{\mu\nu}G^{\nu\kappa}G^{\kappa\lambda}G^{\lambda\mu}+\right. \\
& \left.\hphantom{\frac{1}{24}\Big(}+\frac{1}{420}G^{\mu\nu}G^{\kappa\lambda}G^{\mu\nu}G^{\kappa\lambda}\right),\end{aligned} \\
& \begin{aligned} \ncoverline{b}_{10} \is & \frac{1}{120}\left(\frac{i}{945}G^{\mu\nu}G^{\kappa\lambda}G^{\alpha\mu}G^{\nu\kappa}G^{\lambda\alpha} - \frac{47i}{126}G^{\mu\nu}G^{\mu\nu}G^{\kappa\lambda}G^{\lambda\alpha}G^{\alpha\kappa} +\right. \\
& + \frac{i}{126}G^{\mu\nu}G^{\kappa\lambda}G^{\mu\nu}G^{\lambda\alpha}G^{\alpha\kappa} + \frac{i}{63}G^{\mu\nu}G^{\nu\kappa}G^{\mu\lambda}G^{\lambda\alpha}G^{\alpha\kappa} - \frac{11i}{189}G^{\mu\nu}G^{\kappa\lambda}G^{\lambda\nu}G^{\mu\alpha}G^{\alpha\kappa} + \\
& + \frac{37i}{945}G^{\mu\nu}G^{\nu\kappa}G^{\kappa\lambda}G^{\lambda\alpha}G^{\alpha\mu} + \frac{4}{189}G^{\nu\alpha}G^{\alpha\lambda}G^{\nu\kappa;\mu}G^{\kappa\lambda;\mu} - \frac{2}{63}G^{\kappa\lambda}G^{\nu\alpha;\mu}G^{\nu\alpha}G^{\kappa\lambda;\mu} - \\
& - \frac{2}{189}G^{\kappa\lambda}G^{\nu\alpha;\mu}G^{\alpha\lambda}G^{\nu\kappa;\mu} + \frac{4}{63}G^{\kappa\lambda}G^{\kappa\lambda}G^{\nu\alpha;\mu}G^{\nu\alpha;\mu} + \frac{2}{63}G^{\mu\kappa}G^{\kappa\lambda}G^{\nu\alpha;\mu}G^{\nu\alpha;\lambda} + \\
& \left.+ \frac{4}{189}G^{\kappa\lambda}G^{\lambda\nu}G^{\nu\alpha;\mu}G^{\alpha\kappa;\mu}\right);\end{aligned}
\end{aligned}
\end{equation}
$\ncoverline{b}_{12}$ is given in \cite{heat_kernel_higher_order, a12_coeff} and the $\ncoverline{b}_{2k>12}$ are unknown. The vacuum coefficient is given by $\ncoverline{b}_0$.

\subsection{Heat Kernel Expansion Order by Order -- Numerical Results}
\label{subsec:large_m_results}

Now we plug the heat kernel coefficients (\ref{eq:heat_kernel_coeff_T=0}), together with the caloron field strength (derivatives) derived from (\ref{eq:HS_caloron}), into (\ref{eq:proper_time_rep}) and (\ref{eq:heat_kernel_exp_finite_t})
\begin{equation}
\label{eq:heat_kernel_order_by_order}
\gamma_{\text{s, }-} \is -\int_0^\infty\frac{\ab s}{s}\,\big(e^{-m^2 s}-e^{-\lambda^2 s}\big)\int^1\!\!\ab^4 x\;\tr\Bigg(\sum_{k^{\,}\in^{\,}\mathbb{N}}\!\frac{s^{k-2}}{(4\pi)^2}\,\ncoverline{b}_{2k}\big(A_\text{HS}(x)\big)-\ncoverline{b}_\text{free}\Bigg),
\end{equation}
performing the $s$\dash integrals first. They are of the structure (analogously for $\lambda^2$)
\begin{equation}
\label{eq:I_integral}
I(m^2,k)\is \int_0^\infty\!\!\ab s\,\,e^{-m^2s}\,s^{k-3} \left\lbrace\begin{aligned} 
& \rightarrow\,\infty \quad && \text{: }k\in\lbrace 0,1,2\rbrace \\ & \is \frac{1}{m^{2k-4}}\Gamma(k-2) \is \frac{(k-3)!}{m^{2k-4}} \quad && \text{: }\mathbb{N}\ni k\geq 3
\end{aligned}\right.\,.
\end{equation}
To parametrize the divergences in (\ref{eq:I_integral}), one introduces a small\dash scale cut\dash off $\varepsilon_s$ for the $s$\dash integral and finds
\begin{multline}
\label{eq:I_integral_diverging}
I(m^2,0\leq k\leq 2) \is \lim_{\varepsilon_s^{\,}\rightarrow^{\,}0}\int_{\varepsilon_s}^\infty\!\!\ab s\,e^{-m^2s}\,s^{k-3}\is \\
\is \left\lbrace\begin{aligned} 
& \frac{e^{-m^2\varepsilon_s}(1-m^2\varepsilon_s)}{2\varepsilon_s^2} + \frac{m^4}{2}\left(\ln(m^2\varepsilon_s)+\gamma_\text{E}+\mathcal{O}(m^2\varepsilon_s)\right) \quad && \text{: }k\is 0 \\ & \frac{e^{-m^2\varepsilon_s}}{\varepsilon_s} + m^2\left(\ln(m^2\varepsilon_s)+\gamma_\text{E}+\mathcal{O}(m^2\varepsilon_s)\right) \quad && \text{: }k\is 1 \\ & \lim_{\varepsilon_s^{\,}\rightarrow^{\,} 0}\Gamma(0,m^2\varepsilon_s)\is -\ln(m^2\varepsilon_s)-\gamma_\text{E}+\mathcal{O}(m^2\varepsilon_s) \quad && \text{: }k\is 2 \end{aligned}\right.\,,
\end{multline}
where $\Gamma(0,z)$ is the upper incomplete gamma function.

In the following we consider the different orders of the heat kernel expansion and present our results for the numerical calculation of the resulting caloron density contributions. We obtain explicit functional forms (for numerical $x$\dash integration) for the $\tr(\ncoverline{b}_{2k})$ by performing analytical calculations using the \textsl{OGRe}\dash package \cite{ogre} for \textsl{Mathematica}.\newline

\noindent \textbf{\underline{\textsl{Order $k\is 0^{\,}$}:}}

The $k\is0$\dash coefficients of the caloron is $\ncoverline{b}_0(A_\text{HS}) \is 1$ and multiplies both the quadratically divergent $s$\dash integral $I(m^2,0)$ (\ref{eq:I_integral_diverging}) and $V_4 \is \mathrm{vol}(\spt) \is\int^1\!\ab x^4$. This contribution is cancelled identically by the free term $b_\text{free}\is 1$. The $k\is 0$\dash contribution to $\gammas$ (\ref{eq:heat_kernel_order_by_order}) thus vanishes: $\gamma_0\is 0$.\newline

\noindent \textbf{\underline{\textsl{Order $k\is 1^{\,}$}:}}

For $-D^2_-(A_\text{HS})$ the coefficient $b_2$ vanishes.\newline

\noindent \textbf{\underline{\textsl{Order $k\is 2^{\,}$}:}}

The $k\is 2$\dash proper time integral diverges logarithmically as $I(m^2,2)\propto\Gamma(0)$ (\ref{eq:I_integral_diverging}); this logarithmic divergence is canceled by the Pauli\dash Villars \dash regulator term (see \cite{high_low_m_instantons, det_with_spacetime_mass} for $T\is 0$\dash case). The $\gamma_\text{E}$\dash terms of (\ref{eq:I_integral_diverging}) cancel as well.
The $\gammas$\dash contribution at order $k\is 2$ thus reads
\begin{equation}
\label{eq:b4}
\gamma_4(m,\lambda)\is\ln\Big(\frac{\lambda^2}{m^2}\Big)\frac{1}{(4\pi)^2}\,\text{Tr}\big(\ncoverline{b}_4\big) \is \ln\Big(\frac{m}{\lambda}\Big)\frac{1}{6(4\pi)^2}\int^1\!\!\ab^4 x\,\tr\big(G^{\mu\nu}G^{\mu\nu}\big)\is\frac{1}{6}\ln\!\Big(\frac{m}{\lambda}\Big)\,,
\end{equation}
where we used the definition of the caloron topological charge density (\ref{eq:top_charge_density}) (analogously to $T\is 0$\dash case in \cite{high_low_m_instantons}).

In order to verify our numerical methods, we compute $\frac{1}{(4\pi)^2}\Tr{}^{\!}\big(\ncoverline{b}_4(A_\text{HS})\big)$ numerically for 35 caloron sizes ranging over several orders of magnitude from 0.005 to 4854.
Our numerical results agree with the analytical value up to corrections which remain below $2.3\times 10^{-7}$ (see our data in the ancillary files) which verifies the precision of our numerical approach.

\noindent \textbf{\underline{\textsl{Order $k\is 3^{\,}$}:}}

$I\big(m^2,k\big)$ is finite for $k\geq 3$ (\ref{eq:I_integral}), e.g., $I(m^2,3)\is\frac{1}{m^2}$. For $k\is 3$ we find the $\gammas$\dash term
\begin{equation}
\label{eq:b6_expression}
\frac{\gamma_6(\varrho)}{m^2}\is I(m^2,3,0;0)\,\frac{1}{(4\pi)^2}\, \text{Tr}\big(\ncoverline{b}_6\big) \is -\frac{1}{1440\pi\,m^2}\int_0^1\!\!\ab\tau\!\int_0^\infty\!\!\ab r\,r^2\,\varepsilon^{abc\,}G^{a\,\mu\nu}G^{b\,\mu\kappa}G^{c\,\nu\kappa}\,.
\end{equation}
The corresponding $\lambda$\dash term therefore vanishes as $\lambda\rightarrow\infty$. We calculate $\gamma_6$ for log\dash spaced $\varrho$\dash values 
between 0.001 and 477 and describe the results with a fitting function defined piecewise in the regions of small, intermediate, and large caloron sizes:
\begin{equation}
\label{eq:b6}
\frac{\gamma_6(\varrho)}{m^2}\is \left\lbrace\begin{aligned}
& \frac{0.013}{\tilde{\varrho}^{2.00}m^2} && \text{: }0<\varrho\leq 0.267 \\
& \frac{0.128 \tilde{\varrho}^{1.55}}{m^2} + \frac{0.021}{\varrho^{1.78} m^2} && \text{: } 0.267<\varrho\leq 1.844 \\
& \frac{0.008}{\varrho^{2.00} m^2} + \frac{0.052}{m^2} && \text{: }1.844<\varrho
\end{aligned}\right. .
\end{equation}
\begin{figure}[htb]
\centerline{\includegraphics[width=0.6\textwidth]{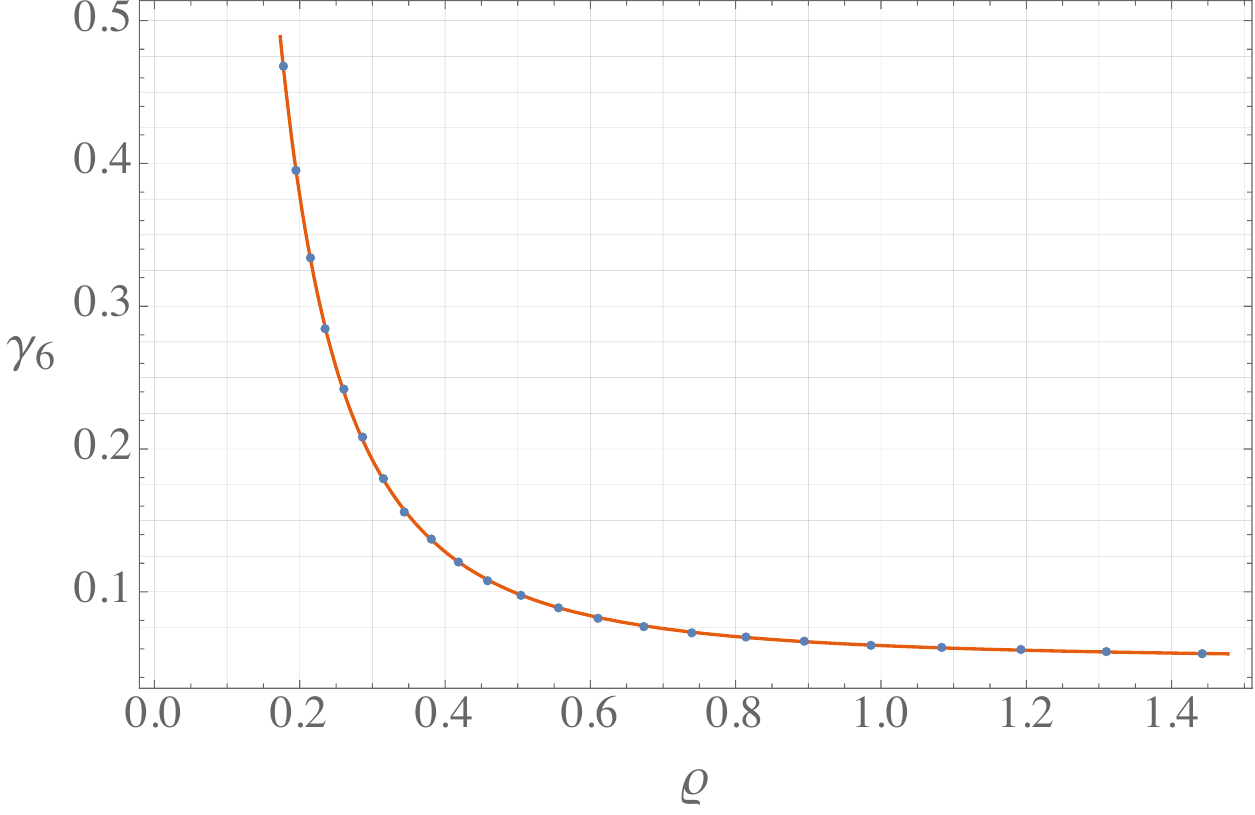}}
    \caption{
    Numerical result for $\gamma_6(\varrho)$, the $m^{-2}$\dash coefficient, as a function of caloron size $\varrho$.}
    \label{fig:gamma6}
\end{figure}
This fit agrees with the exact numerical integration to within (about) $0.3\%$ across all\linebreak caloron sizes.
Note that the $\varrho \ll 1$ regime agrees with the exact result in an instanton background, $\gamma_6(\varrho) \is 1/(75 \varrho^2) \is 0.01\overline{3} / \varrho^2$.
We show the result of the numerical integration in \hyperref[fig:gamma6]{figure \ref{fig:gamma6}} and provide a table of the numerical values as a function of $\varrho$ an the fitting error in the ancillary files.

\noindent \textbf{\underline{\textsl{Order $k\is 4^{\,}$}:}}

\begin{figure}[htb]
    \centerline{\includegraphics[width=0.6\textwidth]{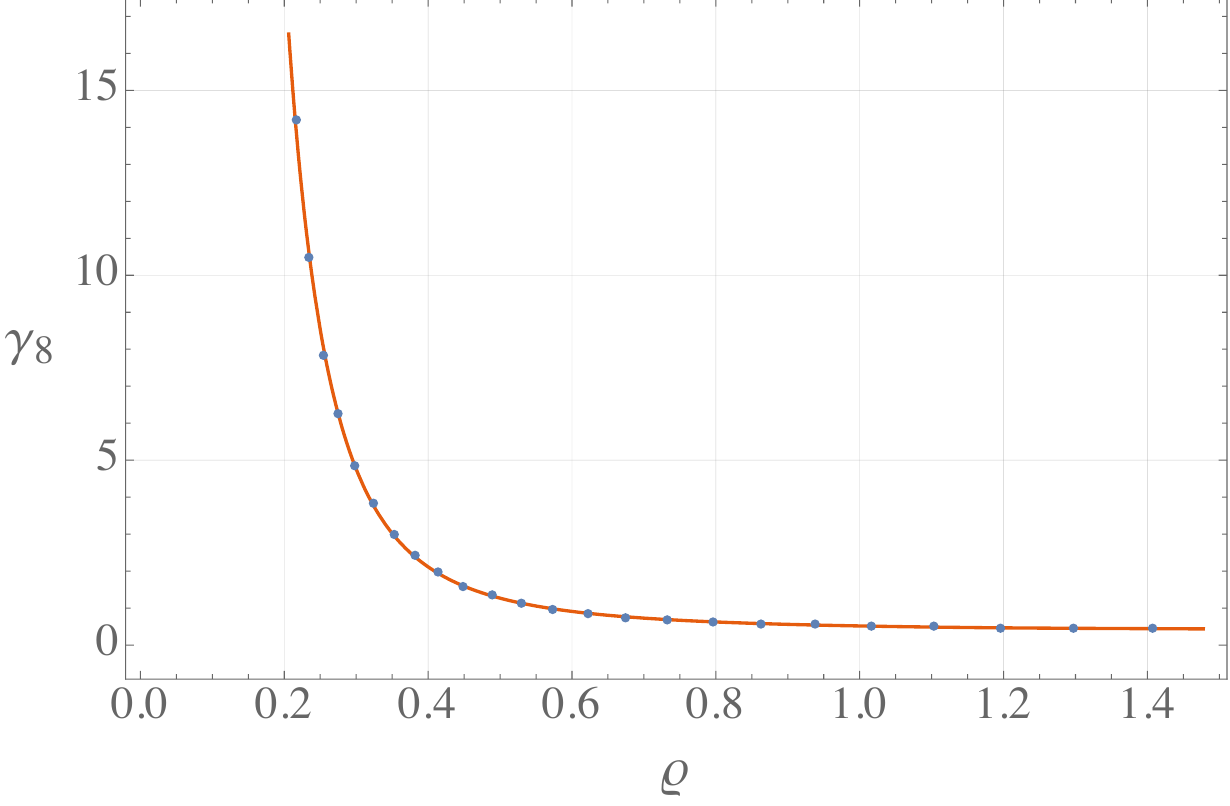}}
    \caption{Coefficient $\gamma_8(\varrho)$, describing $m^{-4}$\dash corrections, as a function of caloron size $\varrho$.}
    \label{fig:gamma8}
\end{figure}

Using (\ref{eq:I_integral}) with $I\big(m^2,4,0;0\big)\is \frac{1}{m^4}$, we calculate the $\gammas$\dash contribution at $k\is 4$ for log\dash spaced caloron sizes between 0.001 and 462:
\begin{equation}
\label{eq:b8}
\begin{aligned} \frac{\gamma_8(\varrho)}{m^4} & \is I(m^2,4,0;0)\frac{1}{(4\pi)^2}\,\text{Tr}\big(\ncoverline{b}_8\big) \is \\
& \is \left\lbrace\begin{aligned}
& \frac{0.023}{\tilde{\varrho}^4 m^4} && \text{: }0<\varrho\leq 0.120 \\
& \frac{0.090}{\tilde{\varrho}^{2.34}m^4} + \frac{0.021}{\varrho^{4.03} m^4} && \text{: } 0.120<\varrho\leq 1.183 \\
& \frac{0.020}{\varrho^{3.11} m^4} + \frac{0.096}{\varrho^{1.99} m^4} + \frac{0.388}{m^4} && \text{: }1.183<\varrho
\end{aligned}\right. ,\end{aligned}
\end{equation}
where the full $\varrho<0.12$\dash coefficient in (\ref{eq:b8}) fits the expected instanton $m^{-4}$\dash coefficient $\frac{17}{735}$ in \cite{high_low_m_instantons} up to $1\%$.
The fitting function is generally accurate to about $1\%$ except around $\varrho\is 0.25$, where it deviates from the numerical results by up to circa $3\%$.
Again, the numerical results are tabulated in the ancillary files.
The result of numerical integration is presented in \hyperref[fig:gamma8]{figure \ref{fig:gamma8}}.
\newline

\noindent \textbf{\underline{\textsl{Order $k\is 5^{\,}$}:}}

\begin{figure}[htbp]
\centerline{\includegraphics[width=0.6\textwidth]{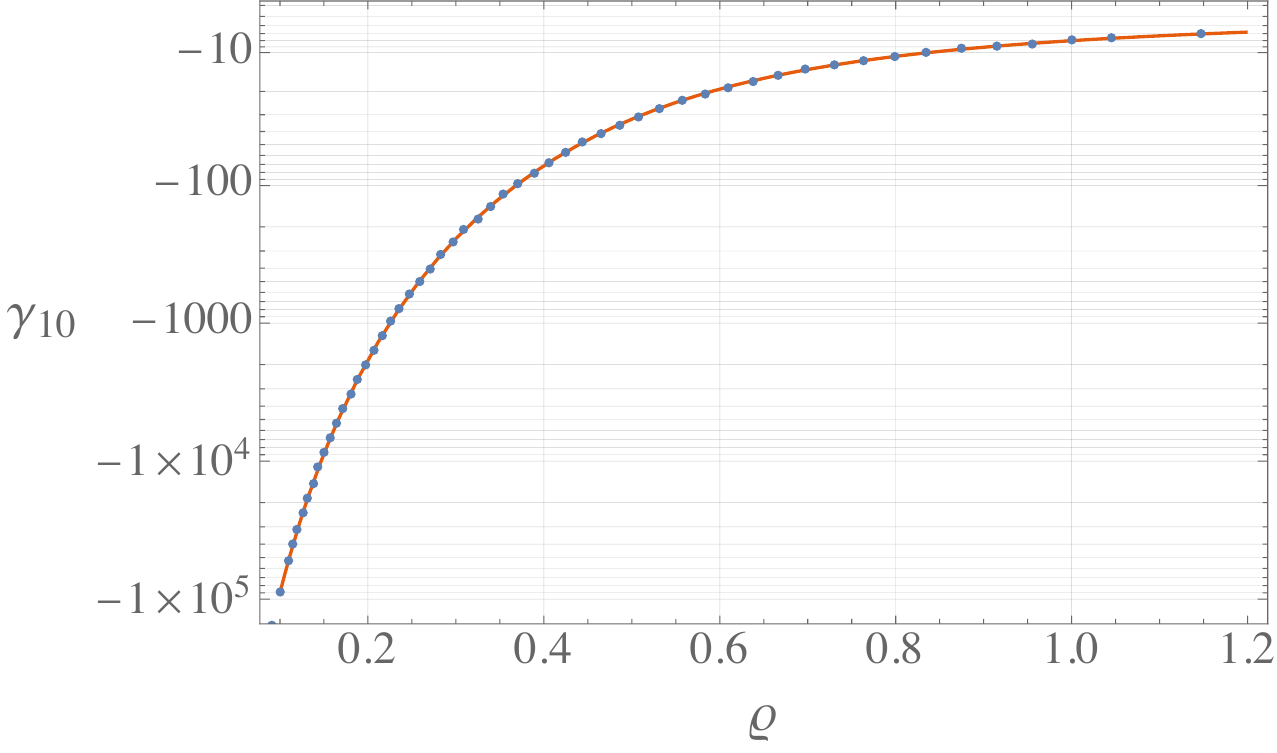}}
\caption{$m^{-6}$\dash coefficient $\gamma_{10}$ (\ref{eq:b10}) of $\gammas$ (\ref{eq:heat_kernel_order_by_order}).\label{fig:gamma10}}
\end{figure}

We calculate the $\mathcal{O}(m^{-6})$\dash contribution to $\gammas$ using (\ref{eq:I_integral}) with $I\big(m^2,5\big)\is \frac{2}{m^6}$ 
for (again log\dash spaced) caloron sizes between $0.001$ and $100$:
\begin{equation}
\label{eq:b10}
\begin{aligned} \frac{\gamma_{10}(\varrho)}{m^6}& \is I(m^2,5,0;0)\frac{1}{(4\pi)^2}\,\text{Tr}\big(\ncoverline{b}_{10}\big) \is \\
& \is \left\lbrace\begin{aligned}
& -\frac{0.082}{\tilde{\varrho}^6 m^6} && \text{: }0<\varrho\leq 0.298 \\
& -\frac{0.463}{\tilde{\varrho}^{3.87}m^6} + \frac{0.267}{\varrho^{5.33} m^6} && \text{: } 0.298<\varrho\leq 1.348 \\
& -\frac{2.280}{\varrho^{2.15} m^6} - \frac{5.106}{m^6} && \text{: }1.348<\varrho
\end{aligned}\right. .
\end{aligned}
\end{equation}
The fitting function has a relative error which is generally $\leq 1\%$.
The small\dash $\varrho$ coefficient here is in good agreement with the instanton coefficient $\frac{232}{2385}$ from \cite{high_low_m_instantons}.
We present our numerical results\footnote{The authors would like to specifically thank Simon Stendebach for setting up the code for the numerical $x$\dash integration of $\tr(\ncoverline{b}_{10})$ in (\ref{eq:b10}) and performing the integrals (using the \textsl{Cubature} package \cite{cubature, cubature_algorithm1, cubature_algorithm2, cubature_parallelization} to handle the highly oscillatory integrand $\tr(\ncoverline{b}_{10})$).} in \hyperref[fig:gamma10]{figure \ref{fig:gamma10}} and tabulate the numerical values in the ancillary materials.\newline

\noindent \textbf{\underline{\textsl{Order $5 < k\in \mathbb{N}^{\,}$}:}}

Due to expected computational cost in calculating the explicit functional form of $\ncoverline{b}_{12}(x,\varrho)$ and numerical difficulties in performing $\Tr{}{\!}(\ncoverline{b}_{12})$, we do not obtain this contribution. For integer $k>6$ no heat kernel coefficients are known.\newline

\noindent \textbf{\underline{\textsl{Overall Result for $m\gg 1^{\,}$}:}}

The regularized, vacuum\dash normalized Klein\dash Gordon operator determinant for a heavy, anti\dash periodic, complex scalar (\ref{eq:heat_kernel_order_by_order}) then reads
\begin{equation}
\label{eq:heat_kernel_result}
\gammas(m\text{ large},\varrho,\lambda) \is \frac{1}{6}\ln\!\Big(\frac{\lambda}{m}\Big) - \frac{\gamma_6(\varrho)}{m^2} - \frac{\gamma_8(\varrho)}{m^4} - \frac{\gamma_{10}(\varrho)}{m^6}\,,
\end{equation}
which is of the structure (\ref{eq:large_mass_exp_general}) with $\gammas^{\text{large, }k}\is\gamma_{2k}$ given by (\ref{eq:b6}) - (\ref{eq:b10}). Using this, we obtain the large\dash mass result for the fermionic correction factor (\ref{eq:correction_factor_general}):
\begin{equation}
\label{eq:correction factor_large_m}
\begin{aligned}
\fferm (m\geq m_\text{large, min}(\varrho),\varrho) & \is \mathpzc{f}_{\,\text{large}}(m,\varrho)\is \\
& \is \frac{e^{-2\alpha\left(\frac{1}{2}\right)}}{(m\varrho)^\frac{1}{3}}\,\exp\!\left(-\frac{2\gamma_6(\varrho)}{m^2} - \frac{2\gamma_8(\varrho)}{m^4} - \frac{2\gamma_{10}(\varrho)}{m^6}\right).
\end{aligned}
\end{equation}

In order to estimate the range of validity of (\ref{eq:heat_kernel_result}) and (\ref{eq:correction factor_large_m}), we demand that successive terms become smaller in magnitude.
For that, we find the ``lightest heavy mass'' so that 1) $\frac{\gamma_6}{m^2}\geq \frac{\gamma_8}{m^4}\,$, 2) $\frac{\gamma_8}{\gamma_6 m^2}\geq \frac{|\gamma_{10}|}{\gamma_6 m^4 + \gamma_8 m^2}\,$, and 3) the finite\dash temperature ambiguities discussed in \hyperref[appendix:finite_T_corrections]{appendix \ref{appendix:finite_T_corrections}} are small compared to (\ref{eq:heat_kernel_result}).
From 1) and 2) we deduce the lower limit $1.4\tilde{\varrho}^{-1}$ on $m$ for $\varrho\lesssim 1$ and $m\gtrsim 2.8$ for larger $\varrho$. The lower mass limit $m_\text{large, min, 3}(\varrho)$ ensuring small $T>0$\dash uncertainties is shown in \hyperref[fig:m_large_min_3]{figure \ref{fig:m_large_min_3}} in \hyperref[appendix:finite_T_corrections]{appendix \ref{appendix:finite_T_corrections}}. All in all, the lower bound for the heavy quark mass expansion is 
\begin{equation}
\label{eq:m_large_min}
m_\text{large, min}(\varrho)\is \max\!\left( \left\lbrace \begin{matrix}& 1.4\tilde{\varrho}^{-1} &\text{: }m\lesssim 1 \\ & 2.8 & \text{: }m\gtrsim 1 \end{matrix}\right\rbrace,m_\text{large, min, 3}(\varrho)\right).
\end{equation}

\section{Interpolation and Application to the Susceptibility}
\label{sec:Pade}

\subsection{Interpolation}
\label{sec:interpolate}

Having performed the small$\,$- and large\dash mass expansions, we can use our results (\ref{eq:correction factor_small_m}) and (\ref{eq:correction factor_large_m}) for the fermionic part of the correction factor $\fferm$ (\ref{eq:correction_factor_general}) to perform the Pad\'e approximation according to (\ref{eq:correction_factor_small_and_large_m}) - (\ref{eq:K_pade_ansatz_general}) as described in \hyperref[sec:strategy]{section \ref{sec:strategy}}.
Since we have an odd $k_\text{max} \is 3$ in (\ref{eq:correction_factor_small_and_large_m}), we set $K\is 1$ (cf. (\ref{eq:K_pade_ansatz_general})) in (\ref{eq:pade_ansatz_general}):
\begin{align}
& \begin{aligned}
& -\ln\!\big(\,\fferm(m,\varrho)\big) \is \gamma_\text{ferm}\is 2\alpha\!\left(\frac{1}{2}\right)+\mathpzc{p}(m,\varrho) \is \\
& \is 2\alpha\!\left(\frac{1}{2}\right)+\left\lbrace \begin{aligned}
& -2\alpha\!\left(\frac{1}{2}\right)+\frac{(\pi\varrho)^2}{3}-2A(\pi\varrho) - 2m^2 \,\gammas^\text{small}(\varrho) &&\text{: }m\leq m_\text{small, max}(\varrho) \\
& \,\frac{1}{3}\ln(m\varrho) + \frac{2\gamma_6(\varrho)}{m^2} + \frac{2\gamma_8(\varrho)}{m^4} + \frac{2\gamma_{10}(\varrho)}{m^6} &&\text{: }m\geq m_\text{large, min}(\varrho) \end{aligned}\right.
\end{aligned} \label{eq:correction_factor_small_and_large_m_final} \\
& \stackrel{k_\text{max}{}^{\,}=^{\,}3\,\Rightarrow \, K^{\,}=^{\,}1}{\Longrightarrow}\; \mathpzc{p}(m,\varrho) \is \frac{\mathpzc{p}_{\,{}^{\!}0}(\varrho) + \mathpzc{p}_{\,{}^{\!}1}(\varrho)\, m^2}{\big(1+\mathpzc{P}_1(\varrho)\,m^2\big)\big(1+\mathpzc{P}_2(\varrho)\, m^2\big)} + \frac{1}{6}\ln\!\left(m^2\varrho^2 + \xi^2(\varrho)\right). \label{eq:pade_ansatz_explicit}
\end{align}

Now we perform the Taylor and Laurent expansions of (\ref{eq:pade_ansatz_explicit}) for small and large masses up to $\mathcal{O}(m^2)$ and $\mathcal{O}(m^{-6})$, respectively, and demand agreement with (\ref{eq:correction_factor_small_and_large_m_final}):
\begin{align}
& \text{\underline{small\dash $m$ Taylor expansion:}} \nonumber \\
& \mathpzc{p}_{\,{}^{\!}0} + \frac{1}{6}\ln(\xi^2) + \left(\mathpzc{p}_{\,{}^{\!}1} - \mathpzc{p}_{\,{}^{\!}0} \mathpzc{P}_1 - \mathpzc{p}_{\,{}^{\!}0} \mathpzc{P}_2 + \frac{\varrho^2}{6\xi^2}\right)m^2 \stackrel{!}{=} -2\alpha\!\left(\frac{1}{2}\right) +\frac{(\pi\varrho)^2}{3} - 2A - 2m^2\,\gammas^\text{small}\,,\\
& \text{\underline{large\dash $m$ Laurent expansion:}} \nonumber\\
& \begin{aligned} & \frac{1}{6} \ln(m^2 \varrho^2) + \frac{6 \mathpzc{p}_{\,{}^{\!}1} \varrho ^2+\mathpzc{P}_1 \mathpzc{P}_2 \xi^2}{6 \mathpzc{P}_1 \mathpzc{P}_2\, m^2 \varrho^2} + \frac{12 \mathpzc{p}_{\,{}^{\!}0} \mathpzc{P}_1 \mathpzc{P}_2 \varrho^4 - 12 \mathpzc{p}_{\,{}^{\!}1} \mathpzc{P}_1 \varrho^4 - 12 \mathpzc{p}_{\,{}^{\!}1} \mathpzc{P}_2 \varrho^4 - \mathpzc{P}_1^2 \mathpzc{P}_2^2 \xi^4}{12 \mathpzc{P}_1^2 \mathpzc{P}_2^2\, m^4 \varrho^4} - \\
& - \frac{18 \mathpzc{p}_{\,{}^{\!}0} \mathpzc{P}_1^2 \mathpzc{P}_2 \varrho^6 + 18 \mathpzc{p}_{\,{}^{\!}0} \mathpzc{P}_1 \mathpzc{P}_2^2 \varrho^6 - 18 \mathpzc{p}_{\,{}^{\!}1} \mathpzc{P}_1^2 \varrho^6 - 18 \mathpzc{p}_{\,{}^{\!}1} \mathpzc{P}_1 \mathpzc{P}_2 \varrho^6 - 18 \mathpzc{p}_{\,{}^{\!}1} \mathpzc{P}_2^2 \varrho^6 - \mathpzc{P}_1^3 \mathpzc{P}_2^3 \xi^6}{18 \mathpzc{P}_1^3 \mathpzc{P}_2^3\, m^6 \varrho^6} \stackrel{!}{=} \\ 
& \stackrel{!}{=} \frac{1}{6} \ln(m^2 \varrho^2) + \frac{2\gamma_6}{m^2} + \frac{2\gamma_8}{m^4} + \frac{2\gamma_{10}}{m^6}\,.
\end{aligned}
\end{align}

\begin{figure}[ht]
\centering
\subcaptionbox{Left: numerical values for the parameter $\xi(\varrho)$ of the ``Pad\'e\dash like'' approximation (\ref{eq:pade_ansatz_explicit}), shown for the physically relevant caloron sizes $\varrho$. Right: $\xi$ shown and for the full set of $\varrho$\dash values.\label{fig:xi_pade_parameter}}[\textwidth]{\includegraphics[scale=0.56]{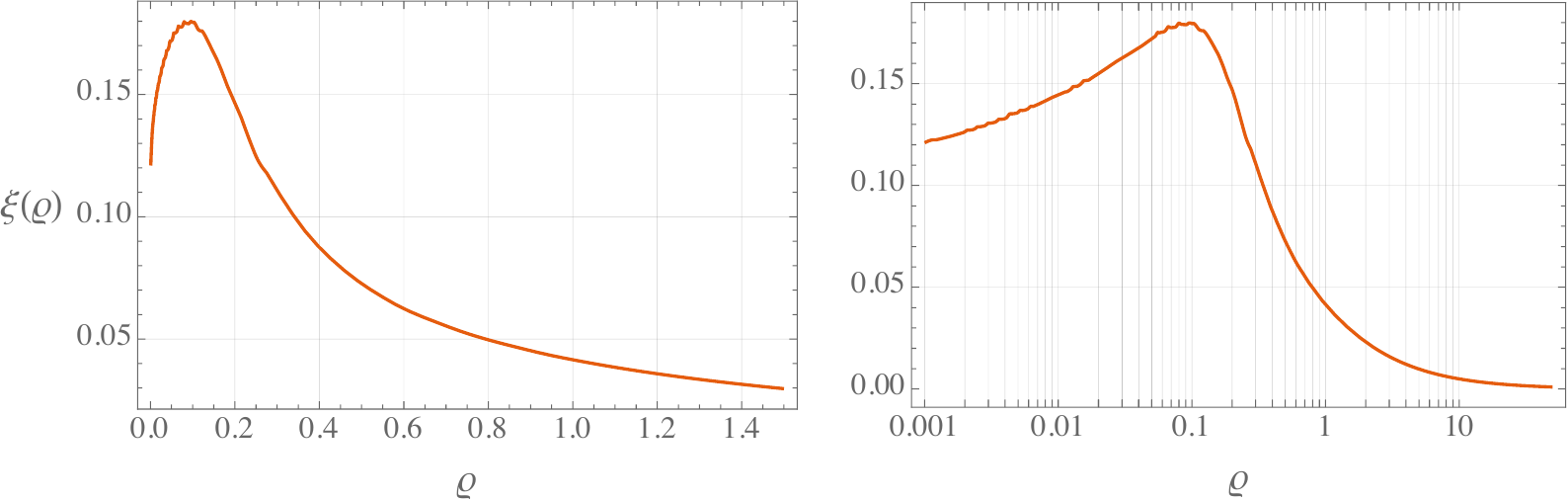}}
\par\bigskip
\subcaptionbox{The parameters $\mathpzc{P}_1(\varrho)$ and $\mathpzc{P}_2(\varrho)$ of (\ref{eq:pade_ansatz_explicit}), shown for the full set of $\varrho$\dash values; the parameters are positive\dash definite functions as demanded in (\ref{eq:pade_ansatz_general}).\label{fig:P1_P2_pade_parameters}}[\textwidth]{\includegraphics[scale=0.53]{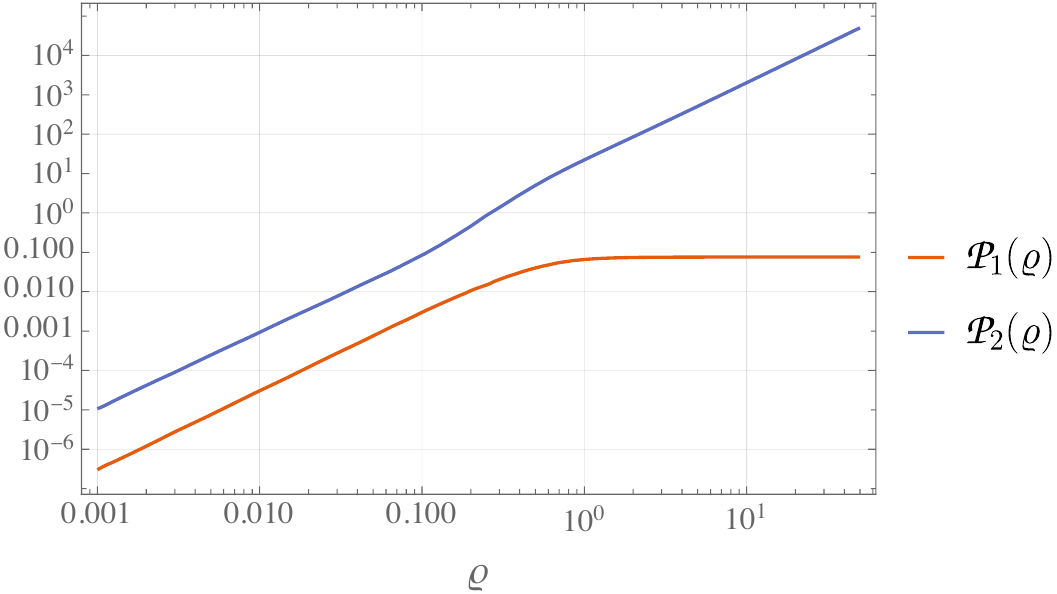}}
\caption{}\label{fig:pade_parameters}
\end{figure}

\begin{figure}[ht]
\centering
\subcaptionbox{\label{fig:pade_agreement_3d_1}}[\textwidth]{\includegraphics[width=\textwidth]{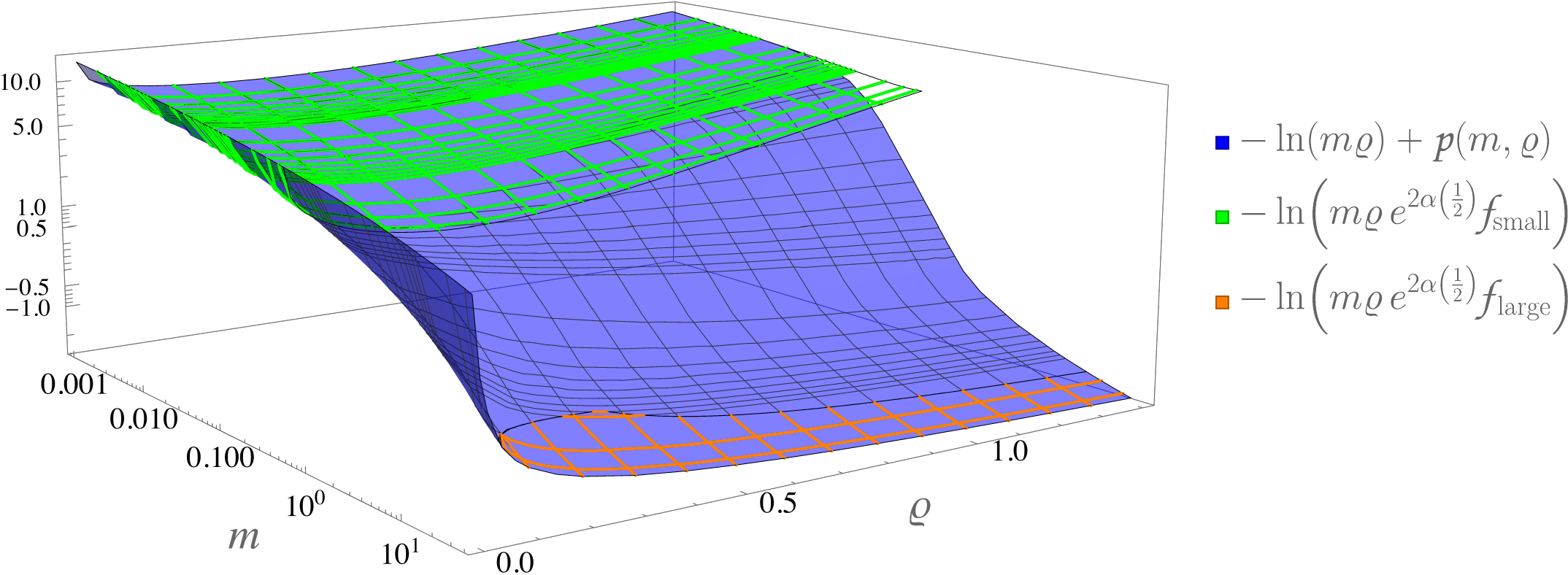}}

\par\bigskip

\subcaptionbox{\label{fig:pade_agreement_3d_2}}[\textwidth]{\includegraphics[width=0.78\textwidth]{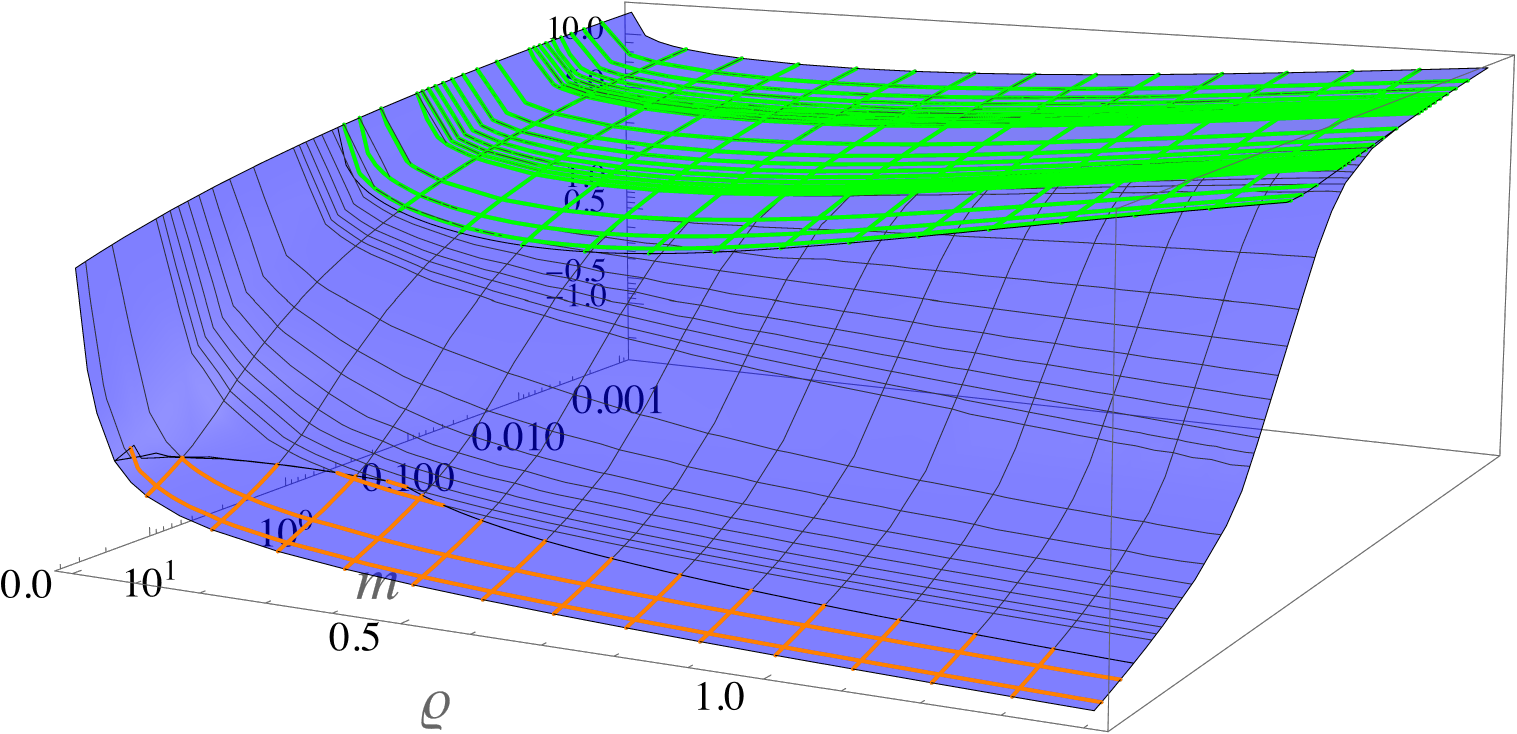}}
\caption{The negative logarithmic caloron density $-\ln(\mathpzc{d})\supset -\ln(m \varrho\, e^{2\alpha\left(\frac{1}{2}\right)} \fferm )$ due to the heavy quark.
\textcolor{green}{Green}:  the small\dash mass expansion.
\textcolor{orange}{Orange}:  the large\dash mass expansion.
\textcolor{blue}{Blue}:  the ``Pad\'e\dash like'' interpolation $-\ln(m\varrho)+\mathpzc{p}(m,\varrho)$, illustrating how well it matches to the two limiting behaviors in the regimes where they are expected to be accurate.
The agreement with the small\dash mass expansion is poor at the largest $\varrho$ values, but otherwise the agreement is good wherever a small$\,$- or large\dash mass expansion is expected to work.}
    \label{fig:pade_agreement_3d}
\end{figure}

\begin{figure}[hb]
\centering

\subcaptionbox{\label{fig:pade_agreement_2d_rho0.1}}{\includegraphics[width=0.32\textwidth]{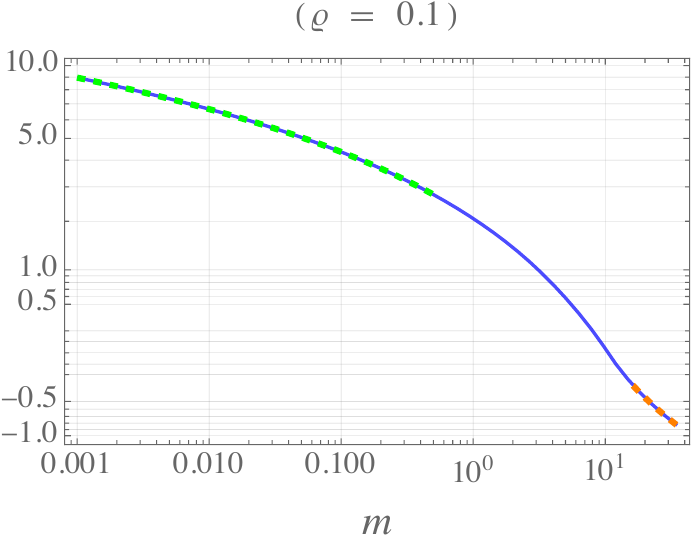}}\hfill
\subcaptionbox{\label{fig:pade_agreement_2d_rho0.3}}{\includegraphics[width=0.32\textwidth]{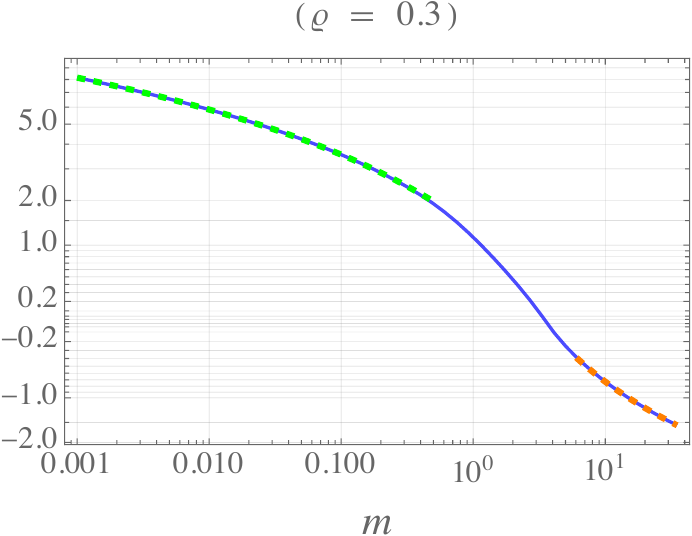}} \hfill
\subcaptionbox{\label{fig:pade_agreement_2d_rho0.5}}{\includegraphics[width=0.32\textwidth]{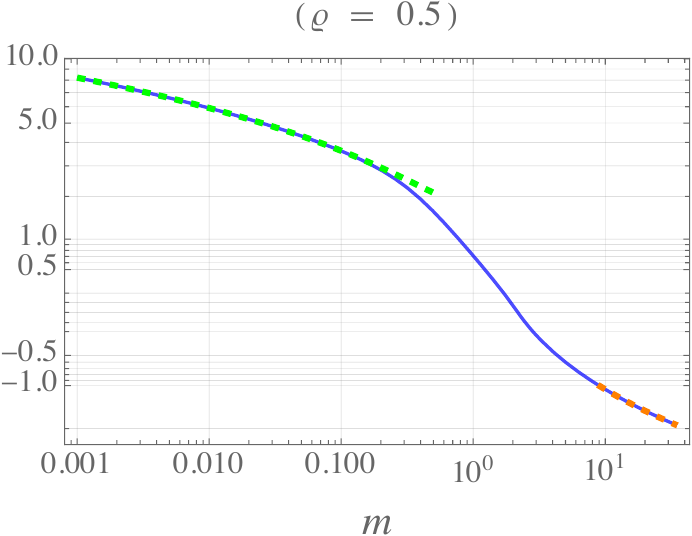}}

\vspace{1ex}

\subcaptionbox{\label{fig:pade_agreement_2d_rho0.75}}{\includegraphics[width=0.32\textwidth]{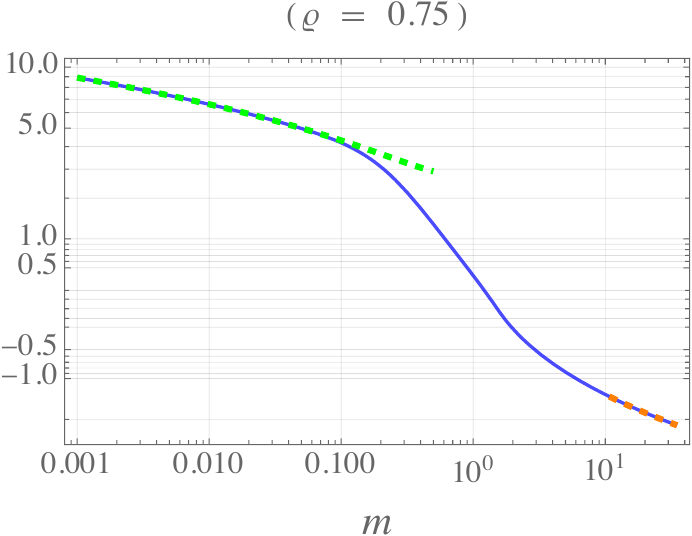}}\hfill
\subcaptionbox{\label{fig:pade_agreement_2d_rho1}}{\includegraphics[width=0.32\textwidth]{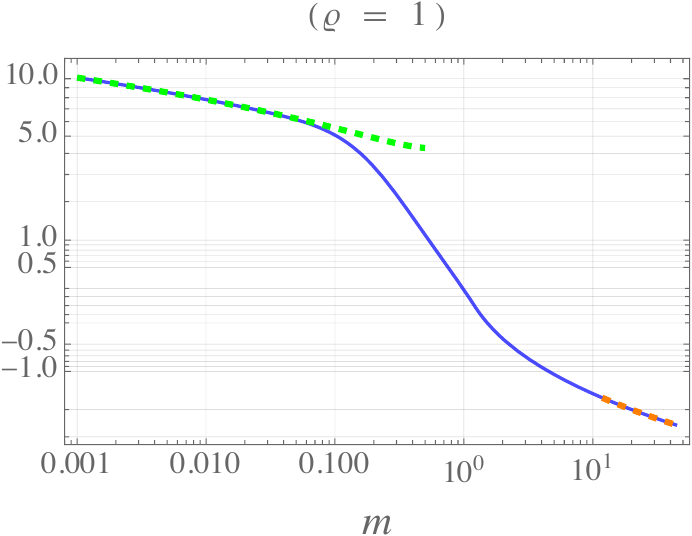}}\hfill
\subcaptionbox{\label{fig:pade_agreement_2d_rho1.25}}{\includegraphics[width=0.32\textwidth]{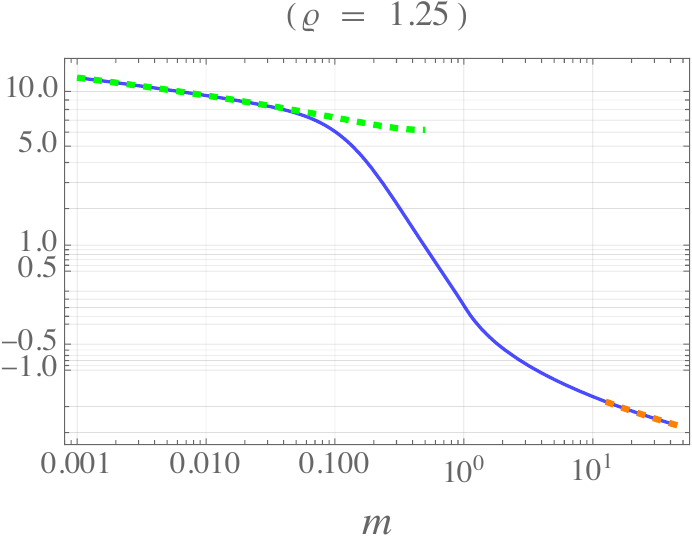}}

\caption{Finite\dash quark mass effect
$-\ln\big(m \varrho\, e^{2\alpha\left(\frac{1}{2}\right)} \fferm\big)$
as a function of $m$ for several $\varrho$ values, corresponding to 2D slices from
\hyperref[fig:pade_agreement_3d]{figure \ref{fig:pade_agreement_3d}}.
The Pad\'e approximation (\textcolor{blue}{\rule[2.3pt]{15pt}{2pt}})
shows good agreement with the large\dash mass result (\textcolor{orange}{\rule[2pt]{5pt}{2pt}~\rule[2pt]{5pt}{2pt}~\rule[2pt]{5pt}{2pt}}) and the small\dash mass result
(\textcolor{green}{\rule[2pt]{5pt}{2pt}~\rule[2pt]{5pt}{2pt}~\rule[2pt]{5pt}{2pt}})
within their expected ranges of validity, except that the small\dash mass expansion breaks down sooner than expected at the largest $\varrho$ values.}
\label{fig:pade_agreement_2d}
\end{figure}

Using the $m^2\,$-, $m^{-2}\,$-, $m^{-4}\,$-, and $m^{-6}$\dash coefficient, we analytically solve for $\mathpzc{p}_{\,{}^{\!}0}\big(\varrho,\xi(\varrho)\big)$, $\mathpzc{p}_{\,{}^{\!}1}\big(\varrho,\xi(\varrho)\big)$, $\mathpzc{P}_1\big(\varrho,\xi(\varrho)\big)$, and $\mathpzc{P}_2\big(\varrho,\xi(\varrho)\big)$ and finally, using the $\mathcal{O}(m^0)$\dash terms, obtain $\xi(\varrho)$ numerically. We find the following analytical results
\begin{align}
& \begin{aligned}
\mathpzc{p}_{\,{}^{\!}0}\big(\varrho,\xi(\varrho)\big) \is & \frac{\sqrt{2}}{12\big(- 18\gamma_{10}\varrho^6+\xi^6\big)}\left(\frac{3}{\sqrt{2}}\big(6\gamma_6\varrho^2-\xi^2\big)\big(12\gamma_8\varrho^4+\xi^4\big)+\frac{1}{\xi\varrho}\sqrt{\mathpzc{q}_1(\varrho,\xi)}\,\times\vphantom{\sqrt{\big(\xi^8}}\right. \\
& \left.\times\sqrt{-6\gammas^\text{small}\big(\xi^8-18\gamma_{10}\xi^2\varrho^6\big) -2\xi^6\varrho^2 + 18\gamma_6\xi^4\varrho^4 - 54\gamma_6^2\xi^2\varrho^6  + 9\gamma_{10}\varrho^8 }\vphantom{\frac{3}{\sqrt{2}}}\,\right),
\end{aligned} \label{eq:pade_coeff_p0}\\
& \begin{aligned}
\mathpzc{p}_{\,{}^{\!}1}\big(\varrho,\xi(\varrho)\big) \is & \frac{6\gamma_6\varrho^2-\xi^2}{3\left(\frac{\mathpzc{q}_5(\varrho,\xi)}{\mathpzc{q}_1}-\frac{\sqrt{3\mathpzc{q}_3(\varrho,\xi)}(12\gamma_8\varrho^4+\xi^4)}{|\mathpzc{q}_1|}\right)}\left(\frac{\mathpzc{q}_2(\varrho,\xi)}{\mathpzc{q}_1}-\frac{\varrho^2\sqrt{3\mathpzc{q}_3}}{|\mathpzc{q}_1|}\right)\times \\
& \times \left(\frac{\mathpzc{q}_4(\varrho,\xi)}{\mathpzc{q}_1}+\frac{\sqrt{3\mathpzc{q}_3}\big(6\gamma_6\varrho^2-\xi^2\big)}{|\mathpzc{q}_1|}\right),
\end{aligned} \label{eq:pade_coeff_p1}\\
& \mathpzc{P}_1\big(\varrho,\xi(\varrho)\big) \is \frac{\mathpzc{q}_2}{\mathpzc{q}_1}-\frac{\varrho^2\sqrt{3\mathpzc{q}_3}}{|\mathpzc{q}_1|}\,, \label{eq:pade_coeff_P1}\\
& \mathpzc{P}_2\big(\varrho,\xi(\varrho)\big) \is \frac{2\varrho^2}{\frac{\mathpzc{q}_5}{\mathpzc{q}_1}-\frac{\sqrt{3\mathpzc{q}_3}(12\gamma_8\varrho^4+\xi^4)}{|\mathpzc{q}_1|}}\left(\frac{\mathpzc{q}_4}{\mathpzc{q}_1}+\frac{\sqrt{3\mathpzc{q}_3}\big(6\gamma_6\varrho^2-\xi^2\big)}{|\mathpzc{q}_1|}\right) \label{eq:pade_coeff_P2}
\end{align}
with
\begin{align}
& \mathpzc{q}_1(\varrho,\xi) \is \xi^8 - 24\gamma_6\xi^6\varrho^2 - 72\gamma_8\xi^4\varrho^4-72\gamma_{10}\xi^2\varrho^6-432\big(\gamma_8^2-\gamma_6\gamma_{10}\big)\varrho^8\,,\label{eq:pade_coeff_func_q1}\\
& \mathpzc{q}_2(\varrho,\xi) \is 3\left(-\big(1+4\mathpzc{p}_{\,{}^{\!}0}^2\big)\xi^6\varrho^2+\gamma_6\xi^4\varrho^4-12\gamma_8\xi^2\varrho^6+72(\gamma_6\gamma_8+\gamma_{10}\mathpzc{p}_{\,{}^{\!}0})\varrho^8\right), \label{eq:pade_coeff_func_q2}\\
& \begin{aligned} \mathpzc{q}_3(\varrho,\xi) \is & \big(7+36\mathpzc{p}_{\,{}^{\!}0}+48\mathpzc{p}_{\,{}^{\!}0}^2\big)\xi^{12} - 36\gamma_6(5+12\mathpzc{p}_{\,{}^{\!}0})\xi^{10}\varrho^2+108\big(13\gamma_6^2-2\gamma_8-4\gamma_8\mathpzc{p}_{\,{}^{\!}0}\big)\xi^8\varrho^4- \\
& - 144\Big(24\gamma_6^3-18\gamma_6\gamma_8(1-2\mathpzc{p}_{\,{}^{\!}0})+\gamma_{10}\big(2+9\mathpzc{p}_{\,{}^{\!}0}+12\mathpzc{p}_{\,{}^{\!}0}^2\big)\Big)\xi^6\varrho^6 - \\
& - 1296\big(6\gamma_6^2\gamma_8-2\gamma_6\gamma_{10}(2+3\mathpzc{p}_{\,{}^{\!}0})+\gamma_8^2(1+12\mathpzc{p}_{\,{}^{\!}0})\big)\xi^4\varrho^8 + \\
& + 15552\big(\gamma_6\gamma_8^2-2\gamma_6^2\gamma_{10}-\gamma_8\gamma_{10}\mathpzc{p}_{\,{}^{\!}0}\big)\xi^2\varrho^{10} - \\
& - 15552\Big(3\gamma_6^2\gamma_8^2-4\gamma_6^3\gamma_{10}-6\gamma_6\gamma_8\gamma_{10}\mathpzc{p}_{\,{}^{\!}0}+\mathpzc{p}_{\,{}^{\!}0}\big(4\gamma_8^2-\gamma_{10}^2\mathpzc{p}_{\,{}^{\!}0}\big)\Big)\varrho^{12}\,,
\end{aligned} \label{eq:pade_coeff_func_q3}\\
& \begin{aligned} \mathpzc{q}_4(\varrho,\xi) \is & -3\Big((1+2\mathpzc{p}_{\,{}^{\!}0})\xi^8 - 12\gamma_6(1-2\mathpzc{p}_{\,{}^{\!}0})\xi^6\varrho^2 + 12\big(3\gamma_6^2+\gamma_8+12\gamma_8\mathpzc{p}_{\,{}^{\!}0}\big)\xi^4\varrho^4 - \\
& - 72(2\gamma_6\gamma_8-\gamma_{10}\mathpzc{p}_{\,{}^{\!}0})\xi^2\varrho^6 + 432\big(\gamma_6^2\gamma_8 + 2\gamma_8^2\mathpzc{p}_{\,{}^{\!}0} - \gamma_6\gamma_{10}\mathpzc{p}_{\,{}^{\!}0}\big)\varrho^8\Big)\,,
\end{aligned} \label{eq:pade_coeff_func_q4}\\
& \begin{aligned} \mathpzc{q}_5(\varrho,\xi) \is & -(5+12\mathpzc{p}_{\,{}^{\!}0})\xi^{10} + 78\gamma_6\xi^8\varrho^2 - 72\big(4\gamma_6^2 - \gamma_8(1-2\mathpzc{p}_{\,{}^{\!}0})\big)\xi^6\varrho^4 - \\
& - 72\big(6\gamma_6\gamma_8-\gamma_{10}(2+3\mathpzc{p}_{\,{}^{\!}0})\big)\xi^4\varrho^6 + 432\big(\gamma_8^2-4\gamma_6\gamma_{10}\big)\xi^2\varrho^8 - \\
& - 2592\big(\gamma_6\gamma_8^2 - 2\gamma_6^2\gamma_{10}-\gamma_8\gamma_{10}\mathpzc{p}_{\,{}^{\!}0}\big)\varrho^{10}\,,
\end{aligned} \label{eq:pade_coeff_func_q5}
\end{align}
and the numerical results for $\xi$ as shown in \hyperref[fig:xi_pade_parameter]{figure~\ref{fig:xi_pade_parameter}}.
In \hyperref[fig:P1_P2_pade_parameters]{figure~\ref{fig:P1_P2_pade_parameters}} we also show the parameters $\mathpzc{P}_1$ and $\mathpzc{P}_2$ and verify that they are indeed positive\dash definite functions as demanded in (\ref{eq:pade_ansatz_general}).
Numerical values are tabulated in the ancillary files.
Finally, in \hyperref[fig:pade_agreement_3d]{figures~\ref{fig:pade_agreement_3d}} and \ref{fig:pade_agreement_2d} we present the full result of the ``Pad\'e\dash like'' interpolation and the agreement with the small$\,$- and large\dash mass results given in (\ref{eq:correction_factor_small_and_large_m_final}).

\subsection{Application to susceptibility}
\label{sec:application}

\begin{figure}
    \centering
    \includegraphics[width=\textwidth]{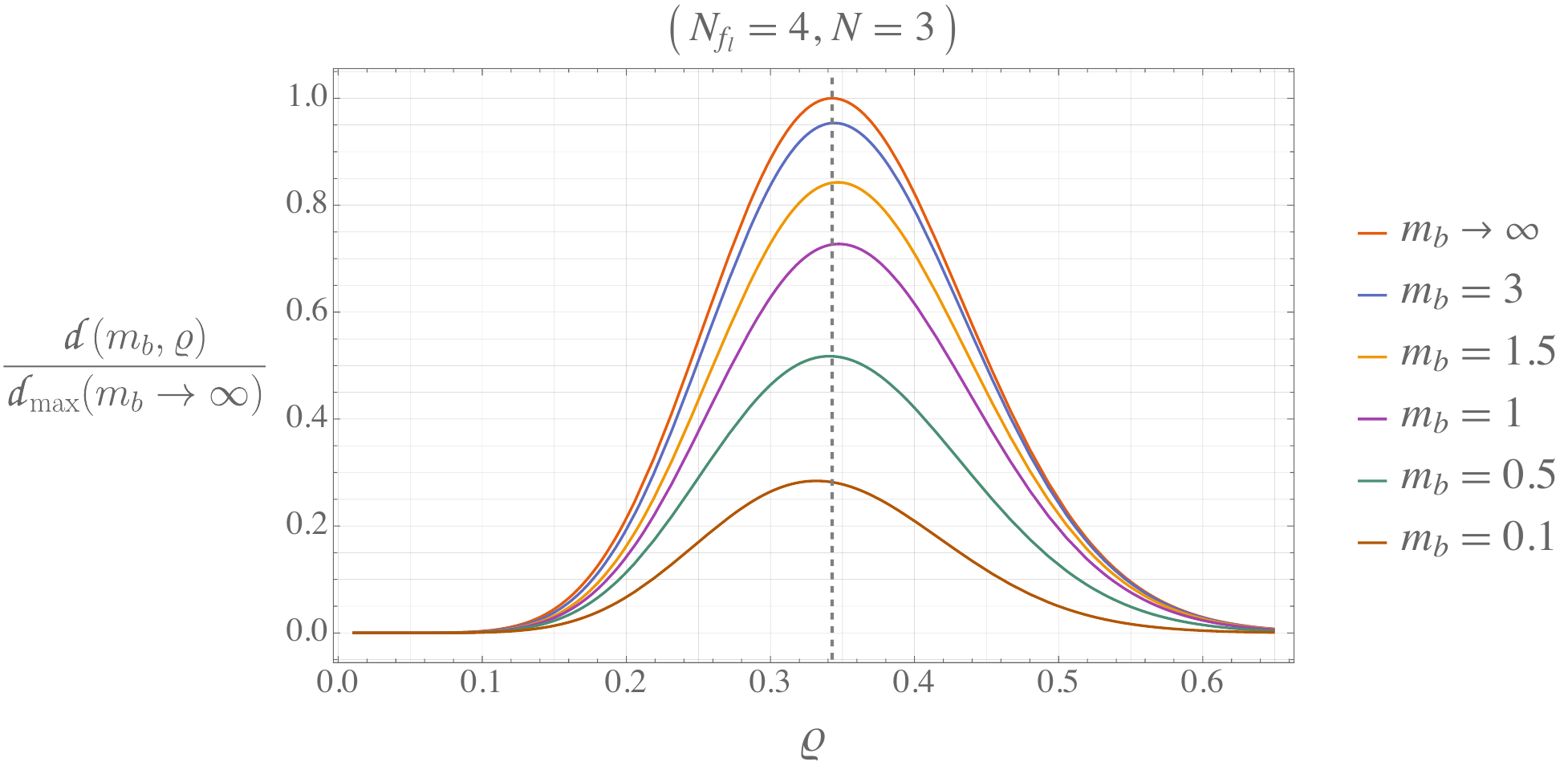}
    \caption{The caloron density $\mathpzc{d}(N_{\! f_l}\!=\! 4,N\!=\! 3,m_b,\varrho,\lambda)$ (\ref{eq:cal_density_with_pade_general}) with the correction factor $e^{-2\alpha\left(\frac{1}{2}\right)-\mathpzc{p}(m,\varrho)}$ (\ref{eq:correction_factor_small_and_large_m_final}).
    We normalize it by the maximum density of the 4\dash flavor theory, represented by $\mathpzc{d}_\text{max}(m_b\rightarrow\infty, g_\text{asy})$ with $g_\text{asy}$ as in (\ref{eq:coupling_modified}), which is located at $\varrho\approx 0.343$ (vertical dashed line).
    Decreasing mass decreases the caloron density and slightly shifts the location of the peak, to larger $\varrho$ at large mass $m>1$ and to smaller $\varrho$ at small mass $m<0.5$.}
    \label{fig:density_ratios}
\end{figure}

Using the full fermionic correction factor $\left.\fferm(m,\varrho)\right|_{T^{\,}>^{\,}0}\is e^{-2\alpha\left(\frac{1}{2}\right)-\mathpzc{p}(m,\varrho)}$ (\ref{eq:correction_factor_small_and_large_m_final}), we obtain the caloron density $\mathpzc{d}$ as given in (\ref{eq:cal_density_with_pade_general}) for the case of $N\is 3$, $N_{\! f_l}\is 4$, and $N_{\! f_\text{h}}\is 1$ (i.e., the heavy quark is the bottom quark). We show the caloron density, normalized by the maximum density of the asymptotic case $m_b\rightarrow\infty$, in \hyperref[fig:density_ratios]{figure \ref{fig:density_ratios}}. Our results shown in \hyperref[fig:density_ratios]{figure~\ref{fig:density_ratios}} also verify the small\dash constituent approximation, as introduced in the \hyperref[sec:prelim]{preliminaries}, for finite $T$ and with heavy quarks.

Lastly, we analyze the topological susceptibility $\sus(m_{f_\text{h}},T)$ and calculate the ratio $\kappa$ (\ref{eq:top_suscep_ratio_general}), i.e., we compare two theories: one with physically heavy ``non\dash light quarks'' - which we analyzed in this work - and one where heavy quark masses are asymptotically large - which is accessible via lattice QCD.
As we discussed in \hyperref[sec:strategy]{section \ref{sec:strategy}}, we modify the coupling constant for the asymptotic heavy mass\dash theory as given in (\ref{eq:coupling_modified}), specifically, we match the theories to have the same IR behavior rather than to have the same UV value of the gauge coupling.
In detail, the $\sus$\dash ratio reads
\begin{equation}
\label{eq:top_suscep_ratio}
\kappa(m_{f_\text{h}},N_{\! f_l},N_{\! f_\text{h}},N ) \is \frac{\mathop{\mathlarger{\int_0^\infty}}\!\!\ab\varrho\, \varrho^{\frac{11N+N_{\!f}}{3}-5}\left.\mathpzc{f}(0,\varrho)\right|_{N_{\! f_l},\, N,\,T^{\,}>^{\,}0}\,\prod_{f_\text{h}}\!\!\sqrt[3]{m_{f_\text{h}}\vphantom{)}}\,e^{-\mathpzc{p}\left(m_{f_\text{h}},\varrho\right)}}{\mathop{\mathlarger{\int_0^\infty}}\!\!\ab\varrho\, \varrho^{\frac{11N+N_{\!f_l}}{3}-5}\left.\mathpzc{f}(0,\varrho)\right|_{N_{\! f_l},\,N,\,T^{\,}>^{\,}0}}\,,
\end{equation}
where the correction factor $\left.\mathpzc{f}(0,\varrho)\right|_{N_{\! f_l},\, N,\,T^{\,}>^{\,}0}$ for a full theory of only light quarks is given in (\ref{eq:correction_factor_massless_quarks}). $N_{\!f_l}\is N_{\! f}-N_{\! f_\text{h}}$ in the denominator $\varrho$\dash exponent is due to $\lim_{\,m^{\,}\rightarrow^{\,}\infty}e^{-\mathpzc{p}} \is (m\varrho)^{-1/3}$.

We calculate (\ref{eq:top_suscep_ratio}) for the physical case $\kappa(m_b,4,1,3)$ and $b$ quark\dash masses between $0.011$ and $25$. As we discussed in \hyperref[sec:intro]{section \ref{sec:intro}}, this is our main result shown in \hyperref[fig:suscep_ratios]{figure~\ref{fig:suscep_ratios}}.

We see that for (low) temperatures $m_b\gtrsim \pi$ we find $\kappa\gtrsim 0.95$, i.e, the difference between the $b$ quark and its infinitely heavy counterpart is less than $5\%$. Only for higher temperatures $m_b\lesssim 2$ do we see $\kappa\lesssim 0.9$ and a more than $10\%$ difference between lattice QCD and physics with a dynamical $b$ quark.

\subsection{Check: comparison with small and large mass asymptotics}
\begin{figure}
    \centering
    \includegraphics[width=\textwidth]{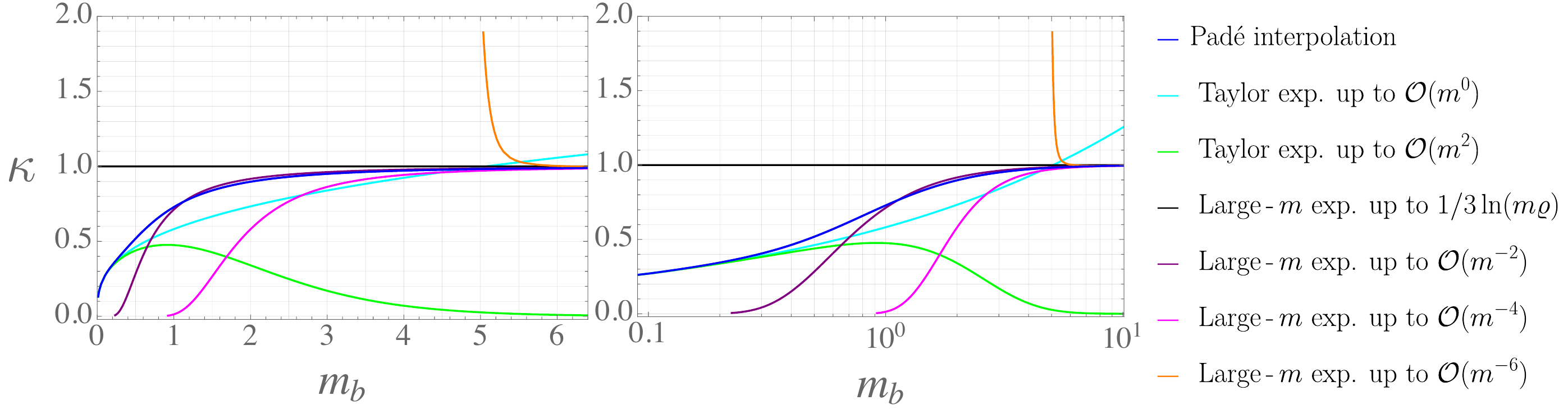}
    \caption{The ratio of topological susceptibilities (\ref{eq:top_suscep_ratio}) for different orders of the small$\,$- and large\dash mass expansions. In comparison with the result for (\ref{eq:top_suscep_ratio}) obtained using the full Pad\'e interpolation (cf. \hyperref[fig:suscep_ratios]{figure \ref{fig:suscep_ratios}}), we show the analogous result obtained by replacing $\mathpzc{p}$ in the exponent by different orders of the expansions in (\ref{eq:correction_factor_small_and_large_m_final}). We see excellent agreement of the Pad\'e and Taylor/heat kernel expansion at small/large $m_b$ and the expected behavior of an asymptotic large\dash $m$ expansion: including higher orders improves the series at large $m_b$, but yields earlier mismatches at small $m_b$.\\ \mbox{} \\
    The Pad\'e result is shown in more detail in \hyperref[fig:suscep_ratios]{figure \ref{fig:suscep_ratios}}.}
    \label{fig:suscep_ratios_all_orders}
\end{figure}

As a check on our results presented in \hyperref[fig:suscep_ratios]{figure~\ref{fig:suscep_ratios}}, we also compute the approximate small$\,$- and large\dash mass expansions of $\kappa$ by repeating the integration in (\ref{eq:top_suscep_ratio}) with the full Pad\'e term $\mathpzc{p}$ replaced by the small$\,$- and large\dash mass expansions of (\ref{eq:correction_factor_small_and_large_m_final}) up to increasing orders, i.e., $\frac{(\pi\varrho)^2}{3}-2A(\pi\varrho)$, $\frac{(\pi\varrho)^2}{3}-2A(\pi\varrho) - 2m^2 \,\gammas^\text{small}$, $\frac{1}{3}\ln(m\varrho)$, $\frac{1}{3}\ln(m\varrho) + \frac{2\gamma_6}{m^2}$, etc. The results for the approximate $\sus$\dash ratios and the comparison with the full Pad\'e expansion are shown in \hyperref[fig:suscep_ratios_all_orders]{figure \ref{fig:suscep_ratios_all_orders}}. 

Let us try to understand the large\dash mass asymptotic, which is the most relevant case for the mass range of interest.
Here it is essential to remember that we match the gauge fields so as to produce the same infrared effective 4\dash quark theory coupling, \textsl{not} the same UV limiting value of the coupling in the 5\dash quark theory.
This shifts the exponent $\exp(-8\pi^2/g^2)$ by a factor of $m^{-2/3}$,
as discussed in (\ref{eq:coupling_modified}) and \hyperref[fig:coupling_modified]{figure~\ref{fig:coupling_modified}}.
The log of the fluctuation determinant contains a term $\propto m$ from the zero mode and an $m^{-1/3}$\dash term from $b_2$, cf. (\ref{eq:b4}, \ref{eq:heat_kernel_result}) for a total behavior proportional to $m^{2/3}$, canceling the $m^{-2/3}$ shift and returning an approximately $m$\dash independent result.
The remaining corrections start at $\mathcal{O}(m^{-2})$ due to $\gamma_6$ of (\ref{eq:b6_expression}).
This explains why the large\dash mass region has a flat asymptote with $m^{-2}$ corrections as one moves towards smaller masses.

However, the series in inverse masses is asymptotic -- after all, the identical series applies for periodic and antiperiodic boundary conditions, even though the results for the two boundary conditions differ as discussed in \hyperref[appendix:finite_T_corrections]{appendix~\ref{appendix:finite_T_corrections}}, indicating a renormalon ambiguity in the resummation of the asymptotic series.
Therefore, while the $m^{-2}$\dash correction is an improvement for a rather broad range of large masses, the higher\dash order terms only help at exceedingly large mass scales, as shown in \hyperref[fig:suscep_ratios_all_orders]{figure \ref{fig:suscep_ratios_all_orders}}.

Next, consider the small-mass region.
Again, the correct prescription for the gauge coupling contributes a factor of $m^{-2/3}$ so that the IR 4\dash quark theory, not the 5\dash quark UV theory, is held fixed.
In addition, the fermionic determinant has a zero mode, contributing a factor of $m$ and giving an overall $m^{1/3}$ behavior at small mass.
Because of the antiperiodic boundary conditions, corrections beyond this are protected from being sensitive to the mass, and therefore represent further corrections $\propto m^2$ to an overall $m^{1/3}$\dash behavior at small mass.

Combining these small$\,$- and large\dash mass expansions, \hyperref[fig:suscep_ratios_all_orders]{figure~\ref{fig:suscep_ratios_all_orders}} shows that our Pad\'e approximant nicely switches from the leading small\dash mass curve to the NLO large\dash mass curve at approximately $m \simeq 0.7$, and is well described by the NLO large\dash mass value in the physically interesting range $3<m<7$.

\section{Conclusions}

We have investigated the dependence of the high\dash temperature topological susceptibility on the presence of an additional heavy quark, phenomenologically motivated by the case of the bottom quark and the temperature range $450\,\mathrm{MeV} < T < 1100\,\mathrm{MeV}$, which is relevant for axion cosmology.
Our results indicate that, in this temperature range, the effects of the heavy bottom quark on the topological susceptibility are below 10\% when compared to working within the 4\dash quark theory with the same infrared coupling strength.
For practical purposes this means that $2+1+1$\dash mass simulations of QCD are sufficient for investigating the hot topological susceptibility for applications to axion cosmology.

\section*{Acknowledgements}
The authors acknowledge the support by the Deutsche Forschungsgemeinschaft (DFG, German Research Foundation) through the CRC\dash TR 211 'Strong\dash interaction matter under extreme conditions'– project number 315477589 – TRR 211.
The authors express particular thanks to Simon Stendebach for his help with numerical calculations in (\ref{eq:b10}) and helpful discussions in general.
The authors also thank Dietrich B\"odeker and Rasmus Nielsen for useful conversations.

\FloatBarrier

\newpage

\appendix

\section{Partial Differential Equation}
\label{appendix:pde}
In order to solve the full problem of the topological susceptibility's quark mass dependence at finite temperatures, one has to compute the determinant ratio $\frac{\det(-D_-^2+m^2_f)\,\det(-\partial_-^2+\lambda^2)}{\det(-D_-^2+\lambda^2)\,\det(-\partial_-^2+m^2_f)}$ (cf. (\ref{eq:cal_density_general})), i.e., one has to solve the eigenvalue problem
\begin{equation}
\label{eq:app_eigenvalue_problem}
(-D^2_- + m^2)\psi_n \is \lambda_n\psi_n
\end{equation}
given in terms of coupled ordinary and partial differential equations (ODEs and PDEs). We derive these differential equations in the following.

\begin{figure}[htbp]
    \centering
    \includegraphics[width=0.51\textwidth]{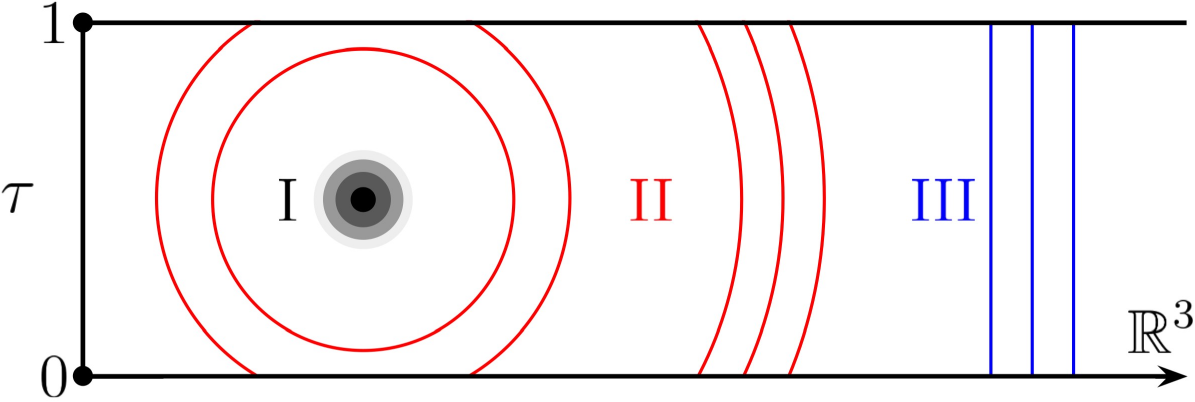}
    \caption{The three spacetime regions important for solving (\ref{eq:app_eigenvalue_problem}). As in \hyperref[fig:large_m_exp]{figure \ref{fig:large_m_exp}}, the caloron is shown as a graded gray sphere. In the ``instanton region'' I the caloron resembles a 4\dash dimensional radially symmetric instanton; in the ``\textcolor{blue}{asymptotic region}'' \textcolor{blue}{III} the caloron is reduced to a 3\dash dimensional, radially symmetric object. The topology of $\spt$ with its distinct, bounded time direction and open space directions and the resulting broken down symmetry group (cf. the discussion at the end of the \hyperref[sec:prelim]{preliminaries}) are relevant in the asymptotic region III as well as the ``\textcolor{red}{transition region}'' \textcolor{red}{II}. This splitting of spacetime was first used in \cite{qcd_at_finite_T}.
    }
    \label{fig:pde_regions}
\end{figure}

Following \cite{qcd_at_finite_T}, the spacetime $\spt$ can be separated into three regions as depicted in \hyperref[fig:pde_regions]{figure~\ref{fig:pde_regions}}: the ``instanton region'' I with $|(\vec{x},\tau)| \is \sqrt{r^2+\tau^2}\ll 1$, the ``\textcolor{blue}{asymptotic region}'' \textcolor{blue}{III} with $r\is\sqrt{\vec{x}^{\,2}}\gg 1$, and the ``\textcolor{red}{transition region}'' \textcolor{red}{II} in between.

In the instanton region I the caloron $A^{a\,\mu}_\text{HS}\is -\ncoverline{\eta}^{\,a\mu\nu}\partial^{\nu\!}\ln(\phi)$ (\ref{eq:HS_caloron}) resembles an instanton of modified size $\tilde{\varrho}$ (cf. (\ref{eq:HS_small_R}) and (\ref{eq:caloron_mod_size})) \cite{qcd_at_finite_T}. 
The behavior of individual solutions to (\ref{eq:app_eigenvalue_problem}) in this region is the same as for an instanton.
This determines the small\dash radius boundary conditions for the solutions in the transition region.

In the \textcolor{blue}{asymptotic region III} we can expand $\phi \stackrel{r^{\,}\gg^{\,}1}{\rightarrow} 1+ \frac{\pi\varrho^2}{r} + \pi^2\varrho^2\,\mathcal{O}(re^{-\pi r})$. This obeys 3\dash dimensional radial symmetry and the resulting asymptotic caloron components read $A^{a\,4}_\text{HS}\stackrel{r^{\,}\gg^{\,}1}{\rightarrow} \frac{1}{1+\frac{r}{\pi\varrho^2}}\frac{x^a}{r^2} \is a^\text{III}_\varrho(r)\frac{x^a}{r^2}$ and $A^{a\,i}_\text{HS}\stackrel{r^{\,}\gg^{\,}1}{\rightarrow} a^\text{III}_\varrho(r)^{\,}\varepsilon^{aij}\frac{x^j}{r^2}$; the caloron is static in $\tau$.\linebreak
We can thus exploit this radial symmetry by adapting the approach developed in \cite{tHooft}: we use the regular angular momentum operators $L^a \is -i^{\,}\varepsilon^{aij}x^i\partial^j$ and define the isospin operators $\mathpzc{T}^a\is \frac{\sigma^a}{2}$ as well as the ``spin $+$ isospin'' operators $J^a\is L^a+\mathpzc{T}^a$.
This means that $L^2$ has eigenvalues $l(l+1)$, $l\in\mathbb{N}\cup\mathbb{N}+\frac{1}{2}$, $\mathpzc{T}^2$ has the eigenvalue $\frac{3}{4}$, and $J^2$ has eigenvalues $j(j+1)$, $j\is |l\pm\frac{1}{2}|$.
Using this, we find the $-D^2_-$\dash operator in region III:
\begin{equation}
\label{eq:app_D2_derivation_region3_2}
\begin{aligned}
-D^2_- & \is -\partial_r^2 -\frac{2}{r}\partial_r+\frac{L^2}{r^2} +2 a^\text{III}_\varrho(r)\frac{\vec{L}^{\!}\cdot^{\!}\vec{\mathpzc{T}}}{r^2}  + \frac{\big(a^\text{III}_\varrho(r)\big)^2}{r^2}\mathpzc{T}^2 - \partial_\tau^2 + \frac{2ia^\text{II}_\varrho(r)}{r}{}^{\,}\e{r}{}^{\!}\cdot^{\!}\vec{\mathpzc{T}}^{\,}\partial_\tau \is \\
& \is -\partial_r^2 -\frac{2}{r}\partial_r+\frac{l(l+1)}{r^2} + a^\text{III}_\varrho\frac{(j-l)(j+l+1)}{r^2} - \frac{3}{4}\frac{\big(a^\text{III}_\varrho\big)^2}{r\pi\varrho^2} - \partial_\tau^2 + \frac{2ia^\text{III}_\varrho}{r}{}^{\,}\e{r}{}^{\!}\cdot^{\!}\vec{\mathpzc{T}}^{\,}\partial_\tau\,,
\end{aligned}
\end{equation}
where we used a separation \textsl{Ansatz} for the eigenfunction $\psi_{n,\,l,\,j}(x) \is \chi_{l,\,j}(\theta,\varphi)\Psi_{n,\,l,\,j}(r,\tau)$ with $\chi_{l,\,j}$ a function and $\widetilde{\Psi}_{n,\,l,\,j}$ a 2\dash spinor.
Furthermore, in the asymptotic limit $r\gg 1$ with $\tau$\dash independent calorons any $\psi_n$ can be expanded in terms of fermionic Matsubara frequencies $p^\text{f}_{\alpha^{\,}\in^{\,}\mathbb{Z}} \is 2\pi\big(\alpha+\frac{1}{2}\big)$, i.e., $\psi_{n,\,l,\,j}(x)\is \chi_{l,\,j}(\theta,\varphi)\sum_{\alpha^{\,}\in^{\,}\mathbb{Z}} \widetilde{\Psi}_{n,\,l,\,j,\,\alpha}(r,p^\text{f}_\alpha)e^{-ip^\text{f}_\alpha\tau}$.
This gives us the $-D^2_-$\dash operator as it acts on $\widetilde{\Psi}_{n,\,l,\,j,\,\alpha}$:
\begin{equation}
\label{eq:app_D2_region3}
-D^2_- \is -\partial_r^2 -\frac{2}{r}\partial_r+\frac{l(l+1)}{r^2} + a^\text{III}_\varrho\frac{(j-l)(j+l+1)}{r^2}+\frac{3}{4}\frac{\big(a^\text{III}_\varrho\big)^2}{r\pi\varrho^2} - (p^\text{f}_\alpha)^2 + \frac{2a^\text{III}_\varrho p^\text{f}_\alpha}{r}{}^{\,}\e{r}\cdot\vec{\mathpzc{T}}\,.
\end{equation}
Note that (\ref{eq:app_D2_region3}) gives a coupled system of two ODEs due to the isospin operators $\mathpzc{T}^a$ (alternatively: $\e{r}{}^{\!}\cdot^{\!}\vec{\mathpzc{T}}\is \frac{\sigma^r}{2}$). Together with (\ref{eq:app_eigenvalue_problem}) this is the eigenvalue problem in the asymptotic region which gives the boundary conditions for the transition region solutions.

For the \textcolor{red}{transition region II} we again introduce $L^a$ and $\mathpzc{T}^a$ together with the separation \textsl{Ansatz} $\psi_{n,\,l,\,j}(x)\is\chi_{l,\,j}(\theta,\varphi)\Psi_{n,\,l,\,j}(r,\tau)$. Additionally, we define the function $\Phi\is \frac{\phi(r,\tau)}{r}$, so that $A^{a\,4}_\text{HS}\is -\left( 1 + r \frac{\partial_r\Phi}{\Phi}\right)\frac{x^a}{r^2}\is -a^\text{II}_\varrho(r,\tau)\frac{x^a}{r^2}$ and $A^{a\,i}_\text{HS}\is -a^\text{II}_\varrho(r,\tau)^{\,}\varepsilon^{aij}\frac{x^j}{r^2} + \delta^{ai}\frac{\partial_\tau\Phi}{\Phi}$. We thus find the differential operator acting on $\Psi_{n,\,l,\,j}$:
\begingroup
\allowdisplaybreaks
\begin{align}
& \begin{aligned}
-D^2_- \is & -\partial^2_r - \frac{2}{r}\partial_r + \frac{L^2}{r^2} + 2 a^\text{II}_\varrho(r,\tau) \frac{\vec{L}^{\!}\cdot^{\!}\vec{\mathpzc{T}}}{r^2} + i^{\,}\partial_r\frac{\partial_\tau\Phi}{\Phi}{}^{\,} \e{r}{}^{\!} \cdot^{\!}\vec{\mathpzc{T}} - i^{\,}\partial_\tau\frac{\partial_r\Phi}{\Phi} + {} \\
& {} + \frac{\big(a^\text{II}_\varrho(r,\tau)\big)^2}{r^2}\mathpzc{T}^2 + \left(\frac{\partial_\tau\Phi}{\Phi}\right)^2\!\mathpzc{T}^2 -\partial_\tau^2 - \frac{2ia^\text{II}_\varrho(r,\tau)}{r^2}{\,}\e{r}{}^{\!}\cdot^{\!}\vec{\mathpzc{T}}^{\,}\partial_\tau \is 
\end{aligned} \nonumber \\
& \begin{aligned}
\hphantom{-D^2_-}\is & -\partial^2_r - \frac{2}{r}\partial_r + \frac{l(l+1)}{r^2} + a^\text{II}_\varrho\frac{(j-l)) (j+l+1)}{r^2} + i^{\,}\partial_r\frac{\partial_\tau\Phi}{\Phi}{}^{\,} \e{r}{}^{\!} \cdot^{\!}\vec{\mathpzc{T}}\, - {} \\
& {} - i^{\,}\partial_\tau\frac{\partial_r\Phi}{\Phi} + \frac{3}{4}\frac{a^\text{II}_\varrho}{r}\frac{\partial_r\Phi}{\Phi} + \frac{3}{4}^{\!}\left(\frac{\partial_\tau\Phi}{\Phi}\right)^2 - \partial_\tau^2 - \frac{2ia^\text{II}_\varrho}{r^2}{\,}\e{r}{}^{\!}\cdot^{\!}\vec{\mathpzc{T}}^{\,}\partial_\tau\,.
\end{aligned}\label{eq:app_D2_region2}
\end{align}
\endgroup
Since $\Psi_{n,\,l,\,j}$ cannot be expanded in terms of Matsubara frequencies in this region, the transition region\dash eigenvalue problem given by (\ref{eq:app_D2_region2}) and (\ref{eq:app_eigenvalue_problem}) is posed in terms of a 2\dash dimensional, coupled system of PDEs.\footnote{If one aims to solve this eigenvalue problem, one could expand $\Phi(r,\tau)$ in terms of radial and temporal variables $u,v$ given by $r\is 1+u$ and $\tau\is 1-v$, respectively. This simplifies the coefficient functions in (\ref{eq:app_D2_region2}) to rational functions of $u$ and $v$, thus possibly simplifying calculations, reducing numerical cost and/or making possible the application of certain theorems from the theory or partial differential equations.}
These are to be matched at small and large $\sqrt{r^2+\tau^2}$ to the asymptotic forms in the other two regions.

\section{Boundary Condition -- Dependence at large Mass}
\label{appendix:finite_T_corrections}

Our calculation of the large\dash mass expansion in \hyperref[subsec:large_m_exp_structure]{section \ref{subsec:large_m_exp_structure}} leads to results which do not depend on the boundary conditions (periodic or anti\dash periodic) of the differential operator $D^2_{\pm}$.
Clearly the full finite\dash mass results do depend on these boundary conditions and therefore the asymptotic large\dash mass expansion must be an asymptotic series with exponentially suppressed trans\dash series corrections of form $m^b e^{-m}$ which do depend on the boundary conditions and which represent an ambiguity in the resummation of the large\dash mass expansion.
Our goal in this appendix is to determine this behavior, which gives the difference between symmetric and anti\dash symmetric boundary conditions.
This difference gives us information on the limitations of the order\dash by\dash order large\dash mass expansion, which helps us understand the range of reliability of said expansion.

The periodicity\dash dependent coefficients depend not only on the chromo\dash electric and chromo\dash magnetic fields $E^i\is G^{i4}$ and $B^i\is \frac{1}{2}\varepsilon^{ijk}G^{jk}$ and their covariant derivatives, but also on dimensionless, matrix\dash valued coefficient functions
\begin{equation}
\label{eq:app_varphi_general}
\varphi_l(\vec{x},\varrho,s)\is\!\sum_{\alpha^{\,}\in^{\,}\mathbb{Z}}\sqrt{4\pi}\, s^{\frac{l+1}{2}}\,\big(i p^\text{b/f}_\alpha -\ln\Omega(\vec{x},\varrho)\big)^l\,e^{s\big(ip^\text{b/f}_\alpha-\ln\Omega(\vec{x},\varrho)\big)^2}\,,
\end{equation}
where $p^\text{b/f}_\alpha$ are again the bosonic/fermionic Matsubara frequencies (we use the $p^\text{f}_\alpha$) and
\begin{equation}
\label{eq:app_Polaykov_loop}
\Omega(\vec{x},\varrho) \is \text{T}\,\exp\!\left(i\!\int_0^1\!\!\ab\tau\, A^4_\text{HS}(\vec{x},\tau)\right) \is \exp\big(i\pi\, \omega(r,\varrho)\,\sigma^r\big) \in SU(2)
\end{equation}
is the Polyakov loop\footnote{The function $-\pi\omega(r,\varrho)$ is given in \cite[eq. (59)]{polyakov_loop_calculation}.} with $A^4_\text{HS}\is  - \frac{\partial_r \phi}{\phi}\,\frac{\sigma^r}{2}$ (note that $[\Omega ]\is [\varphi_l]\is 1$). The known coefficients $b_{2k}$ for the heat kernel $\braket{x|e^{-(-D^2_-)s}|x}$ read:
\begin{equation}
\label{eq:app_heat_kernel_coeff_finite_T_traced}
\begin{aligned}
&b_0(A_\text{HS})\is \varphi_0\,,\qquad b_\text{free}\is b_0(0) \is \varphi_0(\varrho= 0)\,,\quad b_4(A_\text{HS})\is \varphi_0 \ncoverline{b}_4 - \frac{\varphi_0+2\varphi_2}{6}\vec{E}^{\,2}\,,\\
& \begin{aligned} b_6(A_\text{HS})\is & \varphi_0 \ncoverline{b}_6 + \frac{\varphi_0+2\varphi_2}{60}\left(\big(\vec{D}^{\!}\cdot^{\!} \vec{E}^{\,}\big)^2 + 2\big(D_\tau \vec{B}^{\,}\big)^2 + \vec{E}^{\!}\cdot^{\!} \big(\vec{B}^{\!}\times^{\!} \vec{E}^{\,}\big)\right) - \\
&- \left(\frac{\varphi_0}{15} + \frac{\varphi_2}{3} + \frac{2\varphi_4}{15}\right)\big(D_\tau \vec{E}^{\,}\big)^2\,, \end{aligned} \\
& b_{8\,\leq\,2k\,\leq\, 12}(A_\text{HS})\is \varphi_0\ncoverline{b}_{2k} + \text{``unknown''}\,;
\end{aligned}
\end{equation}
the other coefficients either vanish in general ($2k\in\lbrace 1,3\rbrace$) or for $-D^2_-$ ($2k\in\lbrace 2,5\rbrace$), are unknown beyond $\varphi_0\ncoverline{b}_{2k}$ ($6<2k\leq 12$ and $2k\is\text{even}$) or are completely undetermined ($2k>6$ and $2k\is\text{odd}$) \cite{thermal_heat_kernel, thermal_heat_kernel_short, thermal_heat_kernel_alt}.

In \cite{thermal_heat_kernel, thermal_heat_kernel_short} it is shown how $\varphi_0$ can be transformed using Poisson's summation formula for Fourier series: treating $\tilde{f}(p)\is \exp\!\big( (ip-\ln\Omega)^2 \big)$ as a continuous, aperiodic function of $p\in\mathbb{R}$ with a Fourier transform $f(\tau)$, one has $\sum_{p^{\,}\in^{\,}\mathbb{Z}}\tilde{f}(p)\is\sum_{j^{\,}\in^{\,}\mathbb{Z}}f(j)$. This yields
\begin{equation}
\label{eq:app_phi0}
\varphi_0\is \sum_{\alpha^{\,}\in^{\,}\mathbb{Z}}\sqrt{4\pi s}\,e^{s\big(ip^\text{b/f}_\alpha-\ln(\Omega)\big)^2} \is \sum_{j^{\,}\in^{\,}\mathbb{Z}}(\pm 1)^j\,\Omega^j\,e^{-\frac{j^2}{4s}}\
\end{equation}
with $\pm 1$ for the bosonic/fermionic case and $\Omega^j(\vec{x},\varrho) \is \cos\!\big(j\pi\,\omega(r,\varrho)\big) + i\sin\!\big(j\pi\,\omega(r,\varrho)\big)\,\sigma^r$. This Fourier back\dash transformation, which is easier evaluated, connects the momentum space of Matsubara frequencies back to proper time $s$.
The heat kernel $\braket{x|e^{-(-D^2_-)s}|x}$, as a proper time\dash propagator, then yields a closed loop propagator in Euclidean time $\Delta^-(x,x,m^2)$ (cf. (\ref{eq:finite_T_propagator})) via $s$\dash integration (\ref{eq:propagator_from_heat_kernel}).
Here, the $j\is 0$\dash mode corresponds to aperiodic loops in the heat kernel expansion - cf. \hyperref[fig:closed_loops]{figure~\ref{fig:closed_loops}} -, while $j\neq 0$\dash modes correspond to anti\dash periodically closed loops and thus constitute the finite\dash $T$ ambiguities.

All finite\dash $T$ terms with $j\neq 0$ are exponentially suppressed in $s$ compared to the leading contribution $j\is 0$, but appear in the infinite $j$\dash summations. We compute them and show that these ambiguities are also suppressed compared to the large\dash $m$ expansion (\ref{eq:heat_kernel_result}). In (\ref{eq:app_phi_l}) we show that only $\varphi_0$ in (\ref{eq:app_heat_kernel_coeff_finite_T_traced}) contains $j\is 0$\dash terms, i.e., only the $\ncoverline{b}_{2k}$ contribute to the heat kernel expansion which we used in \hyperref[subsec:large_m_results]{section \ref{subsec:large_m_results}}. Everything else, i.e., the $j\neq 0$\dash modes of $\varphi_0$ and the $\varphi_l$\dash combinations, constitute the aforementioned finite\dash temperature uncertainties.

We use the Poisson summation formula\dash trick that gave (\ref{eq:app_phi0}) to calculate higher $\varphi_{l{}^{\,}>{}^{\,}0}$. For that, we write $\varphi_l(a)\is \sum_p \sqrt{4\pi} \,s^{(l+1)/2}\,Q^l e^{sQ^2+aQ}$ with $Q\is ip-\ln(\Omega)$ and obtain the general form $\varphi_l\is s^{l/2}\,(\partial_a^{(l)}{}^{\!}\varphi_0(a))|_{a^{\,}=^{\,} 0}$. Now we Fourier transform $\varphi_0(a)$, employ the Poisson formula, and perform the $a$\dash derivatives before finally setting $a\is 0$.
For the $\varphi_l$\dash combinations appearing in (\ref{eq:app_heat_kernel_coeff_finite_T_traced}) we find:
\begin{equation}
\label{eq:app_phi_l}
\varphi_0 + 2\varphi_2 \is \sum_{j^{\,}\in{}^{\,}\mathbb{Z}}(\pm 1)^j\,\Omega^j\,e^{-\frac{j^2}{4s}}\,\frac{j^2}{2s}\,, \qquad \frac{\varphi_0}{15}+\frac{\varphi_2}{3}+\frac{2\varphi_4}{15}\is \sum_{j^{\,}\in{}^{\,}\mathbb{Z}}(\pm 1)^j\,\Omega^j\,e^{-\frac{j^2}{4s}}\,\frac{j^2(j^2-2s)}{120 s^2}\,.
\end{equation}
All modes $j\neq 0$ are exponentially suppressed and for both combinations in (\ref{eq:app_phi_l}) the $j\is 0$\dash mode vanishes identically.
The limit $s\rightarrow 0$ reproduces the $j\is 0$\dash modes and we thus see that the known finite\dash $T$ terms in (\ref{eq:app_heat_kernel_coeff_finite_T_traced}) vanish exponentially for $T\rightarrow 0 \Rightarrow s\rightarrow 0$, just as expected.

In order to calculate the boundary condition\dash dependent finite\dash $T$ uncertainties, we need to modify the $s$\dash integrals (\ref{eq:I_integral}) to include the $j\neq 0$\dash modes. In general, we consider
\begin{equation}
\label{eq:app_I_integral}
\begin{aligned}
& I\Big(m^2,k,j^2;c\Big)\is \int_0^\infty\!\!\ab s\,\,e^{-m^2s - \frac{j^2}{4s}}\,s^{k-3-c} \is 2^{-(k-3-c)}\left(\frac{|j|}{m}\right)^{k-2-c}\,K_{|k-2-c|}(|j|m) \stackrel{m^{2\,}\gg {}^{\,}1}{\sim} \\
& \sim 2^{-(k-\frac{5}{2}-c)}\sqrt{\pi}\,\frac{|j|^{k-\frac{5}{2}-c}}{m^{k-\frac{3}{2}-c}}e^{-|j|m}\left(1+\sum_{u^{\,}\in^{\,}\mathbb{N}^+}\frac{\prod^u_{v\is 1} \big(4(k-2-c)^2-(2v-1)^2\big)}{u!(8|j|m)^u}\right)
\end{aligned}
\end{equation}
with $c\in\mathbb{N}$ resulting from the $\varphi_l$\dash combinations (\ref{eq:app_phi_l}) and $K_\alpha(x)$ the modified Bessel function of the second kind. Note that for the $j$\dash summation $I\big(m^2,k,j^2;c\big)$ has to be multiplied by the corresponding factor $\propto j^{2c}$ from the terms (\ref{eq:app_phi_l}).
Since the $K_\alpha$\dash expansion is an asymptotic one, it is justified for large (enough) $|j|m$ to keep only the first few terms.
These integrals (\ref{eq:app_I_integral}) are $e^{-m}$\dash damped, but part of the infinite $j$\dash summation. Therefore, we compute the total finite\dash $T$ terms to quantify the large enough masses required to keep these terms suppressed compared to (\ref{eq:heat_kernel_result}).\newline

\noindent \textbf{\underline{\textsl{Order $k\is 0^{\,}$}:}}

For the $j\neq 0$\dash modes of $b_0(A_\text{HS})$ and $b_\text{free}$ we find finite $s$\dash integrals $I(m^2,0,j^2;0)$. Performing the $j$\dash summation of $\varphi_0$ (\ref{eq:app_phi0}), we obtain

\begin{equation}
\sum_{j^{\,}\in{}^{\,}\mathbb{Z}\setminus\lbrace 0\rbrace}\!\!\!(-1)^j \Omega^j \,2^\frac{5}{2}\sqrt{\pi}\frac{m^\frac{3}{2}}{|j|{}^\frac{5}{2}}e^{-|j|m}\Big(1+\frac{15}{8|j|m}+ \mathcal{O}(m^{-2})
\Big)\,.
\label{eq:app_j_sum_k=0_droping_rule}
\end{equation}
Here we neglect all contributions which are mass damped by inverse powers of $m$ with respect to the $j\is 0$\dash mode: for $k\is 0$ the ``leading'' term is $0$ and thus $\mathcal{O}(m^0)$. Therefore, we drop all terms $\mathcal{O}(m^{-1/2})$ and lower in (\ref{eq:app_j_sum_k=0_droping_rule}) (neglecting the overall $e^{-m}$\dash damping).
We plug in $\Omega^j$ and keep only the traceful part $\cos(j\pi\,\omega)$:
\begin{align}
& \sum_{j^{\,}\in{}^{\,}\mathbb{Z}\setminus\lbrace 0\rbrace}\!\!\!(-1)^j \Omega^j \,2^\frac{5}{2}\sqrt{\pi}\frac{m^\frac{3}{2}}{|j|{}^\frac{5}{2}}e^{-|j|m}\Big(1+\frac{15}{8|j|m}\Big) \,=_\tr \nonumber\\
& =_\tr \sum_{j^{\,}\in{}^{\,}\mathbb{N}^+}(-1)^j \cos(j\pi\,\omega) \,2^\frac{7}{2}\sqrt{\pi}\frac{m^\frac{3}{2}}{j^\frac{5}{2}}e^{-j m}\Big(1+\frac{15}{8 j m}\Big)\is \nonumber\\
& \is  2^\frac{7}{2}\sqrt{\pi}\Big( m^\frac{3}{2}\,\text{Re}\big(\text{Li}_\frac{5}{2}(-e^{i\pi\omega-m})\big) + \frac{15\, m^\frac{1}{2}}{8}\,\text{Re}\big(\text{Li}_\frac{7}{2}(-e^{-i\pi\omega-m})\big)\Big) \is \nonumber\\
& \is 2^\frac{7}{2}\sqrt{\pi}\Big(\mathrm{L}_{\frac{3}{2},\frac{5}{2}}\big(m,\omega(r,\varrho)\big)+\frac{15}{8}\,\text{L}_{\frac{1}{2},\frac{7}{2}}\big(m,\omega(r,\varrho)\big)\Big)\,, \label{eq:app_j_sum_k=0_derivation}
\end{align}
where $\text{Li}_b(z)$ is the polylogarithm and we introduced the function
\begin{equation}
\label{L_function}
\mathrm{L}_{a,b}\big(m,\omega(r,\varrho)\big)\is m^a\,\text{Re}\!\left(\text{Li}_b\big(-e^{i\pi\omega(r,\varrho)-m}\big)\right).
\end{equation}
For general heat kernel contributions $\mathcal{O}(m^d)$ our rule about dropping mass damped terms thus translates to neglecting all contributions $\mathrm{L}_{a,b}$ with $a<d$.\footnote{The pattern of $a$ decreasing and $b$ increasing in units steps in $L_{a,b}$ as shown in (\ref{eq:app_j_sum_k=0_derivation}) is universal to the finite\dash $T$ terms due to (\ref{eq:app_I_integral}).}
The free case is calculated analogously with $\Omega\is 1\Leftrightarrow\omega\is 0$ and we have
\begin{equation}
2^\frac{7}{2}\sqrt{\pi}\,\Big(m^\frac{3}{2}\,\text{Li}_\frac{5}{2}(-e^{-m}) + \frac{15\,m^\frac{1}{2}}{8}\,\text{Li}_\frac{7}{2}(-e^{-m})\Big)\,.
\end{equation}

\begin{wrapfigure}[12]{l}{0.37\textwidth}
\centering
\includegraphics[width=0.35\textwidth]{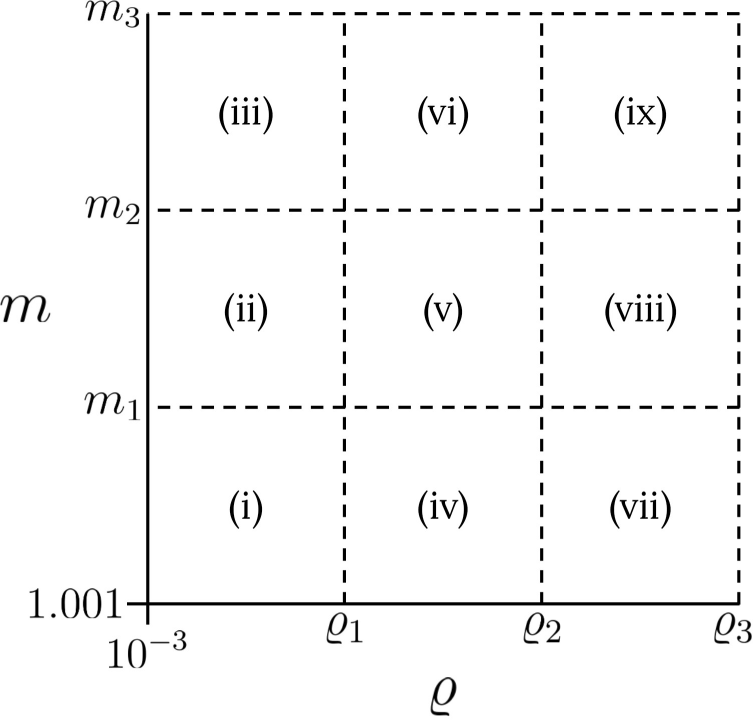}
\caption{}
\label{fig:m_rho_plane}
\end{wrapfigure}

The overall finite\dash $T$ term at $k\is 0$ is thus given by\footnote{Including an additional factor of $(2\pi)^{-1}$ due to a factor of 2 from $\tr\big(1\!\in\!\mathfrak{su}(2)\big)$, the heat kernel ex- pansion prefactor $(4\pi)^{-2}$, and the volume element $4\pi r^2\,\ab r$; the integral $\int_0^1\ab\tau$ is trivial and yields $1$.}
\begin{equation}
\label{eq:app_B0}
\begin{aligned}
\gamma_0^T(m,\varrho)\is \frac{2^{\frac{5}{2}}}{\sqrt{\pi}}\int_0^\infty\!\!\ab r\, r^2\Big(& \mathrm{L}_{\frac{3}{2},\frac{5}{2}}\big(m,\omega(r,\varrho)\big)+\frac{15}{8}\,\mathrm{L}_{\frac{1}{2},\frac{7}{2}}\big(m,\omega(r,\varrho)\big) - \\
& - m^\frac{3}{2}\,\text{Li}_\frac{5}{2}(-e^{-m}) - \frac{15\,m^\frac{1}{2}}{8}\,\text{Li}_\frac{7}{2}(-e^{-m})\Big)\,.
\end{aligned}
\end{equation}
We compute $\gamma_0^T$ on a grid in the $m$\dash$\varrho$\dash plane with the plane split in nine sectors (i) \dash (ix) as shown in \hyperref[fig:m_rho_plane]{figure \ref{fig:m_rho_plane}}. For that we use the \textsl{SymPy} \cite{sympy} and \textsl{mpmath} \cite{mpmath} packages.\footnote{These packages allows us to handle polylogarithms in symbolic and numerical \textsl{Python}\dash calculations, respectively.}

It is difficult to provide a good functional fit to the $\gamma_0^T$\dash data (this is indeed the case for all finite\dash temperature terms); the best approximation is given by the function
\begin{equation}
\label{eq:app_finite_T_heat_kernel_general}
\zeta(m,\tilde{\varrho},\varrho^{\,} ;\,\tilde{a},\tilde{b},\tilde{c},a,b,c) \is e^{-m}\left(\tilde{a}\, m^{\tilde{b}}\,\tilde{\varrho}^{\,\tilde{c}} + a\,m^b\varrho^c\right),
\end{equation}
which serves to provide an intuition for the general functional form.

The coefficients $\tilde{a}$, $\tilde{b}$, $\tilde{c}$, $a$, $b$, $c$ for $\gamma_0^T$ and the sector boundaries are given in \hyperref[table:k=0_finite_T_table]{table~\ref{table:k=0_finite_T_table}}.
The boundaries in \hyperref[table:k=0_finite_T_table]{table \ref{table:k=0_finite_T_table}} are roughly set by hand to minimize the fitting error: the fit agrees with our numerical data to within about 5\%. As a consequence of (\ref{eq:app_finite_T_heat_kernel_general}), $\gamma_0^T(\lambda,\varrho)$ vanishes exponentially as $\lambda\rightarrow\infty$ (analogously for $k>0$).

Our $\gamma_0^T$\dash data and the corresponding fitting error\dash data as well as the data for $k\is 2$ and $k\is 3$ (see below) can be found in the ancillary files. We showed that despite the infinite $j$\dash summation, the finite\dash $T$ uncertainties reflecting the boundary conditions are exponentially small as $m^b e^{-m}$, $b\sim 1$ for large enough masses.

\begin{table}[H]
\begin{tabularx}{\textwidth}{r|X|X|X|X|X|X|X|X|X}
 & (i) & (ii) & (iii) & (iv) & (v) & (vi) & (vii) & (viii) & (ix) \\
\hline
$\tilde{a}$ & 8.878 & 8.107 & 5.000 & $\tilde{a}_0^{(1)^{\vphantom{\dagger}}}(\varrho)$ & $\tilde{a}_0^{(2)}(\varrho)$ & $\tilde{a}_0^{(3)}(\varrho)$ & 0 & 0 & 0 \\
\hline
$\tilde{b}^{\vphantom{\big(}}$ & 1.03 & 1.17 & 1.42 & 1.03 & 1.17 & 1.42 & 0 & 0 & 0 \\
\hline
$\tilde{c}$ & 3.00 & 3.00 & 3.00 & 3.13 & 3.13 & 3.13 & 0 & 0 & 0 \\
\hline
$a$ & 0 & 0 & 0 & $a_0^{(1)^{\vphantom{\dagger}}}(\varrho)$ & $a_0^{(2)}(\varrho)$ & $a_0^{(3)}(\varrho)$ & 3.075 & 2.823 & 1.700\\
\hline
$b$ & 0 & 0 & 0 & 1.04 & 1.18 & 1.42 & 1.04 & 1.18 & 1.43 \\
\hline
$c$ & 0 & 0 & 0 & 2.25 & 2.26 & 2.26 & 2.00 & 2.00 & 2.00
\end{tabularx}
\caption{The sector boundaries for $\gamma_0^T$ are $m_1\approx 2$, $m_2\approx 8$, $\varrho_1\approx 0.22$, $\varrho_2\approx 1.2$; $m_3\is\varrho_3\is 200$ .
The functions $\tilde{a}(\varrho)$ and $a(\varrho)$ in sectors (iv) - (vi) are required due to the rapid transition from small\dash caloron ($\tilde{\varrho}$) to large\dash caloron description ($\varrho$). They contain step functions $\Theta_l(\varrho,u)\is \big(1+e^{-2u\varrho}\big)^{-1}$: $\tilde{a}_0^{(1)}\is 11.24\big(1-\Theta_l(\varrho-0.4,7.5)\big)$, $\tilde{a}_0^{(2)}\is 10.29(1-\Theta_l)$, $\tilde{a}_0^{(3)}\is 6.350(1-\Theta_l)$, $a_0^{(1)}\is 2.904\,\Theta_l$, $a_0^{(2)}\is 2.657\,\Theta_{l\,}$, $\tilde{a}_0^{(3)}\is 1.650\,\Theta_l$.}
\label{table:k=0_finite_T_table}
\end{table}

\noindent \textbf{\underline{\textsl{Order $k> 0^{\,}$}:}}

For $k\is 2$ we have $j\neq 0$\dash terms $\propto (\mathrm{L}_{-\frac{1}{2},\frac{1}{2}} + ...)\ncoverline{b}_4$ from $\varphi_0$, which we neglect as mass damped, and the following boundary condition term from $\frac{1}{6}(\varphi_0+2\varphi_2)$:
\begin{equation}
\label{eq:app_B4}
\gamma_4^T(m,\varrho) \is -\frac{1}{2^\frac{5}{2}\cdot 3\sqrt{\pi}}\int_0^1\!\!\ab\tau\int_0^\infty\!\!\ab r\, r^2\,\mathrm{L}_{\frac{1}{2},-\frac{1}{2}}\big(m,\omega(r,\varrho)\big)(E^{a\,i} E^{a\,i})(r,\tau,\varrho)\,.
\end{equation}

At heat kernel order $k\is 3$ the finite\dash temperature terms read (one has to be careful about the $\mathfrak{su}(2)$\dash traces; note that, e.g., $E^{a\,i;i}\is (\vec{D}^{\!}\cdot^{\!}\vec{E}^{\,})^a$ and $E^{a\,i;4} \is (D_\tau E^{i})^a$)
\begin{equation}
\label{eq:app_B6}
\begin{aligned}
\gamma_6^T(m,\varrho) \is & \frac{1}{2^{\frac{7}{2}}\cdot 15\sqrt{\pi}}\int_0^1\!\!\ab\tau\int_0^\infty\!\!\ab r\, r^2\left[\frac{1}{6}\,\mathrm{L}_{-\frac{3}{2},-\frac{1}{2}}\,\varepsilon^{abc}G^{a\,\mu\nu}G^{b\,\mu\kappa}G^{c\,\nu\kappa} + \vphantom{\left(\frac{1}{2}\right)} \right. \\
& + \left(\frac{1}{2}\,\mathrm{L}_{-\frac{1}{2},-\frac{3}{2}} - \frac{1}{16}\,\mathrm{L}_{-\frac{3}{2},-\frac{1}{2}}\right)\left((E^{a\,i;i})^2 + 2(B^{a\,i;4})^2 + \varepsilon^{abc}\varepsilon^{ijk}\,E^{a\,i}B^{b\,j}E^{c\,k}\right) - \\
& - \left. \left(\mathrm{L}_{\frac{1}{2},-\frac{5}{2}} - \frac{5}{8}\,\mathrm{L}_{-\frac{1}{2},-\frac{3}{2}} + \frac{1}{128}\,\mathrm{L}_{-\frac{3}{2},-\frac{1}{2}}\right)(E^{a\,i;4})^2\right] .
\end{aligned}
\end{equation}

\begin{figure}[htb]
    \centering
    \subcaptionbox{The leading heat kernel expansion shown in orange compared to 1.2 times the finite\dash temperature uncertainties depicted in blue (the factor 1.2 allows for a conservative restriction). The $\varrho\is 0.31$\dash plane shows that for smaller caloron sizes the boundary condition\dash corrections are always smaller, while for $\varrho\gtrsim 0.31$ a large enough quark mass is required.\label{fig:finite_T_damping_3D}}[\textwidth]{\includegraphics[width=0.76\textwidth]{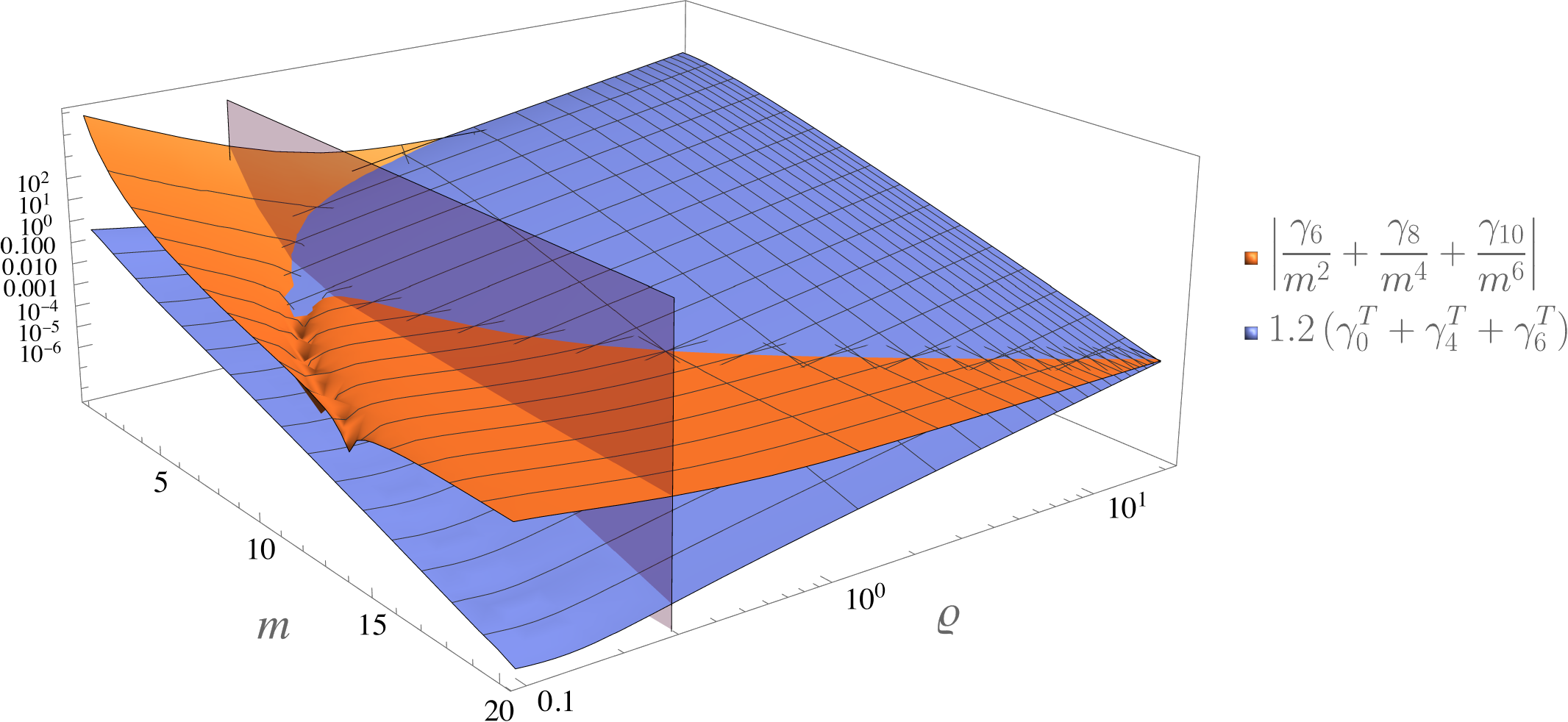}}

\par\bigskip

    \subcaptionbox{The minimal, i.e., lightest, possible heavy mass $m_\text{large, min, 3}(\varrho)$ determined from \hyperref[subsec:large_m_results]{condition 3)}. We choose it conservative so that $\left|\frac{\gamma_6}{m^2} + \frac{\gamma_8}{m^4} + \frac{\gamma_{10}}{m^6}\right|$ exceeds $\gamma_0^T + \gamma_4^T+\gamma_6^T$ by at least $20\%$. The blue lines mark data points and fill the area of allowed masses. We identify a roughly logarithmic growth of $m_\text{large, min, 3}(\varrho)$.\label{fig:m_large_min_3}}[\textwidth]{\includegraphics[width=0.615\textwidth]{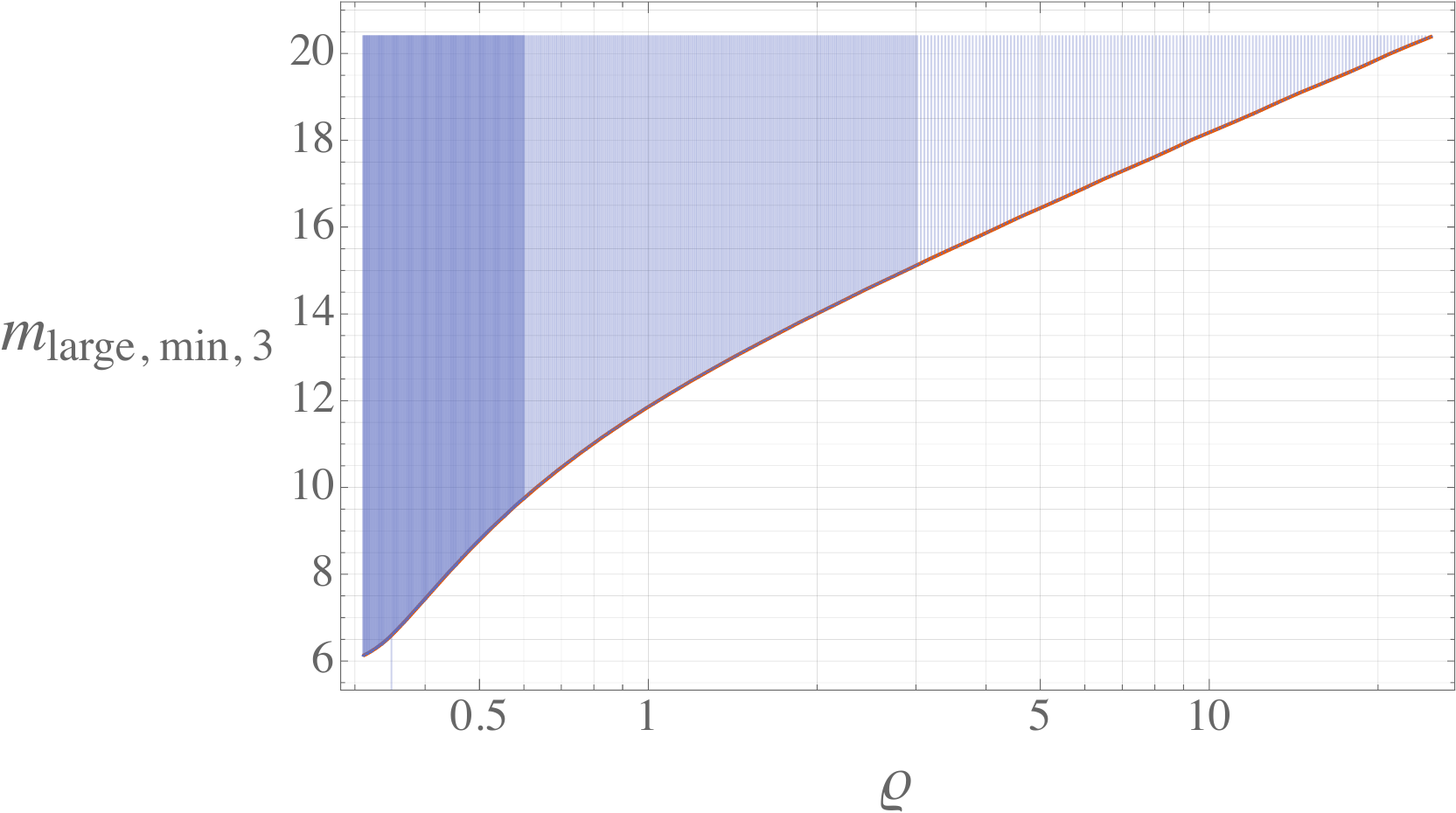}}
\caption{}\label{fig:finite_T_damping_mass_condition}
\end{figure}

\hyperref[fig:finite_T_damping_mass_condition]{Figure \ref{fig:finite_T_damping_mass_condition}} shows the strength of the trans\dash series corrections due to boundary conditions compared to the order\dash by\dash order large\dash mass expansion as well as the resulting large\dash mass restriction according to condition 3) in \hyperref[subsec:large_m_results]{section \ref{subsec:large_m_results}}.

\section{Diagonal Parts of the massless scalar Propagators}
\label{appendix:propagators}

Here we show how to generalize the calculation of \hyperref[subsec:propagator]{section \ref{subsec:propagator}} to general (anti$\,$-)periodic scalar propagators $\Delta^\pm(x,y,m=0) \is \Delta^\pm(x,y)$ using (\ref{eq:finite_T_propagator}), (\ref{eq:aperiodic_propagator_via_F}), and (\ref{eq:F_function}). We show this for the traceful, i.e., diagonal parts. The traceless, off\dash diagonal parts can be obtained analogously by considering the terms $\propto\vec{\sigma}$ in (\ref{eq:F_function}).

First, we calculate (\ref{eq:F_function}) for the general case, denoting $\ncoverline{t}_x-\ncoverline{t}_y\is \Delta\ncoverline{t}$ and $\ncoverline{x}-\ncoverline{y} \is\ncoverline{\Delta}$:
\begingroup
\allowdisplaybreaks
\begin{align}
& F(\ncoverline{x},\ncoverline{y})\istr 1+ \varrho^{2\!} \sum_{k^{\,}\in^{\,}\mathbb{Z}} \frac{\vec{\ncoverline{x}}\cdot\vec{\ncoverline{y}} + (\ncoverline{t}_x-k)(\ncoverline{t}_y-k)}{\vec{\ncoverline{x}}{}^{\,2}\, \vec{\ncoverline{y}}{}^{\,2} + \vec{\ncoverline{x}}{}^{\,2}(\ncoverline{t}_y-k)^2 + \vec{\ncoverline{y}}{}^{\,2}(\ncoverline{t}_x-k)^2+(\ncoverline{t}_x-k)^2(\ncoverline{t}_y-k)^2} \is \nonumber\\
& \is 1 + \varrho^2\,\text{Re}\!\left(\frac{i\,\vec{\ncoverline{x}}\cdot\vec{\ncoverline{y}}-|\vec{\ncoverline{x}}|(i\,|\vec{\ncoverline{x}}|+\Delta\ncoverline{t})}{|\vec{\ncoverline{x}}|\left(\vec{\ncoverline{y}}{}^{\,2}-(|\vec{\ncoverline{x}}|-i\Delta\ncoverline{t})^2\right)}\big(\psi(-\ncoverline{t}_x-i\,|\vec{\ncoverline{x}}|) -\psi(1+\ncoverline{t}_x+i\,|\vec{\ncoverline{x}}|)\big)\right) + (\ncoverline{x}\leftrightarrow \ncoverline{y}) \is \nonumber\\
& \is 1 + \pi\varrho^2\,\text{Re}\!\left(\frac{i\,\vec{\ncoverline{x}}\cdot\vec{\ncoverline{y}}-|\vec{\ncoverline{x}}|(i\,|\vec{\ncoverline{x}}|+\Delta\ncoverline{t})}{|\vec{\ncoverline{x}}|\left(\vec{\ncoverline{y}}{}^{\,2}-(|\vec{\ncoverline{x}}|-i\Delta\ncoverline{t})^2\right)}\,\cot\!\big(\pi(\ncoverline{t}_x+i\,|\vec{\ncoverline{x}}|)\big)\right) + (\ncoverline{x}\leftrightarrow \ncoverline{y}) \is \nonumber\\
& \is 1 + \frac{\pi\varrho^2}{c_1(|\vec{\ncoverline{x}}|,\ncoverline{t}_x)^{\,} d(|\vec{\ncoverline{x}}|,|\vec{\ncoverline{y}}|,\ncoverline{t}_x,\ncoverline{t}_y)}\Big(\!-\Delta\ncoverline{t}^{\,}\ncoverline{\Delta}^2 \sin(2\pi\,{}^{\!} \ncoverline{t}_x)\, + \nonumber\\
& \phantom{\is} + \left(|\vec{\ncoverline{x}}|\ncoverline{\Delta}^2 + \e{\vec{\ncoverline{x}}}\cdot\vec{\ncoverline{y}}^{\,}\ncoverline{\Delta}^2 + 2|\vec{\ncoverline{x}}|(\e{\vec{\ncoverline{x}}}\cdot\vec{\ncoverline{y}}{}^{\,})^2-2|\vec{\ncoverline{x}}|^{\,}\vec{\ncoverline{y}}{}^{\,2}\right)\sinh(2\pi |\vec{\ncoverline{x}}|) \Big) + (\ncoverline{x} \leftrightarrow \ncoverline{y})\,, \label{eq:app_F_function_derivation}
\end{align}
\endgroup
where $\psi(z)$ is the digamma function, we used $\psi(1-z) \is \psi(z)+\pi\cot(\pi z)$ (reflection identity \cite{handbook}) and shortened the notation by defining $c_1(z_1,z_2)\is \cosh(2\pi z_1)-\cos(2\pi z_2)$ as well as $d(|\vec{\ncoverline{x}}|,|\vec{\ncoverline{y}}|,\ncoverline{t}_x,\ncoverline{t}_y)\is \big((|\vec{\ncoverline{x}}|-|\vec{\ncoverline{y}}|)^2+\Delta\ncoverline{t}^{\,2}\big)\big((|\vec{\ncoverline{x}}|+|\vec{\ncoverline{y}}|)^2+\Delta\ncoverline{t}^{\,2}\big)$.

For calculating the periodic and anti\dash periodic propagators according to (\ref{eq:finite_T_propagator}) and (\ref{eq:aperiodic_propagator_via_F}), we identify three types of time copy\dash summations in (\ref{eq:app_F_function_derivation}):
\begin{align}
& (1)^\pm_{\hphantom{x,y}}\is \sum_{j{}^{\,}\in{}^{\,}\mathbb{Z}}\frac{(\pm 1)^j}{4\pi^2(\ncoverline{\Delta}-j\e{4})^2}\,, \label{eq:app_sum1}\\
& (2)^\pm_{x,y}\is \sum_{j{}^{\,}\in{}^{\,}\mathbb{Z}}\frac{-(\pm 1)^j(\Delta\ncoverline{t}-j)}{4\pi\,d(|\vec{\ncoverline{x}}|,|\vec{\ncoverline{y}}|,\ncoverline{t}_x,\ncoverline{t}_y+j)}\,, \label{eq:app_sum2}\\
& (3)^\pm_{x,y}\is \sum_{j{}^{\,}\in{}^{\,}\mathbb{Z}}\frac{(\pm 1)^j\left(|\vec{\ncoverline{x}}|(\ncoverline{\Delta}-j\e{4})^2 + \e{\vec{\ncoverline{x}}}\cdot\vec{\ncoverline{y}}^{\,}(\ncoverline{\Delta} - j\e{4})^2 + 2|\vec{\ncoverline{x}}|(\e{\vec{\ncoverline{x}}}\cdot\vec{\ncoverline{y}}{}^{\,})^2-2|\vec{\ncoverline{x}}|^{\,}\vec{\ncoverline{y}}{}^{\,2}\right)}{4\pi (\ncoverline{\Delta} - j\e{4})^2\,d(|\vec{\ncoverline{x}}|,|\vec{\ncoverline{y}}|,\ncoverline{t}_x,\ncoverline{t}_y+j)}\, , \label{eq:app_sum3}
\end{align}
where $(2)^\pm_{x,y}$ and $(3)^\pm_{x,y}$ are not (explicitly) $x\leftrightarrow y$\dash symmetric and a factor of $\pi$ has been absorbed into these terms. The full propagator then reads

\begin{equation}
\label{eq:app_finite_T_propagator_via_sums}
\begin{aligned}
\Delta^\pm(x,y) & \istr \!\left[(1)^\pm + \varrho^2\left((2)^\pm_{x,y}\frac{\sin(2\pi\tau_x)}{c_1(r_x,\tau_x)} + (2)^\pm_{y,x}\frac{\sin(2\pi\tau_y)}{c_1(r_y,\tau_y)}\right. + \right. \\
&\phantom{\istr} \left. \left. \hphantom{\Big[(1)^\pm + \varrho^2\Big(}\, +\, (3)^\pm_{x,y}\frac{\sinh(2\pi r_x)}{c_1(r_x,\tau_x)} + (3)^\pm_{y,x}\frac{\sinh(2\pi r_y)}{c_1(r_y,\tau_y)}\right) \right]\frac{1}{\sqrt{\phi(x)\phi(y)}}\,.
\end{aligned}
\end{equation}
Here we dropped the barred notation, because performing the $j$\dash summation describes the transition from $\mathbb{R}^4$ to $\spt$.

Explicitly, we find for the periodic case, denoting similarly to above $\tau_x-\tau_y\is \Delta\tau$, $r_x-r_y\is\Delta r$, and $\vec{x}-\vec{y}\is\vec{\Delta}$:
\begin{align}
 (1)^+_{\hphantom{x,y}} & \is \frac{\sinh(2\pi|\vec{\Delta}|)}{4\pi|\vec{\Delta}|\,c_1(|\vec{\Delta}|,\Delta\tau)}\,, \label{eq:app_sum1_per} \\
 (2)^+_{x,y} & \is -\frac{\sinh(2\pi r_x)\sinh(2\pi r_y)\sin(2\pi\Delta\tau)}{8r_x r_y\,c_1(r_x+r_y,\Delta\tau)\,c_1(\Delta r,\Delta\tau)}, \label{eq:app_sum2_per} \\
 (3)^+_{x,y} & \is \frac{1}{16r_x}\left(\frac{\sinh(2\pi\Delta r)}{r_y\,c_1(\Delta r,\Delta\tau)} - \frac{\sinh(2\pi(r_x+r_y))}{r_y\,c_1(r_x+r_y,\Delta\tau)} + \frac{2\sinh(2\pi |\vec{\Delta}|)}{|\vec{\Delta}|\,c_1(|\vec{\Delta}|,\Delta\tau)}\right). \label{eq:app_sum3_per}
\end{align}
Plugging our results (\ref{eq:app_sum1_per}) - (\ref{eq:app_sum3_per}) into (\ref{eq:app_finite_T_propagator_via_sums}) yields the traceful (diagonal) part of the full massless, periodic propagator.\footnote{In the periodic propagator according to (\ref{eq:app_finite_T_propagator_via_sums}) we find one noteworthy simplification:
\begin{align*} & (2)^+_{x,y}\frac{\sin(2\pi\tau_x)}{c_1(r_x,\tau_x)} + (2)^+_{y,x}\frac{\sin(2\pi\tau_y)}{c_1(r_y,\tau_y)} \is \\ & \is \frac{\big(\cosh(2\pi r_x)\sin(2\pi\tau_y) - \cosh(2\pi r_y)\sin(2\pi\tau_x) + \sin(2\pi\Delta\tau)\big)\sinh(2\pi r_x)\sinh(2\pi r_y)\sin(2\pi\Delta\tau)}{8r_xr_y\,c_1(r_x,\tau_x)\,c_1(r_y\tau_y)\,c_1(r_x+r_y,\Delta\tau)\,c_1(\Delta r,\Delta\tau)}\,.\end{align*}}

For the anti\dash periodic propagator we proceed analogously. For numerical reasons, we split up the sum $(3)^-_{x,y}$ into three parts in doing so, $(3.1)^-_{x,y}\is \sum_{j{}^{\,}\in{}^{\,}\mathbb{Z}}\frac{(-1)^j\big(|\vec{\ncoverline{x}}| + \,\e{\vec{\ncoverline{x}}}^{\,}\cdot^{\,}\vec{\ncoverline{y}}\,{}^{\!}\big)}{4\pi \,d(|\vec{\ncoverline{x}}|,|\vec{\ncoverline{y}}|,\ncoverline{t}_x,\ncoverline{t}_y+j)}$, $(3.2)^-_{x,y}\is \sum_j \frac{(-1)^j|\vec{\ncoverline{x}}| \big(\e{\vec{\ncoverline{x}}}^{\,}\cdot^{\,}\vec{\ncoverline{y}}\,{}^{\!} \big)^2}{2\pi (\ncoverline{\Delta} - j\e{4})^2\,d(|\vec{\ncoverline{x}}|,|\vec{\ncoverline{y}}|,\ncoverline{t}_x,\ncoverline{t}_y+j)}$, and $(3.3)^-_{x,y} \is -\sum_j \frac{(-1)^j\,|\vec{\ncoverline{x}}|^{\,}\vec{\ncoverline{y}}{}^{\,2}}{2\pi (\ncoverline{\Delta}-j\e{4})^2\,d(|\vec{\ncoverline{x}}|,|\vec{\ncoverline{y}}|,\ncoverline{t}_x,\ncoverline{t}_y+j)}$. Also, we introduce the notation $c_2(z_1,z_2)\is \cosh^2(\pi z_1)-\cos^2(\pi z_2)$:
\begin{align}
& (1)^-_{\hphantom{x,y}}\is \frac{\sinh(\pi|\vec{\Delta}|)\cos(\pi\Delta\tau)}{2\pi|\vec{\Delta}|\,c_1(|\vec{\Delta}|,\Delta\tau)}\,, \label{eq:app_sum1_antiper} \\
& (2)^-_{x,y} \is -\frac{\big(\cosh(2\pi r_x) + \cosh(2\pi r_y) + \cos(2\pi\Delta\tau) +1\big)\sinh(\pi r_x)\sinh(\pi r_y)\sin(\pi\Delta\tau)}{16r_x r_y\, c_2(r_x+r_y,\Delta\tau)\,c_2(\Delta r,\Delta\tau)}\,, \label{eq:app_sum2_antiper} \\
& \begin{aligned}
(3)^-_{x,y} \is -\Bigg(&\frac{(1 - \e{\vec{x}}\cdot\e{\vec{y}})\sinh(\pi(r_x+r_y))}{(r_x+r_y)c_2(r_x+r_y,\Delta\tau)} + \frac{(1 + \e{\vec{x}}\cdot\e{\vec{y}})\sinh(\pi\Delta r)}{\Delta r\,c_2(\Delta r, \Delta\tau)} - \frac{2\sinh(\pi|\vec{\Delta}|)}{|\vec{\Delta}|\,c_2(|\vec{\Delta}|,\Delta\tau)}-  \\
& - (r_x+r_{y\,}\e{\vec{x}}\cdot\e{\vec{y}})\left(\frac{\sinh(\pi(r_x+r_y))}{r_y(r_x+r_y)\,c_1(r_x+r_y,\Delta\tau)} + \frac{\sinh(\pi|\Delta r|)}{c_1(\Delta r,\Delta\tau)}\right)\!\!\Bigg) \frac{\cos(\pi\Delta\tau)}{16r_x}\,.\end{aligned} \label{eq:app_sum3_antiper}
\end{align}
Using again (\ref{eq:app_finite_T_propagator_via_sums}) gives the massless, anti\dash periodic scalar propagator's diagonal part.\footnote{We again give the one important simplification:
\begin{align*}
& (2)^-_{x,y}\frac{\sin(2\pi\tau_x)}{c_1(r_x,\tau_x)} + (2)^-_{y,x}\frac{\sin(2\pi\tau_y)}{c_1(r_y,\tau_y)} \is \big(\cosh(2\pi r_x)\sin(2\pi\tau_y) - \cosh(2\pi r_y)\sin(2\pi\tau_x) + \sin(2\pi\Delta\tau)\big) \times \\ & \qquad\times \frac{\big(\cosh(2\pi r_x) + \cosh(2\pi r_y)+\cos(2\pi\Delta\tau)+1\big)\sinh(\pi r_x)\sinh(\pi r_y)\sin(\pi\Delta\tau)}{16r_xr_y\,c_1(r_x,\tau_x)\,c_1(r_y,\tau_y)\,c_2(r_x+r_y,\Delta\tau)\,c_2(\Delta r,\Delta\tau)}\,.
\end{align*}}

\newpage

\bibliographystyle{JHEP2}
\bibliography{refs}

\end{document}